 \documentstyle[12pt,dina4p,titlepage]{report}

 \newcommand{\la}{\langle}
 \newcommand{\ra}{\rangle}
 \newcommand{\r}{\right}
 \renewcommand{\l}{\left}
 \newcommand{\beq}{\begin{equation}}
 \newcommand{\eeq}{\end{equation}}
 \newcommand{\beqa}{\begin{eqnarray}}
 \newcommand{\eeqa}{\end{eqnarray}}
 \renewcommand{\d}[1]{\frac{\partial}{\partial #1}}
 \newcommand{\dd}[1]{\frac{\partial^2}{\partial #1^2}}
 \newcommand{\ab}[2]{\frac{\partial #1}{\partial #2}}
 
 \newcommand{\co}[1]{{{\cal C}^\infty _o (#1)}}
 
 \newcommand{\rf}[1]{(\ref{#1})}
 
 \newcommand{\D}[1]{{\cal D}(#1)}
 \newcommand{\Rn}{{{\bf R}^n}}
 \newcommand{\M}{{\cal M}}
 \newcommand{\Dp}[1]{{\cal D}'(#1)}
 \newcommand{\x}{(x_1,\xi_1;x_2,\xi_2)}
 \newcommand{\Ht}{{{\cal H}^t}}
 \newcommand{\lt}{\Lambda^{(2)}_t}
 \newcommand{\dm}{\int d\mu(\vec{k})}
 \newcommand{\ph}{\phi_{\vec{k}}}
 \newcommand{\ol}{\overline}
 \newcommand{\vk}{(\vec{k})}
 \newcommand{\A}{{\cal A}}
 \newtheorem{thm}{Theorem}[chapter]
 \newtheorem{lemma}[thm]{Lemma}
 \newtheorem{dfn}[thm]{Definition}
 \newtheorem{cor}[thm]{Corollary}
\begin{document}
\title{Adiabatic Vacua and Hadamard States for Scalar Quantum Fields
       on Curved Spacetime}
\author{\vspace*{3cm}\\
        Dissertation\\
        zur Erlangung des Doktorgrades\\
        des Fachbereichs Physik\\
        der Universit\"at Hamburg\\[3.5cm]
        vorgelegt von\\
        Wolfgang Junker\\
        aus Weilheim/Obb.\\[3cm]
        Hamburg\\
        1995
        }
\date{ }
\maketitle
\thispagestyle{empty}
hep-th/9507097\\
DESY 95-144
\vspace*{18cm}
\begin{tabbing}
  Sprecher des Fachbereichs Physik \qquad \= Prof.~Dr.~B. Kramer      \kill{}
  Gutachter der Dissertation: \> Prof.~Dr.~K. Fredenhagen      \\
                              \> Prof.~Dr.~D. Buchholz         \\[0.2cm]
  Gutachter der Disputation:  \> Prof.~Dr.~K. Fredenhagen      \\
                              \> Prof.~Dr.~G. Mack             \\[0.2cm]
  Datum der Disputation:      \> 13.~Juli~1995                  \\[0.2cm]
  Sprecher des Fachbereichs Physik \>                          \\
  und Vorsitzender des        \>                               \\
  Promotionsausschusses:      \>  Prof.~Dr.~B. Kramer
\end{tabbing}
%
\vspace*{\fill}
\begin{abstract}
  Quasifree states of a linear Klein-Gordon quantum field on globally
  hyperbolic spacetime manifolds are considered. Using techniques from
  the theory of pseudodifferential operators and wavefront sets on
  manifolds a criterion for a state to be an Hadamard state is
  developed. It is shown that ground- and KMS-states on certain
  static spacetimes and adiabatic vacuum states on Robertson-Walker
  spaces are Hadamard states. Finally, the problem of constructing
  Hadamard states on arbitrary curved spacetimes is solved in
  principle.\\[1cm]
  \language 1
 \centerline{\bf Zusammenfassung}\\
     \\
  Es werden quasifreie Zust{\"a}nde eines
  quantisierten linearen Klein-Gordon-Feldes auf global hyperbolischen
  Raumzeit-Mannigfaltigkeiten betrachtet. Unter Verwendung von
  Methoden aus der Theorie der Pseudodifferentialoperatoren und
  Wellenfrontenmengen auf Mannigfaltigkeiten wird ein Kriterium
  entwickelt, das es erm{\"o}glicht, die Hadamard-Eigenschaft von
  Zust{\"a}nden nachzuweisen. Es wird gezeigt, da{\ss} Grund- und
  KMS-Zust{\"a}nde auf gewissen statischen Raumzeiten und adiabatische
  Vakuumzust{\"a}nde auf Robertson-Walker-Raumzeiten
  Hadamard-Zu\-st{\"a}n\-de sind. Zu guter Letzt wird ein
  Konstruktionsverfahren f{\"u}r Hadamard-Zu\-st{\"a}n\-de auf beliebigen
  gekr{\"u}mmten Raumzeiten angegeben.
\end{abstract}
%
\language 0
\tableofcontents
\chapter{Introduction}
Hawking's remarkable discovery twenty years ago \cite{Hawking75} that the
gravitational collapse of a star to a black hole is accompanied by a
thermal radiation of quantum fields at the temperature $T=1/8\pi M$
(in natural units, $M$ the mass of the black hole) was an essential
stimulus for the investigation of quantum field theory in curved
spacetime. This is a semiclassical theory in so far as the
gravitational field is assumed to be given as a classical background
field wherein the quantized matter fields act dynamically (for good
introductions into the subject see \cite{Fulling89,Kay87,Wald84}).
The
backreaction of the matter fields on the gravitational field, i.e.~the
spacetime metric $g_{\mu\nu}$, occurs via the semiclassical Einstein
equations
\[ R_{\mu\nu}-\frac{1}{2}g_{\mu\nu} R= 8\pi \la
\hat{T}_{\mu\nu}(x)\ra_\omega, \]
where $\la \hat{T}_{\mu\nu}\ra_\omega$ is the expectation value of the
energy-momentum tensor of the matter fields in a quantum state
$\omega$, $R_{\mu\nu}$ the Ricci tensor of $g_{\mu\nu}$ and
$R=g^{\mu\nu}R_{\mu\nu}$ the Ricci scalar. As an effect of the
backreaction the thermal radiation draws its energy from the
gravitational field, which causes the black hole to evaporate. In a
fundamental theory unifying gravity and the other forces of nature it
is expected that also the metric field must be ``quantized'' (in some
still undefined sense), nevertheless the semiclassical approximation
should have validity in a large range up to the Planck scale.\\
One must admit that the expected effects of quantum fields in a
gravitational background are very small: a black hole of ten solar
masses emitts radiation at a temperature of about $10^{-8}$K (which is
of course concealed behind the cosmic background radiation), so only
small primordial black holes or effects in the early epoch of the
universe can be expected to yield observable phenomena.\\
On the other hand, the study of quantum field theory in curved
spacetimes has already given deep insights into the interplay between
quantum theory and spacetime geometry, and we think this source of
physical understanding is not yet exhausted. One example is the
thermodynamic behaviour of black holes which had already been noted
but not understood physically before Hawking's discovery. Another is
the question which r\^{o}le the global features of a spacetime play in
the setting of quantum field theory. What are the local states in the
absence of Poincar\'{e} symmetry? It is this question which we will
mainly pursue in this work. \\
To this end, we consider the model of a linear scalar quantum field
coupled to a gravitational background.
Such a system possesses infinitely many
degrees of freedom. Hence there are unitarily inequivalent
representations of the canonical commutation relations and one has to
pick out the physically interesting ones. For quantum fields in
Minkowski space one uses the Poincar\'{e} group to specify a ``vacuum
state'': it is the (usually unique) Poincar\'{e}-invariant state such
that the translations are unitarily implementable in its
GNS-Hilbertspace with a positive energy-momentum operator (spectral
condition). It is the state of lowest energy, the particle states are
local excitations of this vacuum state.\\
If we take external gravitational fields into consideration in which
the quantum fields propagate one has to replace the Minkowski space by
a curved spacetime manifold that does in general not possess any
symmetries and we cannot expect that there exists a preferred vacuum
state (this was the first time realized in \cite{Fulling73}). The
particle concept itself becomes dubious and it is probably sensible
only asymptotically in regions of weak gravitational fields. In this
situation the algebraic approach to quantum field theory
\cite{HK64,Haag92} is the appropriate frame to describe the physical
situation. Here, one starts with a net of local algebras that contain
the local observables and fields. A concrete physical realization of
the system is determined by a state, i.e.~a positive, linear,
normalized functional on the observable algebra. It gives the
expectation values of all physical quantities. Each state fixes via
the GNS-construction a representation of the observable algebra on a
Hilbertspace, in which it acts as a cyclic (``vacuum''-like) vector. The
``folium'' of a state is the set of all vector- and density
matrix-states in its GNS-Hilbertspace. However, not every
(mathematical) state is physically realizable. One has to select among
the many unitarily inequivalent representations those which describe
physically sensible situations. It was observed by Haag, Narnhofer and
Stein \cite{HNS84,Haag92} that in Minkowski space all physically
realizable states -- when restricted to a finite, contractible
spacetime region -- belong to the same unique (primary) folium (namely
that of the vacuum state), and they adopted this for quantum field
theory on curved spacetime as a hypothesis, called the ``principle of
local definiteness''. There is then a unique von Neumann algebra (with
trivial center) which is the weak closure of all observable algebras
belonging to finite, contractible regions, and the physical states are
the normal states of this algebra. Two states are called locally
quasiequivalent if they are normal w.r.t.~this algebra, i.e.~if they
determine the same local folium. \\
Let us summarize: To make quantum field theory on curved spacetime as
well defined a theory as on Minkowski space we have to specify --
besides commutation relations and field equations -- the folium (=
quasiequivalence class) of physical states, but this folium cannot be
defined as easily as in Minkowski space by appealing to the vacuum
state, which does in general not exist in the presence of a
gravitational field.\\
A second constraint on the choice of physical states is the
requirement that the expectation value of the energy-momentum tensor
$T_{\mu\nu}$ in a state can be regularized to become a tensorfield at
a single spacetime point. This is necessary in order that the
semiclassical Einstein equations make sense. In Minkowski space it is
achieved by Wick ordering the fields w.r.t.~the Minkowski vacuum. In
curved spacetimes one has to choose the states such that certain
regularization procedures can be applied.\\
We are concerned here with the linear Klein-Gordon quantum field on
globally hyperbolic spacetimes. The corresponding algebra (with the
canonical commutation relations imposed) was constructed by Dimock
\cite{Dimock80}. There are three classes of (quasifree) states on this
algebra which have been considered so far. These are\\
1.) the set $S_1$ of adiabatic vacua\\
2.) the set $S_2$ of quasifree Hadamard states\\
3.) the set $S_3$ of quasifree states possessing a scaling limit at
each spacetime point that satisfies the spectral condition in the
tangent space.\\
Let us shortly comment on their relation (the classes $S_1$ and $S_2$
will be discussed in detail in sections 3.2 and 3.4). $S_3$ was
introduced by Haag, Narnhofer and Stein \cite{HNS84} (see also
\cite{FH87}), they show that KMS-states of this type have the correct
Hawking temperature in spacetimes with horizons. Although the scaling
limit assumption works even for interacting theories, the states in
$S_3$ are in general not quasiequivalent \cite{HNS84}, i.e.~the
condition is not restrictive enough. $S_2$ is a (proper) subset of $S_3$.
It is known to be a local quasiequivalence class \cite{Verch94}, but
is only well defined for linear fields. It is in general very
difficult, to construct states of $S_2$, up to now examples have only
been known for
spacetimes with certain symmetries. In contrast, $S_1$ is a large
class of explicitly constructed states on cosmological spacetime
models. It was shown by L{\"u}ders and Roberts \cite{LR90} that $S_1$
forms a local folium of states. The exact relation between $S_1$ and
$S_2$ has not been investigated so far. It is the aim of this work to
show that in fact all adiabatic vacua are Hadamard states
(i.e.~$S_1=S_2$ for linear Klein-Gordon fields on Robertson-Walker
spaces) and to combine the physical ideas behind the Hadamard states
and the adiabatic vacuum states to produce a construction scheme for
Hadamard states on arbitrarily curved globally hyperbolic spacetimes.\\
We have organized this work as follows:\\
Chapter 2 contains a mathematical introduction into the techniques of
pseudodifferential operators and wavefront sets. Since these do
presently not belong to the daily applied equipment of the theoretical
physicist we found it useful to give a short collection of (nearly)
all mathematical material that is needed to understand our arguments
in chapter 3. Most of the material is taken from
\cite{Taylor81,Horm71,DH72}, only Theorem \ref{theorem1.17} and
Corollary \ref{cor1.18} are not contained in the literature in this
form. Although we tried to concentrate the facts which are spread over
the literature as much as possible we did not retain from giving some
of the (easier) proofs for the pedagogical benefit of the
reader. Nevertheless, whoever is not interested in the mathematical
side of the physical problems may skip chapter 2 altogether without
hesitation and may look up a theorem or a definition when arriving at
a point in chapter 3 where we refer back to it. \\
Chapter 3 contains the physical part of this work. In section 3.1 we
present the basic setting, namely the theory of the scalar
Klein-Gordon quantum field in globally hyperbolic spacetimes. In
section 3.2 we review the definition of Hadamard states, its physical
relevance and the new results due to Radzikowski \cite{Radz92} that
give a local characterization of Hadamard states by the wavefront set
of their two-point distributions. This is the main technical
ingredient which we use in section 3.3 to prove (Theorem
\ref{theorem3.9}) that certain quasifree states of the Klein-Gordon
quantum field are Hadamard states. As a first application we show that
ground- and KMS-states on ultrastatic spacetimes are Hadamard states
(Corollary \ref{cor3.10} and Corollary \ref{cor3.10a}). This can
easily be generalized  to static spacetimes possessing a timelike
Killing vectorfield the norm of which is bounded from below by a
positive constant. In section 3.4 we introduce -- following
\cite{LR90} -- the adiabatic vacuum states on cosmological spacetime
models (Robertson-Walker spaces) and apply the techniques developed
so far to prove that these all are Hadamard states (Theorem
\ref{theorem3.15}). So the essential physical statement is (Corollary
\ref{cor3.16}) that on these spacetimes Hadamard states and adiabatic
vacua define the same local folium of states.\\
The problem of constructing physical states on arbitrarily curved
globally hyperbolic spacetimes has found much attention in the
literature, but no solution. In section 3.5 we give a counterexample
to certain methods proposed in the literature by choosing a typical
state from such a sample and showing (Theorem \ref{theorem3.19}) that it does
not lie in the local folium of Hadamard states. At last, in section
3.6, we solve the problem ``in principle'' by presenting an iteration
procedure which produces an asymptotic expansion of
Hadamard states in
close analogy to the adiabatic vacua (Theorem \ref{theorem3.22}).
\chapter{Mathematical preliminaries}
\section{Pseudodifferential operators on manifolds}
Following H\"ormander \cite{Horm71} and Taylor \cite{Taylor81}
we first introduce pseudodifferential operators on $\Rn$ and
later, by localization, on curved manifolds.
The idea is to generalize linear differential operators with variable
coefficients. If $p(x,D) :=
\sum_{|\alpha|\le k}a_\alpha (x) D_x^\alpha$ is a differential operator
with coefficients depending on $x\in {\bf R}^n$, then
\beqa
p(x,D)u(x) &=& \frac{1}{(2\pi)^{n/2}}p(x,D)\int_{{\bf R}^n}d^n\xi \;
           \hat{u}(\xi)e^{ix\xi} \nonumber\\
       &=&\frac{1}{(2\pi)^{n/2}}\int_{{\bf R}^n}d^n\xi\;p(x,\xi)
           \hat{u}(\xi)e^{ix\xi}, \label{1.1}
\eeqa
with $u\in \D{{\bf R}^n},\; \hat{u}$ its Fourier transform,
$p(x,\xi)= \sum_{|\alpha|\le k}a_\alpha
(x)\xi_\alpha$ (for basic definitions and notation see the appendix).\\
If we replace in this expression the polynomial $p(x,\xi)$ by
suitable functions $a(x,\xi)$, called {\it symbols}, we obtain
a pseudodifferential operator. We first introduce the relevant
symbol classes:
\begin{dfn}
Let $X$ be an open subset of $\Rn$. Let $m,\rho,\delta$ be real
numbers with $0\le\delta,\rho\le 1$.\\
Then we define the {\rm \bf symbols of order $m$ and type $\rho,
\delta$} to be the set
\beqa
S^m_{\rho,\delta} (X\times \Rn)&:=& \l\{a\in {\cal C}^\infty
(X\times \Rn);\;\mbox{for every compact $K\subset X$ and for all
multiindices}\; \alpha, \beta \r.\nonumber\\
& & \exists C_{\alpha, \beta, K}\in
{\bf R}:
 \l|D_x^\beta D_\xi^\alpha a(x,\xi)\r|\leq C_{\alpha, \beta, K}
  (1+|\xi|)^{m-\rho|\alpha| +\delta|\beta|} \label{1.2}\\
& &\l. \mbox{for}\; x\in K,\;\xi \in \Rn\r\}. \nonumber
\eeqa
(We also simply write $S^m_{\rho,\delta}$ if no confusion is
possible).\\
$S^{-\infty}:=\bigcap_m S^m_{\rho,\delta} = \bigcap_m S^m_{1,0}.$
\end{dfn}
In the physical applications in the next chapter we only have to deal
with symbols (and pseudodifferential operators) of type
1,0. Nevertheless, we keep the discussion in this mathematical
introduction more general because it costs no more effort and makes
the comparison with the mathematical literature easier.
\begin{lemma}\label{lemma1.2}
Let $a\in S^m_{\rho,\delta}(X\times\Rn), b\in S^{m'}_{\rho',
\delta'}(X\times\Rn)$. Then \\
a) $ab\in S^{m+m'}_{\rho'',\delta''},$
where $\rho'':=\min(\rho,\rho'),\delta'':=
\max(\delta,\delta')$,\\
b) $D^\beta_xD^\alpha_\xi a \in S^{m-\rho|\alpha|+\delta|\beta|}
_{\rho,\delta}$, \\
c) If $|a(x,\xi)|\le C (1+|\xi|)^{-m},$ then $a(x,\xi)^{-1}\in
S^{-m}_{\rho,\delta}.$\\
d) $\chi\in \co{\Rn} \Rightarrow \chi a\in S^{-\infty}$. This implies
that a symbol changes only by a term in $S^{-\infty}$ if we modify it
in a compact domain in $\xi$.
\end{lemma}
The proof follows easily from the chain rule.\\
Before introducing pseudodifferential operators we give the notion of
the {\it \bf asymptotic expansion} of a symbol. It is an important
tool for the construction of certain pseudodifferential operators
(as used e.g.~in Theorem \ref{theorem1.11}) and will have an essential
application in section \ref{section2.7}:
\begin{lemma}\label{lemma1.3}
Suppose $a_j\in S^{m_j}_{\rho,\delta}(X\times\Rn), m_j\downarrow -\infty
\;(j=0,1,2,\ldots)$.\\
Then there exists $a \in S^{m_o}_{\rho,\delta}(X\times\Rn)$ such that
for all $N>0$:
\beq
a-\sum_{j=0}^{N-1} a_j \in S^{m_N}_{\rho,\delta}(X\times\Rn).\label{1.2a}
\eeq
The function $a$ is uniquely determined modulo
$S^{-\infty}(X\times\Rn)$.\\
If \rf{1.2a} holds we write $a \sim \sum_{j\geq 0} a_j.$
\end{lemma}
{\sc Proof}:\\
i) Pick compact sets $K_i$ with $K_1\subset K_2 \subset \ldots \to
X$.\\
Take $\psi \in {\cal C}^\infty(\Rn)$ with $\psi (\xi)=\l\{
\begin{array}{cc}0,& |\xi|\leq 1 \\1, & |\xi|\geq 2
\end{array}
\r.,\; 0\leq \psi(\xi)\leq 1. $\\
Choose $\epsilon_j, j=0,1,2,\ldots,$ such that $1\geq \epsilon_o>\epsilon_1
\ldots >\epsilon_j \to 0\; (j\to \infty)$ and set
\beq
a(x,\xi):=\sum_{j=0}^\infty \psi(\epsilon_j \xi)a_j(x,\xi). \label{1.2b}
\eeq
Note that
\beq
\psi(\epsilon\xi)=\l\{
\begin{array}{cc} 0, &|\xi|\leq 1/\epsilon\\
                  1, &|\xi|\geq 2/\epsilon
\end{array} \r. , \label{1.2c}
\eeq
hence, for $|\xi|\leq 1/\epsilon$ or $ |\xi|\geq2/\epsilon,\; D^\alpha_\xi
\psi(\epsilon\xi)=0\; (\alpha \neq 0)$, whereas, if $1/\epsilon <|\xi|<2/
\epsilon$ for $0<\epsilon\leq 1$, then $\epsilon \leq 2/|\xi|\leq
4(1+|\xi|)^{-1}$ and, since $\psi$ varies only over a compact
interval,
\[
|D^\alpha_\xi\psi(\epsilon\xi)|\leq C_\alpha \epsilon^{|\alpha|}\leq
C_\alpha' (1+|\xi|)^{-|\alpha|}
\]
for $\alpha \neq 0$ ($C_\alpha$ independent of $\epsilon$).\\
Thus, $\psi(\epsilon\xi)\in S^0_{1,0}\subset S^0_{\rho,\delta}$
for $ 0<\epsilon\leq 1$.\\
So, by Lemma \ref{lemma1.2},
for any $i, j$ and any $0<\epsilon\leq 1$ we have for all $x \in K_i$:
\beqa
|D^\alpha_\xi D^\beta_x \psi(\epsilon\xi) a_j(x,\xi)| &\leq&
C_{i,j,\alpha,\beta}(1+|\xi|)^{m_j-\rho|\alpha|+\delta|\beta|} \nonumber\\
&\leq& \l[C_{i,j,\alpha,\beta}(1+|\xi|)^{-1}\r]
       (1+|\xi|)^{m_j+1-\rho|\alpha|+\delta|\beta|}. \label{1.2d}
\eeqa
Now determine $\epsilon_j >0$ such that $C_{i,j,\alpha,\beta}\epsilon_j
\leq 2^{-j}$ for $|\alpha| +|\beta|+i \leq j$.\\
If $(1+|\xi|)^{-1}\geq \epsilon_j$, then $|\xi|\leq 1+|\xi|\leq 1/
\epsilon_j$ and, by \rf{1.2c}, $\psi(\epsilon_j\xi)=0$.\\
On the other hand, if $ (1+|\xi|)^{-1}\leq \epsilon_j$, we have from
\rf{1.2d}
\beq
|D^\alpha_\xi D^\beta_x (\psi(\epsilon_j\xi) a_j(x,\xi))| \leq
2^{-j}(1+|\xi|)^{m_j+1-\rho|\alpha|+\delta|\beta|} \label{1.2e}
\eeq
for $x\in K_i$ and $|\alpha|+|\beta|+i\leq j$.
The sum in \rf{1.2b} is finite for any $(x,\xi)$, and since
$\sum_{j=0}^\infty |D^\alpha_\xi D^\beta_x (\psi(\epsilon_j \xi)
a_j(x,\xi))| < \infty$ by \rf{1.2e}, we have $a(x,\xi)\in {\cal C}
^\infty (X\times \Rn)$.\\
ii) Given $\alpha,\beta$ and $x\in K_i$ we choose $k$ so large that
$|\alpha|+|\beta|+i\leq k$ and $m_k+1\leq m_o$ and write
\begin{eqnarray*}
|D^\alpha_\xi D^\beta_x a(x,\xi)|&\leq& \l|D^\alpha_\xi D^\beta_x
\sum_{j=0}^{k-1}\psi(\epsilon_j\xi)a_j(x,\xi)\r| +
\l|D^\alpha_\xi D^\beta_x\sum_{j=k}^\infty\psi(\epsilon_j\xi)a_j(x,\xi)
\r| \\
&\stackrel{\rf{1.2d}\rf{1.2e}}{\leq}&
C_{\alpha,\beta,i}(1+|\xi|)^{m_o-\rho|\alpha|
+\delta|\beta|}+ \underbrace{\l(\sum_{j=k}^\infty2^{-j}\r)}_{\leq 1}
(1+|\xi|)^{m_o-\rho|\alpha|+\delta|\beta|} \\
&\leq& C_{\alpha,\beta,i}' (1+|\xi|)^{m_o-\rho|\alpha|+\delta|\beta|}.
\end{eqnarray*}
Since this holds for any $\alpha, \beta$, we have $a(x,\xi)\in S^{m_o}
_{\rho,\delta}$.\\
iii) Similarly, for any $N\in{\bf N}$ we obtain
$\sum_{j=N}^\infty \psi(\epsilon_j\xi)a_j \in S^{m_N}_{\rho,\delta}$
and $\sum_{j=0}^{N-1}(\psi(\epsilon_j\xi)-1)a_j \in S^{-\infty}$, since
$\psi(\epsilon_j\xi)-1 =0$ for $j\leq N-1$ and $|\xi|\geq 2/\epsilon
_{N-1}$ (equ.~\rf{1.2c}), and hence
\begin{eqnarray*}
a-\sum_{j=0}^{N-1}a_j &=& \sum_{j=0}^{N-1}(\psi(\epsilon_j\xi)-1)
a_j + \sum_{j=N}^\infty\psi(\epsilon_j\xi)a_j \\
&\in& S^{m_N}_{\rho,\delta}\quad\mbox{for any $N$},
\end{eqnarray*}
which proves \rf{1.2a}.\\
iv) Let $b \in S^{m_o}_{\rho,\delta}(X\times\Rn)$ be another symbol
with property \rf{1.2a}, then
\[
a-b=\l(a-\sum_{j=0}^{N-1}a_j\r)-\l(b-\sum_{j=0}^{N-1}a_j\r)\in S^{m_N}_{
\rho,\delta},
\]
for all $N>0$, hence $a=b\: (mod\: S^{-\infty})$.\hfill${\bf\Box}$
\begin{dfn}
If $a(x,\xi)\in S^m_{\rho,\delta}(X\times\Rn)$ the operator
\beq
Au(x):=\frac{1}{(2\pi)^{n/2}}\int_\Rn e^{ix\eta}a(x,\eta)\hat{u}(\eta)
\;d^n\eta, \label{1.3}
\eeq
$u\in {\cal S}(\Rn), x\in X$, is said to belong to $L^m_{\rho,\delta}
(X)$, the {\rm\bf pseudodifferential operators} of type $\rho, \delta$
(we drop the $X$ and write $L^m_{\rho,\delta}$ when the context is clear).
\end{dfn}
{\bf Examples}:
\begin{enumerate}
\item Let $A:=\sum_{|\alpha|\leq m}a_{\alpha}(x) D^\alpha_x, a_\alpha\in
{\cal C}^\infty(X)$, be a linear partial differential operator of
order $m$ on $X\subset \Rn$.
Then $A\in L^m_{1,0}(X)$.
\item Let $(Au)(x):=\int_\Rn K(x,y)u(y)\;d^n y$ with $K\in {\cal C}^\infty
(X\times X)$ such that $supp K(x,\cdot)$ is compact for each $x\in X$.\\
Then $A\in L^{-\infty}(X)$. [The symbol is $a(x,\xi) = \int_\Rn K(x,y)
e^{i(y-x)\xi}\; d^n y$, i.e.~the Fourier transform of a function with
compact support and hence rapidly decreasing in $\xi$.]\\
In particular the convolution $u\mapsto u\ast\varphi :=\int_\Rn \varphi(x-y)
u(y)\;d^n y $ with $\varphi\in {\cal C}^\infty_o(X)$ is such a
pseudodifferential operator. We remark that in this case $supp\: Au =
supp\: u + supp \:\varphi$.
\item Let $A$ denote the multiplication with $\chi\in{\cal C}^\infty_o(X):
\quad (Au)(x):=\chi(x)u(x)$.\\
Then $A\in L^0_{1,0}(X)$.
\end{enumerate}
The properties of $A$ as an operator are given in the following
theorem.
\begin{thm}\label{theorem1.4}
a) $A\in L^m_{\rho,\delta}(X)$ is a continuous operator $A: \D{X}\to
{\cal C}^\infty(X)$.\\
b) If $\delta <1$, then the map can be extended to a continuous map
$A: {\cal E}'(X) \to {\cal D}'(X)$.
\end{thm}
{\sc Proof}:\\
a) Let $a\in S^m_{\rho,\delta}(X\times\Rn), u\in \D{X}$. Since
$\hat{u}\in{\cal S}(X)$ the integral \\$Au(x)= (1/2\pi)^{n/2}
\int a(x,\xi)\hat{u}(\xi)
e^{ix\xi}\;d^n \xi$ is absolutely convergent, and one can differentiate
under the integral sign, obtaining always absolutely convergent integrals
because of \rf{1.2}.\\
b) We show that the functional $v \mapsto \la Au,v\ra, \;v\in\D{X},$
is well defined for $u\in {\cal E}'(X)$: Formally
\beqa
\la Au,v\ra &=& \frac{1}{(2\pi)^{n/2}}
\int\!\!\!\int v(x) a(x,\xi)\hat{u}(\xi)e^{ix\xi}\;d^n\xi\,
d^n x \nonumber\\
&=& \frac{1}{(2\pi)^{n/2}}
\int a_v(\xi)\hat{u}(\xi)\;d^n \xi \label{1.4}\\
\mbox{with}\quad a_v(\xi)&:=& \int v(x)a(x,\xi)e^{ix\xi}\;d^n x.
 \nonumber
\eeqa
Since the Fourier transform $\hat{u}$ of a distribution $u$ with compact
support can at most grow polynomially \rf{1.4} is well defined for any
$u\in {\cal E}'(X)$ if $a_v(\xi)$ is rapidly decreasing:\\
Integration by parts yields for $\eta,\xi\in \Rn$:
\begin{eqnarray*}
\l|\eta^\alpha\int v(x)a(x,\xi)e^{ix\eta}\;d^n x\r| &=&
\l|\int D^\alpha_x (v(x)a(x,\xi))e^{ix\eta}\;d^n x\r| \\
&\leq& \int \l|D^\alpha_x(v(x)a(x,\xi))\r|\;d^n x \\
&\leq& C_\alpha (1+|\xi|)^{m+\delta |\alpha|}\quad\mbox{by
  \rf{1.2}}\\
\Rightarrow \l|\int v(x)a(x,\xi)e^{ix\eta}\;d^n x\r| &\leq&
C_N (1+|\xi|)^{m+\delta N}(1+|\eta|)^{-N},\\
\mbox{hence, for $\xi=\eta$:}\quad |a_v(\xi)|&\leq&
C_N (1+|\xi|)^{m+(\delta -1)N}.
\end{eqnarray*}
If $\delta <1$, this implies the rapid decrease of $a_v(\xi)$.\hfill
{$\bf\Box$}
 \\  \\
Since $A$ is continuous it is given by a distribution kernel $K_A\in
{\cal D}'(X\times X)$ via $\la Au,v\ra = \la K_A,u\otimes v\ra$ for
$u,v \in \D{X}$ (Schwartz' kernel theorem).
\begin{lemma}\label{lemma1.5}
a) If $A\in L^m_{\rho,\delta}(X)$
for $\rho >0$, then $K_A$ is ${\cal C}^\infty$
off the diagonal in $X\times X$.\\
b) If $A \in L^{-\infty}(X)$, then $K_A$ is smooth everywhere in
$X\times X$ (which is the converse of Example 2) above).
\end{lemma}
{\sc Proof}:\\
Let $u,v\in \D{X}$. We have
\begin{eqnarray*}
\la K_A,u\otimes v\ra &=& \la Au,v\ra = \int v(x)Au(x)\;d^n x = \\
&=& \frac{1}{(2\pi)^{n/2}}\int\!\!\!\int a(x,\xi)e^{ix\xi}v(x)\hat{u}(\xi)
\;d^n\xi \,d^n x = \\
&=&
\frac{1}{(2\pi)^{n}}\int\!\!\!\int\!\!\!\int a(x,\xi)e^{i(x-y)\xi}
v(x)u(y)\;d^ny\, d^n\xi\,d^nx.
\end{eqnarray*}
Thus
\[
K_A(x,y)=\frac{1}{(2\pi)^n}\int a(x,\xi)e^{i(x-y)\xi}\;d^n\xi
\]
as a distribution integral. In this sense it follows after
partial integration
\[
\l|(x-y)^\alpha K_A(x,y)\r| = \l|\frac{1}{(2\pi)^n}\int e^{i(x-y)\xi}
D^\alpha_\xi a(x,\xi)\;d^n\xi\r|,
\]
which converges absolutely by \rf{1.2} if $m-\rho|\alpha|<-n$.
Furthermore,
\[
\l|D^\beta_x D^\gamma_y (x-y)^\alpha K_A(x,y)\r| = \l|\frac{1}{(2\pi)^n}
\int D^\beta_x D^\gamma_y e ^{i(x-y)\xi} D^\alpha_\xi a(x,\xi)\;d^n\xi\r|
\]
converges absolutely if $m-\rho |\alpha| +\delta|\beta|< - n-|\beta|-|\gamma|
$. Since $|\alpha|$ can be made arbitrarily large this shows that
$K_A(x,y)$ is smooth for $x \neq y$. If $m$ can be chosen arbitrarily
negative, absolute convergence holds even for $|\alpha| =0$, i.e.~$K_A$
is smooth everywhere.
\hfill${\bf\Box}$\\
\begin{dfn}
A distribution $u\in \Dp{X\times X}$ is called {\rm \bf properly supported}
if $\{(x,y)\in supp\: u;\; x\in K\;\mbox{or}\;
 y\in K\}$ is compact for every compact
set $K\subset X$, i.e.~$supp\: u$ has compact intersection with
$K\times X$ and $X\times K$. Equivalently, $u$ is properly supported
if for each compact $K\subset X$ there exists a compact $K'
\subset X$ such that
\beqa
supp\: f\subset K &\Rightarrow& supp\: uf \subset K' \label{1.5a}\\
\mbox{and $f=0$ on $K'$} &\Rightarrow& uf =0\;\mbox{on}\; K.\nonumber
\eeqa
A pseudodifferential operator $A$ is called properly supported if its
distribution kernel $K_A$ is properly supported.
\end{dfn}
{}From Theorem~\ref{theorem1.4} it follows that, if $A$ is properly
supported, then $A:{\cal C}^\infty(X)\to {\cal C}^\infty(X)$ and
$A:{\cal D}'(X)\to {\cal D}'(X)$ (for $\delta < 1$).\\
Every pseudodifferential operator A can be written as the sum
of one with a ${\cal C}^\infty$-kernel and one which is properly
supported. To this end, we choose a $\chi \in {\cal C}^\infty (X\times X)$
such that $\chi =1$ in a neighborhood of the diagonal and $\chi$ is
properly supported. By writing $K_A = (1-\chi)K_A + \chi K_A$ we
obtain the desired splitting: $(1-\chi)K_A$ is smooth because of
Lemma \ref{lemma1.5}a) and $\chi K_A$ is properly supported since $\chi$
is so.\\
Thus, in the following, we can always assume that pseudodifferential
operators are properly supported. This has the advantage that for
these there exists a simple calculus:\\
Two pseudodifferential operators $A\in L^m_{\rho,\delta}$ and $B\in
 L^{m'}_{\rho,\delta}$ (with symbols $a(x,\xi)$ resp.~$b(x,\xi)$, say)
can be composed yielding again a
pseudodifferential operator (in $L^{m+m'}_{\rho,\delta}$) with symbol
\beq
\sigma_{AB}(x,\xi) \sim \sum_{\alpha\geq 0}\frac{i^{|\alpha|}}{\alpha
  !} (D_{\xi}^\alpha a(x,\xi))(D_x^\alpha b(x,\xi)), \label{1.5b}
\eeq
and there exists the adjoint $A^t$ resp.~$A^*\in L^m_{\rho,\delta}$ of a
pseudodifferential operator $A\in L^m_{\rho,\delta}$
(defined by $(Au,v)=(u,A^t v)$
resp.~$(Au,v)=(u,A^* v)$ for the scalar product in $L^2_{\bf
  R}(\Rn,\sqrt{h}d^nx)$ resp.~$L^2_{\bf C}(\Rn,\sqrt{h}d^nx)$, $h\in
{\cal C}^\infty (\Rn)$) with symbol
\beqa
\sigma_{A^t}(x,\xi) &\sim& \sum_{\alpha\geq 0}\frac{i^{|\alpha|}}{\alpha
  !h(x)^{1/2}} D_\xi^\alpha D_x^\alpha
\l[h(x)^{1/2}a(x,-\xi)\r]\nonumber\\
\sigma_{A^*}(x,\xi) &\sim& \sum_{\alpha\geq 0}\frac{i^{|\alpha|}}{\alpha
  !h(x)^{1/2}} D_\xi^\alpha D_x^\alpha
\l[h(x)^{1/2}\ol{a(x,\xi)}\r]. \label{1.5c}
\eeqa
Note in particular that
\beqa
\sigma_{AB}(x,\xi)-a(x,\xi) b(x,\xi) &\in&
S_{\rho,\delta}^{m+m'-(\rho-\delta)} \nonumber\\
\sigma_{A^t}(x,\xi)-a(x,-\xi) &\in&
S^{m-(\rho-\delta)}_{\rho,\delta} \label{1.5d}\\
\sigma_{A^*}(x,\xi)-\ol{a(x,\xi)} &\in&
S^{m-(\rho-\delta)}_{\rho,\delta}.\nonumber
\eeqa
For details see \cite[section II\S 4]{Taylor81} or \cite[section 2.1]
{Horm71}.\\
For us, the most important aspect is the effect of a change of
variables on a properly supported pseudodifferential operator.
It will allow to give these operators a well defined meaning on a
curved manifold.\\
Let $X$ and $Y$ be open regions in $\Rn$ and $\kappa:X\to Y$ a
diffeomorphism. Let $A\in L^m_{\rho,\delta}(X)$ with symbol
$a(x,\xi)$ and set
\[ \tilde{A} u :=(A(u\circ \kappa))\circ\kappa^{-1}, \quad u\in
{\cal C}^\infty_o(Y),
\]
so $\tilde{A}: {\cal C}^\infty_o(Y)\to {\cal C}^\infty(Y)$.\\
The following main theorem shows that $\tilde{A}$ is also a
pseudodifferential operator and gives the transformation law for the
symbol:
\begin{thm}\label{theorem1.6}
If $A\in L^m_{\rho,\delta}(X)$ is properly supported and if $\rho >1/2$
and $\rho +\delta\geq 1$, then $\tilde{A} \in L^m_{\rho,\delta}(Y)$
with symbol
\beq
\tilde{a}(\kappa(x),\xi)\sim\sum_{\alpha \geq 0}\frac{1}{\alpha!}
\varphi_\alpha(x,\xi)D^\alpha_\xi a(x,{^t\kappa'}(x)\xi),\label{1.6}
\eeq
where $\varphi_\alpha (x,\xi) := D^\alpha_y \l.\exp{i\la(\kappa(y)-\kappa(x)
-\kappa'(x)(y-x)),\xi\ra}\r|_{x=y}$ is a polynomial in $\xi$ of
degree $\leq |\alpha|/2$, in particular
\begin{eqnarray*}
\varphi_o(x,\xi)&=&1,\quad\varphi_\alpha(x,\xi)=0\quad\mbox{for}\;|\alpha|=1,\\
\varphi_\alpha (x,\xi) &=&i D^\alpha_x \la\kappa(x),\xi\ra \quad\mbox{for}\;
|\alpha|=2.
\end{eqnarray*}
Here $\kappa'$ denotes the Jacobian $\frac{D\kappa(x)}{Dx}$ of
$\kappa$ and $^t\kappa'$ its transpose.
\end{thm}
The proof uses as a main technical tool a new integral representation
for pseudodifferential operators, which we do not want to introduce
here. Therefore we refer the interested reader to \cite[II\S 5]{Taylor81}.\\
Up to now we have considered pseudodifferential operators on $\Rn$
as determined modulo operators in $L^{-\infty}$ and symbols in $S^{-\infty}$.
Formula \rf{1.6} however suggests a different point of view if we are
on a manifold:
The terms of index $\alpha\neq 0$ in the sum of equ.~\rf{1.6} are of order
$\leq m-\rho|\alpha|+|\alpha|/2 =m-|\alpha|(\rho -\frac{1}{2})\leq m-
(2\rho-1) <m$ for $\rho > 1/2$ and $\rho+\delta\geq 1$.
We can define equivalence classes $L^m_{\rho,\delta}(X)/L^{m-(2\rho-1)}
_{\rho,\delta}(X)$ of pseudodifferential operators, i.e.~we consider
two operators as equivalent if they differ by an operator of order
$\leq m-(2\rho-1)$.
\begin{dfn}
If $A\in L^m_{\rho,\delta}(X)$ we define the {\rm\bf principal symbol}
of $A$ to be a member in the corresponding equivalence class
$S^m_{\rho,\delta}(X\times \Rn)/S^{m-(2\rho-1)}_{\rho,\delta}(X\times
\Rn)$.
\end{dfn}
The decisive point now is the following: From equ.~\rf{1.6} we observe
that if $\tilde{A}$ is obtained from $A$ (having symbol $a(x,\xi)$) by
a change $\kappa$ of coordinates as discussed above, then a principal
symbol of $\tilde{A}$ is given by
\beq
a(\kappa^{-1}(x),{^t\kappa'}(x)\xi),\label{1.6a}
\eeq
i.e.~the principal symbol is a well defined function on the cotangent bundle
$T^\ast {\cal M}$ of a manifold ${\cal M}$.\\
In a similar way we obtain from \rf{1.5d}
the principal symbol of the adjoint $A^t$ resp.~$ A^*$ as $a(x,-\xi)$
resp.~$\ol{a(x,\xi)}$ and if $b(x,\xi)$ is a principal symbol of $B$ then
$a(x,\xi)b(x,\xi)$ is a principal symbol of $AB$. (This, by the way,
means that pseudodifferential operators commute in highest order.)\\
The fact that the principal symbols form a well defined $\ast$-algebra
of functions on the cotangent bundle $T^\ast{\cal M}$ makes
pseudodifferential operators such a useful tool for analysis on curved
manifolds. We are now naturally led to define:
\begin{dfn}
Let $\M$ be a ${\cal C}^\infty$ paracompact manifold of dimension $n$.\\
For $\rho>1/2$ and $\rho+\delta\geq 1$ we define {\bf $ L^m_{\rho,\delta}
(\M)$} to be the space of continuous linear operators
$A:{\cal C}^\infty_o(\M)\to{\cal C}^\infty(\M)$ with the property that
for each diffeomorphism $\kappa$ of a coordinate patch
$X_\kappa\subset\M $ to an open set $\kappa X_\kappa\subset\Rn$ we
have $A_\kappa \in L^m_{\rho,\delta}(\kappa X_\kappa)$, where
$A_\kappa u :=(A(u\circ\kappa))\circ\kappa^{-1}$ for $u\in {\cal C}
^\infty_o(\kappa X_\kappa)$.
\end{dfn}

It is sufficient to require that this condition is verified for a
covering of $\M$ by coordinate patches if in addition we require that
$K_A$ is smooth off the diagonal in $\M\times\M$. It is also
equivalent to the following condition: if $x^1,\ldots,x^n$ are local
coordinates in an open coordinate patch $X\subset\M$ and if
$v\in {\cal C}^\infty_o(X)$, then
\beq
e^{-i x\xi}A\l(ve^{ix\xi}\r)\in S^m_{\rho,\delta}(X\times \Rn),\label{1.7}
\eeq
where $\xi\in\Rn,\; x\xi:=x^1\xi_1+\ldots+ x^n\xi_n$. \\
Of course, the theorems and lemmata proven above remain valid locally
on a  manifold.
\section{Wavefront sets of distributions}\label{section1.2}
The notion of the wavefront set of a distribution has been introduced
by H\"ormander \cite{Horm71}. It will be the main tool to
characterize two-point functions of quasifree states of a quantum
field, as will be explained in the next chapter. Therefore we have
to introduce this concept, its connection with
pseudodifferential operators and the calculus related to it.
\begin{dfn}
Let $X\subset \M$ be a coordinate patch with coordinates $(x,\xi)$ of
$T^\ast\M$.\\
If $u\in{\cal D}'(X)$ the {\rm\bf wavefront set} $WF(u)$ is the set
\beq
WF(u):=\bigcap_{\begin{array}{c}
A\in L^0_{1,0}\\ Au\in {\cal C}^\infty
\end{array}}
char\: A \label{1.13}
\eeq
where
\beq
char\:A :=\{(x,\xi)\in T^\ast X\setminus\{0\}; \liminf_{t\to\infty}
|a(x,t\xi)|=0\} \label{1.14}
\eeq
is the {\rm\bf characteristic set} of a properly supported
pseudodifferential operator $A$ with principal symbol $a(x,\xi)$
(the choice of principal symbol is irrelevant in the definition).
\end{dfn}
The most important properties of wavefront sets are the following: \\
1) Let $u\in\Dp{X}.\;
\forall \varphi\in {\cal C}^\infty_o(X):\;WF(\varphi u)\subset
WF(u)$, and $(x_o,\xi_o)\in WF(u)\Leftrightarrow (x_o,\xi_o)\in
WF(\varphi u)$ when $\varphi(x_o)\not= 0$.  \\
This shows that the wavefront set is a local object depending only on
arbitrarily small neighborhoods of points $x_o\in X$.\\
2) We observed in the last section that a principal symbol of a
pseudodifferential operator transforms covariantly under
diffeomorphisms. Therefore by the Definition \rf{1.13} $WF(u)$
is a well defined subset of the cotangential bundle $T^\ast\M$ of a
manifold $\M$, i.e. if $\kappa:X\to Y$ is a diffeomorphism between
open coordinate patches $X,Y\subset\Rn$ of a manifold $\M$,
$u\in{\cal D}'(Y)$ and $\tilde{u}\in{\cal D}'(X)$ the distribution
with $\tilde{u}(f):=u(f\circ\kappa^{-1})$ for $f\in{\cal D}(X)$ then
\[ WF(\tilde{u}) =\kappa_\ast WF(u):=\{(\kappa^{-1}
(x),{^t \kappa'}(x)\xi);\;(x,\xi)\in WF(u)\}.
\]
Because of property 1) we can define the wavefront set of a
distribution on a manifold just by localization on coordinate patches.\\
3) $WF(u)$ is a closed cone in $T^\ast\M\setminus\{0\}$, i.e.
$(x,\xi)\in WF(u)$ implies $(x,t\xi)\in WF(u)$ for all $t>0$.\\
4) The wavefront set is a refinement of the notion of singular support
of a distribution in the following sense:
\begin{thm}\label{theorem1.9}
Let $\pi:T^\ast\M\to\M$ denote the projection of $T^\ast\M$ onto its
base space. Then
\beq
\pi (WF(u)) = singsupp\:u. \label{1.15}
\eeq
In particular, the wavefront set is empty if u is smooth.
\end{thm}
{\sc Proof}:\\
i) $x_o \not\in singsupp\:u \Rightarrow \exists \varphi\in {\cal C}^\infty
_o(X),\;\varphi=1$ near $x_o$ such that $\varphi u\in {\cal C}^\infty_o(X).$
\\ Clearly $(x_o,\xi)\not\in char\:\varphi \supset WF(u)$ for any $\xi
\not=0$, hence $\pi(WF(u))\subset singsupp\:u$.\\
ii) $x_o \not\in \pi(WF(u)) \Rightarrow \forall \xi\not=0\;\exists A\in
L^0_{1,0}$ such that $(x_o,\xi)\not\in char\: A$ and $Au\in {\cal C}^\infty.
$\\ Thus there exist finitely many $A_j\in L^0_{1,0}$ such that
$A_ju\in{\cal C}^\infty$ and each $(x_o,\xi),\;|\xi|=1$, is noncharacteristic
for some $A_j$.\\
Let $B:=\sum_j{}A_j^* A_j\in L^0_{1,0}.$ Then $B$ is elliptic near $x_o$
and $Bu\in {\cal C}^\infty$, so -- by the following Theorem~\ref{theorem1.11}
-- $u$ is ${\cal C}^\infty$ near $x_o$, which shows that
$singsupp\:u\subset \pi(WF(u))$.\hfill{$\bf\Box$}\\
5)
\beq
WF(u_1+u_2)\subset WF(u_1)\cup WF(u_2) \label{1.15a}
\eeq
6) Differential operators $P$ are in an essential way characterized by
their locality property, namely that always $supp\:(Pu)\subset supp\:u$.
For pseudodifferential operators this is in general no longer true
(see Example 2) in the previous section), but there is a remnant of
it, the so-called {\it\bf pseudolocal property}:
\begin{thm}[Theorem 1.6 of \cite{Taylor81}]\label{theorem1.10}
Let $A\in L^m_{\rho,\delta}(\M)$ for $\rho>0$ and $u\in{\cal D}'(\M)$.
Then
\beq
WF(Au)\subset WF(u). \label{1.16}
\eeq
\end{thm}
An important application of pseudodifferential operators is the
treatment
of elliptic (differential) equations. One result in this direction
which
we need later on is easy to state and very instructive to prove:
\begin{dfn}
An operator $A\in L^m_{\rho,\delta}(X)$ is {\rm\bf elliptic of order
  $m$} if on each compact $K\subset X$ there are constants $C_K$ and
$R$ such that for its symbol
\beq
|a(x,\xi)|\geq C_K|\xi|^m \quad\mbox{for}\;
x\in K,\;|\xi|>R.\label{1.17}
\eeq
Because of the transformation law \rf{1.6a} of the principal symbol of
$A$ under diffeomorphisms the property of ellipticity is invariant under
diffeomorphisms, and we define a pseudodifferential operator on a
manifold $\M$ to be elliptic if it is so in any local chart.
\end{dfn}
{\bf Example}:\\
The Laplace-Beltrami operator on a Riemannian manifold is an elliptic
(pseudo-)differential operator of order 2.\\
The next theorem states that such operators can be inverted ``up to
$L^{-\infty}$''. This is done by the construction of so-called parametrices.
\begin{dfn}
Let $A$ be a properly supported pseudodifferential operator on a
manifold
$\M$. If $Q$ is a continuous mapping ${\cal C}^\infty_o(\M) \to
{\cal C}^\infty(\M)$ such that i) $QA=I+R_1$, ii) $AQ=I+R_2$ or
iii) $QA=AQ=I+R_3$, where $R_i$ have smooth kernels and $I$ is the
identity operator, then we call $Q$ a i) {\rm\bf left}, ii) {\rm\bf right}
or iii) {\rm\bf two-sided parametrix} of $A$. (For a two-sided parametrix
we often simply say ``parametrix''.)
\end{dfn}
\begin{thm}\label{theorem1.11}
If $A \in L^m_{\rho,\delta}(X),\;\rho>\delta$, is a properly supported
elliptic pseudodifferential operator of order $m$, then there is a
properly supported parametrix $Q\in L^{-m}_{\rho,\delta}(X)$ which is
elliptic of order $-m$. It follows that
\beq
WF(u)=WF(Au)\quad\mbox{for}\;u\in {\cal D}'(X). \label{1.18}
\eeq
\end{thm}
{\sc Proof}:\\
If $Q$ has the desired properties we have
\[ u= (I-QA)u+QAu=Ru+QAu, \]
where $R$ has smooth kernel, and, using Theorem~\ref{theorem1.10},
\[ WF(u)\subset WF(QAu)\subset WF(Au)\subset WF(u), \]
from which \rf{1.18} follows.\\
Therefore it remains to construct $Q$ which we do by successive
approximations.
If $a(x,\xi)$ is the symbol of $A$ we set
\[q_o(x,\xi):=\chi(x,\xi) a(x,\xi)^{-1} \]
where $\chi=0$ in a neighborhood of the zeros of $a$ and $\chi$ is
identically 1 for large $\xi$ (there, $a$ cannot have zeros because of
condition \rf{1.17}). Hence, because of Lemma \ref{lemma1.2}c)
\[ q_o\in S^{-m}_{\rho,\delta}(X\times \Rn).\]
Let $Q_o \in L^{-m}_{\rho,\delta}(X)$ with symbol $q_o$, then
$Q_oA$ has symbol $\chi(x,\xi)+r(x,\xi)$ with $r(x,\xi)\in S^{-(\rho
-\delta)}_{\rho,\delta}(X\times\Rn),$ and since $\chi -1 \in
S^{-\infty}$ we have
\[ Q_oA = I+R, \quad R\in L^{-(\rho-\delta)}_{\rho,\delta}(X). \]
Now we define $E\in L^0_{\rho,\delta}(X)$ to have the asymptotic
expansion
\[ E\sim I-R+R^2-R^3+\ldots \]
(in the sense of Lemma \ref{lemma1.3}), then
\[(EQ_o)A=I+K_1,\quad K_1\in L^{-\infty}.\]
Consequently, $Q:=EQ_o\in L^{-m}_{\rho,\delta}$ is a left parametrix
of $A$.\\
Similarly we can construct a right parametrix $\tilde{Q}$ of $A$,
 namely, with
\[ AQ_o=I+\tilde{R}, \quad \tilde{R}\in L^{-(\rho-\delta)}_{\rho,\delta},
\]
take
\[\tilde{E}\sim I-\tilde{R}+\tilde{R}^2-\tilde{R}^3+\ldots,\]
hence
\[A(Q_0\tilde{E})=I+K_2,\quad K_2\in L^{-\infty},\]
and let $\tilde{Q}:=Q_o\tilde{E}\in L^{-m}_{\rho,\delta}(X).$\\
Therefore we have
\begin{eqnarray*}
QA\tilde{Q}&=&(I+K_1)\tilde{Q}=\tilde{Q}+K_1\tilde{Q}\quad\mbox{and}\\
QA\tilde{Q} &=& Q(I+K_2)=Q+QK_2,\quad\mbox{hence}\\
Q-\tilde{Q} &=& K_1\tilde{Q}-QK_2\in L^{-\infty}(X),
\end{eqnarray*}
i.e. $Q=\tilde{Q}\;mod\:L^{-\infty}(X)$ is a two-sided parametrix.\\
If $q$ is a principal symbol of $Q$, then $QA=\chi$ is a principal
symbol of $QA$. Since $a\in S^m_{\rho,\delta}$ we have for $x\in
K\subset X$ and large $\xi$
\[|q(x,\xi)|=|\chi(x,\xi)a(x,\xi)^{-1}|\geq C_K(1+|\xi|)^{-m}, \]
i.e. $Q$ is elliptic of order $-m$. \hfill{$\bf\Box$}\\
\\
\rf{1.18} is a special property of elliptic operators. For hyperbolic
operators (e.g. the Klein-Gordon operator, which plays a prominent
r\^{o}le in this work) the behaviour of wavefront sets is more complicated.
It is determined by the following important theorem on the
{\it\bf propagation of singularities} which we will repeatedly apply
in connection with the Klein-Gordon operator.
\begin{thm}[Theorem 6.1.1. of \cite{DH72}]\label{theorem1.11a}
Let $A\in L^m_{1,0}(\M)$ be a properly supported pseu\-do\-dif\-ferential
operator with real principal symbol $a$ which is homogeneous of
degree $m$.\\
If $u\in {\cal D}'(\M)$ and $Au=f$ it follows that
\beq
WF(u)\setminus WF(f) \subset a^{-1}(0)\setminus\{0\} \label{1.18a}
\eeq
and $WF(u)\setminus WF(f)$ is invariant under the Hamiltonian
vector field $H_a$ given by
\beq
H_a:=\sum_{i=1}^n \l[\ab{a(x,\xi)}{x^i}\d{\xi_i}-\ab{a(x,\xi)}
{\xi_i}\d{x^i}\r] \label{1.18b}
\eeq
in local coordinates.
\end{thm}
{\bf Remarks}:\\
This theorem contains as special cases two properties which we have
already learnt of:
\begin{enumerate}
\item If $a^{-1}(0)\setminus\{0\} =\emptyset$, i.e. $A$ is elliptic, then
$WF(u)\subset WF(Au) \subset WF(u)$ by \rf{1.18a} and \rf{1.16},
which was the result of Theorem~\ref{theorem1.11}.
\item If $Au$ is a smooth function, i.e. $WF(Au)=\emptyset$, then by
\rf{1.18a} $WF(u)\subset a^{-1}(0)\setminus\{0\}$,
which is already contained in
the Definition \rf{1.13} of the wavefront set. In this definition
one can replace $A\in L^0_{1,0}$ by $A\in L^q_{1,0}$ for any $q\in
{\bf R}$, because for a pseudodifferential operator $A\in L^0_{1,0}$
and an elliptic one $B\in L^q_{1,0}$, $char\:(BA)=char\:(A)$ and
$Au\in {\cal C}^\infty \Leftrightarrow BAu\in {\cal C}^\infty$.
\end{enumerate}
A distribution $u\in {\cal D}'(X)$ need not possess a Fourier
transform. But if we localize $u$ with a function $\varphi\in {\cal
  C}^\infty_o(X)$ with compact support, then $\widehat{\varphi u}$
is an analytic function which grows at most polynomially (see e.g.
\cite[Theorem IX.12]{RSII}). If $supp\:\varphi \cap singsupp\: u
=\emptyset$, then $\widehat{\varphi u}$ even decays rapidly (i.e.
faster than any inverse power). The next theorem gives a complete
characterization of the wavefront set of a distribution via the
decay properties of its Fourier transform.
\begin{thm}[Theorem 1.8. of \cite{Taylor81}]\label{theorem1.12}
\begin{eqnarray*}
(x_o,\xi_o)\not\in WF(u) &\Leftrightarrow &
\exists \varphi \in {\cal C}^\infty_o,\;\varphi(x_o)\not=0,
 \exists \;\mbox{conic neighborhood $\Gamma$ of $\xi_o$ s.th.}\;
\forall N\in{\bf N}: \\
& & |(\widehat{\varphi u})(\xi)|\leq C_N(1+|\xi|)^{-N} \quad\mbox{for all}\;
\xi \in \Gamma.
\end{eqnarray*}
\end{thm}
{\bf Example}:\\
For the $\delta$-distribution in ${\cal D}'(\Rn)$ one easily
calculates from the criterion of the theorem
\[ WF(\delta) = \{(0,\xi);\; \xi \in\Rn\setminus \{0\}\}.
\]
The last theorem allows to derive the elegant calculus of wavefront
sets which is taken without proofs from \cite[section 2.5]{Horm71}
 in those parts that we need for our purposes. From Theorem
 \ref{theorem1.12} one again sees that the wavefront set is a local
concept, therefore in the following we restrict ourselves to open
sets $X\subset \Rn$, but all results are equally valid on manifolds.\\
First we state a lemma which is a simple consequence of Theorem
\ref{theorem1.12}, but which we need later on:
\begin{lemma}\label{lemma1.12a}
a) Let $u\in \Dp{X}$ and $\bar{u}$ its complex conjugate.
Then
\beq
WF(\bar{u}) =\{(x,\xi)\in T^*X;\;(x,-\xi)\in WF(u)\}=:-WF(u). \label{1.19a}
\eeq
b) Let $v\in \Dp{X\times X}$ with $v(\bar{f_1},\bar{f_2})=\overline
{v(f_2,f_1)}$, i.e.~$v(x_1,x_2) =\overline{v(x_2,x_1)}$.\\
Then $WF(v) = - WF(v)$.
\end{lemma}
\begin{dfn}
The {\rm\bf product} of two distributions $u_1, u_2$ -- if it exists --
is defined by convolution of Fourier transforms as the distribution
$v\in {\cal D}'(X)$ such that $\forall x\in X \;\exists f\in {\cal D}(X)$
with $f=1$ near $x$ such that for all $\xi \in \Rn$:
\beq
\widehat{f^2 v}(\xi) = \frac{1}{(2\pi)^{n/2}}\int_\Rn \widehat{fu_1}
(\eta)\widehat{fu_2}(\xi-\eta)\; d^n\eta. \label{1.20}
\eeq
\end{dfn}
For a detailed discussion of this definition see \cite[IX.10]{RSII},
heuristically it means that for a test function $h\in \D{X}$
\beq
v(h)=\int_X u_1(x)u_2(x)h(x)\;d^nx. \label{1.21}
\eeq
According to the remark before Theorem~\ref{theorem1.12} $\widehat{fu_1}(\eta)$
 and $\widehat{fu_2}(\xi-\eta)$ are still polynomially bounded, if $\eta$
and $\xi -\eta$ are contained in the resp.~wavefront sets, but decay
rapidly, if not (because of Theorem~\ref{theorem1.12}). Therefore,
for the integral \rf{1.20} to converge (for all $\xi$) we would expect
 that it is enough that either $\eta$ or $-\eta$ is not contained in
 the resp.~wavefront set for all directions $\eta$. This is stated in the
next theorem.
\begin{thm}\label{theorem1.13}
Let $u_1, u_2\in \Dp{X}$. Suppose that for all $x\in X$:
\[ (x,0)\not\in WF(u_1)\oplus WF(u_2):=\{(x,\xi_1+\xi_2);\;(x,\xi_i)\in
WF(u_i),i=1,2\}.
\]
Then the product $u_1u_2$ exists and
\beq
WF(u_1u_2)\subset WF(u_1)\cup WF(u_2)\cup [WF(u_1)\oplus WF(u_2)].\label{1.19}
\eeq
\end{thm}
Now we consider the restriction of distributions to submanifolds.\\
Let $\M$ be an $n$-dimensional manifold and $\Sigma$ an
$(n-1)$-dim. hypersurface (i.e. there exists a ${\cal C}^\infty$-imbedding
$\varphi: \Sigma\to\M$) with normal bundle
\beq
N_\varphi
:=\{(\varphi(y),\xi)\in T^\ast\M;\;y\in\Sigma,\varphi_\ast(\xi):=
{^t\varphi'}(y)\xi =0\} \label{1.22}
\eeq
 in local coordinates (notation as in Theorem~\ref{theorem1.6}).\\
Let $u\in \Dp{\M}$.\\
If we could, we would naturally define the restriction $u_\Sigma\in
\Dp{\Sigma}$ of $ u$ to $\Sigma$ by
\beqa
u_\Sigma: f&\mapsto& (u\cdot(f\delta_\Sigma))({\bf 1})\quad\mbox{for}\;
f\in \co{\Sigma} \label{1.23}\\
\mbox{where}\;f\delta_\Sigma:g &\mapsto&\int_\Sigma fg\;d^{n-1}\sigma
\quad\mbox{for}\;g\in\co{\M},\nonumber
\eeqa
$d^{n-1}\sigma$ is the volume element of $\Sigma$, and ${\bf 1}\in \co{\M}$
is a function equal to 1 in a neighborhood of $\{\varphi (y);\;y\in
supp\:f\}.$\\
In \rf{1.23} we must multiply the distributions $u$ and $\delta_\Sigma$.
If $\Sigma$ is locally given by $t=0$, then $\delta_\Sigma = \delta(t)$
is the delta-function in the $t$-variable whose wavefront set is
according to the example above
\begin{eqnarray*}
WF(\delta(t)) &=&\{(0,\vec{y};\lambda,\vec{0})\in T^\ast\M;\;
\vec{y}\in \Sigma,\lambda \not= 0\}\quad\mbox{or invariantly}\\
WF(\delta_\Sigma) &=&\{(x,\xi)\in T^\ast\M;\;x=\varphi(y) \;\mbox{for
some}\;y\in \Sigma, \varphi_\ast(\xi)=0,\xi\not= 0\}\\
&=& N_\varphi\setminus \{0\},
\end{eqnarray*}
hence, by Theorem~\ref{theorem1.13}, we would expect that formula
\rf{1.23} holds for $WF(u)\cap N_\varphi =\emptyset.$
This is indeed confirmed by
\begin{thm}\label{theorem1.14}
Let $\M,\; \Sigma$ be as above.\\
Let $u\in \Dp{\M}$ with $WF(u)\cap N_\varphi =\emptyset.$\\
Then the restriction $u_\Sigma$ of $u$ defined by \rf{1.23} is a well
defined distribution in $\Dp{\Sigma}$ and
\[ WF(u_\Sigma)\subset \varphi_\ast WF(u):=
\{(y,\varphi_\ast(\xi))\in T^\ast\Sigma;\;(\varphi(y),\xi)\in WF(u)\}.
\]
\end{thm}
The last formula means that the wavefront set becomes projected tangentially
onto the surface. The theorem can easily be generalized to arbitrary
submanifolds of $\M$.\\  \\
If we have two properly supported distributions $K_1\in \Dp{X_1\times X_2},
\; K_2\in\Dp{X_2\times X_3},\; X_i\subset {\bf R}^{n_i},\; i=1,2,3,$ and
if we can compose the corresponding continuous maps to a continuous
map
\[K=K_1\circ K_2:\co{X_3}\to \Dp{X_1}, \]
then its kernel distribution is given by
\beq
K(x_1,x_3)=\int_{X_2} K_1(x_1,x_2)K_2(x_2,x_3)\;dx_2. \label{1.24}
\eeq
In view of Theorem~\ref{theorem1.13} and the fact that $WF(K_2u)\subset
WF_{X_2}(K_2)$ for any $u\in \D{X_3}$ this is well defined if
\beq
WF'_{X_2}(K_1)\cap WF_{X_2}(K_2) =\emptyset, \label{1.25}
\eeq
where
\beqa
WF'(K_1)&:=&\{(x_1,\xi_1;x_2,-\xi_2)\in T^\ast X_1\times T^\ast X_2;\;
(x_1,\xi_1;x_2,\xi_2)\in WF(K_1)\} \nonumber\\
WF_{X_2}(K_2) &:=&\{(x_2,\xi_2)\in T^\ast X_2;\;(x_2,\xi_2;x_3,0)\in
WF(K_2)\;\mbox{for some}\; x_3\in X_3\} \label{1.26}\\
WF'_{X_2}(K_1) &:=& \{(x_2,\xi_2)\in T^\ast X_2;\;(x_1,0;x_2,\xi_2)\in
WF'(K_1)\;\mbox{for some}\; x_1\in X_1\}, \nonumber
\eeqa
and we have
\begin{thm}\label{theorem1.15}
If \rf{1.25} holds for two properly supported distributions $K_1\in
\Dp{X_1\times X_2}$ and $K_2\in \Dp{X_2\times X_3}$, then the
composition \rf{1.24} is a well defined distribution in $\Dp{X_1\times
X_3}$ and we have
\beq
WF'(K_1\circ K_2)\subset WF'(K_1)\circ WF'(K_2)\cup (WF_{X_1}(K_1)\times X_3)
\cup (X_1\times WF'_{X_3}(K_2)), \label{1.27}
\eeq
where $X_1$ and $X_3$ are shorthand for $X_1\times\{0\}$,
resp.~$X_3\times \{0\}$, and $WF'(K_1)\circ WF'(K_2)$ means the obvious
composition of the two sets.
\end{thm}
\section{The Laplace-Beltrami operator}
Now we introduce the Laplace-Beltrami operator on an $n$-dimensional
Riemannian manifold $(\Sigma,h)$ and cite a result to the effect
 that certain functions
of it (in particular the square-root) are pseudodifferential
operators.\\
If $h_{ij}$ is a positive definite Riemannian metric on a manifold
$\Sigma$ we define the {\it\bf Laplace-Beltrami operator} by
\beqa
^{(n)}\Delta_h u &:=& \nabla^i\nabla_i u = h^{ij}\nabla_i\nabla_j u
\nonumber\\
&=& \frac{1}{\sqrt{h}}\partial_i\l(\sqrt{h}h^{ij}\partial_j u\r)
\label{1.8}
\eeqa
for $u\in {\cal C}^\infty (\Sigma)$,
where $h^{ij}$ is the inverse matrix of $h_{ij}$, $\nabla_i$ is the
covariant derivative w.r.t.~$h_{ij}$, $h$ is the determinant of $h_{ij}$
and $\partial_i$ are the partial derivatives in some coordinate system
(we simply write $\Delta$ when no confusion is possible). It is a
positive symmetric operator w.r.t.~the natural scalar product
$(u_1,u_2):=\int_\Sigma u_1 u_2\,d^n\sigma$ on $\co{\Sigma}$
($d^n\sigma := \sqrt{h} d^nx$), since, using Stokes' theorem and the
fact that $h_{ij}$ is covariantly constant (i.e.~$\nabla_i h_{jk}=0$)
\beqa
(u_1,\Delta u_2)&=& \int_\Sigma u_1 h^{ij}\nabla_i \nabla_j u_2
\,d^n\sigma \nonumber\\
&=&\int_\Sigma h^{ij}(\nabla_iu_1)(\nabla_ju_2)\,d^n\sigma
 \label{1.9}
\eeqa
for $u_1,u_2\in\co{\Sigma}$.
A theorem due to Chernoff \cite{Chernoff73} shows that $\Delta$
(and also all its powers) are essentially selfadjoint on
$L^2(\Sigma,d^n\sigma)$ if $\Sigma$ is a geodesically complete
manifold:
\begin{thm}[Chernoff \cite{Chernoff73}]\label{theorem1.7}
Let $(\Sigma,h)$ be a complete $n$-dimensional Riemannian manifold
with Laplace-Beltrami operator $^{(n)}\Delta_h$ as given by
\rf{1.8} and measure $d^n\sigma$ as in \rf{1.9}.\\
Then for $\mu\geq 0$ the operator $-^{(n)}\Delta_h+\mu^2:
\co{\Sigma}\to L^2(\Sigma,d^n\sigma)$ and all its natural powers
are essentially selfadjoint.
\end{thm}
Taking the closure $\overline{-\Delta +\mu^2}$ we obtain the unique
selfadjoint extension of $-\Delta+\mu^2$ on $L^2(\Sigma,d^n\sigma)$
which we will denote again by $-\Delta +\mu^2$ for simplicity. It
is also a positive operator and we can form the square-root
$(-\Delta+\mu^2)^{1/2}$ yielding a well defined positive selfadjoint
operator on $L^2(\Sigma,d^n\sigma)$, which is even strictly positive
(i.e.~has no eigenvalue zero) and hence invertible if $\mu>0$.\\
On a compact manifold there is a functional calculus for certain
pseudodifferential operators (in particular for the square-root of the
Laplace-Beltrami operator) due to Seeley \cite{Seeley68}
and Taylor \cite{Taylor81} which we will have the opportunity to use
in section~\ref{section2.4}:
\begin{thm}\label{theorem1.8}
a) Let $A\in L^1_{1,0}(\M)$ be an elliptic positive selfadjoint
operator on a compact manifold $\M$ with real principal symbol
$a(x,\xi)$ which is homogeneous of degree $\leq 1$.\\
Let $p(\lambda)\in S^m_{\rho,0}({\bf R})$ be a Borel function with
$1/2<\rho\leq 1$.\\
Then $p(A)\in L^m_{\rho,1-\rho}(\M)$ with principal symbol
$p(a(x,\xi))$.\\
b) $A:=(-\Delta+\mu^2)^{1/2}$ ($\mu\geq 0$) on a compact manifold $\M$
is a pseudodifferential operator in $L^1_{1,0}(\M)$ with principal
symbol $a(x,\xi)=(h^{ij}\xi_i\xi_j)^{1/2}$.
\end{thm}
\section{Parametrices of the Klein-Gordon operator}\label{section1.3}
In this section we consider the Klein-Gordon operator $P$ on a
4-dimensional globally hyperbolic and time-orientable spacetime-manifold
$(\M,g)$ and ask for the existence of parametrices of $P$ and their
wavefront sets. The following analysis has been carried out by
Radzikowski \cite{Radz92} with the methods provided by the general
theorems of Duistermaat and H\"ormander \cite{DH72}. \\
Let the Klein-Gordon operator $P$ be given by
\beqa
P &:=& \Box_g+\mu^2 \equiv g^{\mu\nu}\nabla_\mu\nabla_\nu +\mu^2
\nonumber\\
&=& \frac{1}{\sqrt{|g|}}\partial_\mu\l(\sqrt{|g|} g^{\mu\nu}\partial_\nu
\cdot \r) +\mu^2, \label{1.28}
\eeqa
where $\mu\geq 0$ represents the mass of a
scalar field, $\nabla_\mu$ is the covariant derivative defined by
the metric $g_{\mu\nu},\;g:=\det g_{\mu\nu},\;g^{\mu\nu}$ is the
inverse matrix of $g_{\mu\nu}$
and $\partial_\mu =\partial/\partial x^\mu$ the
partial derivative in local coordinates.\\
$P$ is properly supported (by criterion \rf{1.5a}) and has the real
principal symbol $p(x,\xi) :=g^{\mu\nu}(x)\xi_\mu\xi_\nu$, which
is homogeneous of degree 2. Put
\beq
N:=p^{-1}(0)\setminus \{0\} =\{(x,\xi)\in T^\ast\M\setminus\{0\};\;
p(x,\xi)=0\} \label{1.28a}
\eeq
and consider the Hamiltonian vector field \rf{1.18b} of $p$
\beqa
H_p &=& \sum_{\mu=0}^3\l[\ab{p(x,\xi)}{x^\mu}\d{\xi_\mu} -
 \ab{p(x,\xi)}{\xi_\mu}\d{x^\mu}\r]  \nonumber\\
&=& \ab{g^{\kappa\lambda}}{x^\mu}\xi_\kappa \xi_\lambda\d{\xi_\mu}-
 2 g^{\mu\lambda}\xi_\lambda \d{x^\mu}, \label{1.29}
\eeqa
which is tangential to $N$.
\begin{dfn}
The {\rm\bf bicharacteristic strips} of $P$ are the integral curves
of $H_p$ in $N$. The {\rm\bf bicharacteristic curves} of $P$ are the
projections of these strips onto $\M$. The {\rm\bf bicharacteristic
relation} of $P$ is the set
\[C:=\{(x_1,\xi_1;x_2,\xi_2)\in N\times N;\;(x_1,\xi_1) \;\mbox{and}\;
(x_2,\xi_2)\;\mbox{lie on the same bicharacteristic strip}\}.
\]
\end{dfn}
{}From \rf{1.29} one finds \cite{Radz92} that the bicharacteristic
curves of $P$ are the null geodesics of $(\M,g)$ and that
\beq
C=\{(x_1,\xi_1;x_2,\xi_2)\in N\times N;\;(x_1,\xi_1)\sim (x_2,\xi_2)\},
\label{1.30}
\eeq
where $(x_1,\xi_1)\sim(x_2,\xi_2)$ means that there is a null geodesic
$\gamma:\tau\mapsto x(\tau)$ such that $x(\tau_1)=x_1,\;x(\tau_2)
=x_2$ and $\xi_{1\nu}= \dot{x}^\mu(\tau_1) g_{\mu\nu}(x_1),\;
\xi_{2\nu} = \dot{x}^\mu(\tau_2) g_{\mu\nu}(x_2),$ i.e. $\xi_1^\mu,\;
\xi_2^\mu$ are tangent vectors to the null geodesic $\gamma$, and
hence parallel transports of each other along $\gamma$.
(For $x_1=x_2$ we mean by $(x_1,\xi_1)\sim(x_2,\xi_2)$ that
$\xi_1=\xi_2$.)\\
Let $\Delta_N$ be the diagonal of $N\times N$:
\beq
\Delta_N := \{(x_1,\xi_1;x_2,\xi_2)\in N\times N; \;x_1=x_2, \xi_1
=\xi_2\}. \label{1.31}
\eeq
Then $C\setminus \Delta_N$ decomposes into the open and disjoint sets
\beqa
C^+ &:=& \{(x_1,\xi_1;x_2,\xi_2)\in N\times N;\;(x_1,\xi_1)\succ
(x_2,\xi_2)\} \label{1.32}\\
&=& \{\x\in C;\;x_1^0>x_2^0\;\mbox{if $\xi_1^0>0$ or $x_1^0<x_2^0$ if
$\xi_1^0<0$}\} \nonumber\\
C^- &:=& \{\x \in N\times N;\; (x_1,\xi_1)\prec (x_2,\xi_2)\}
\label{1.33}\\
&=& \{\x \in C;\;x_1^0>x_2^0\;\mbox{if $\xi_1^0<0$ or $x_1^0<x_2^0$
if $\xi_1^0>0$}\},\nonumber
\eeqa
where $(x_1,\xi_1)\prec\:(\succ)\:(x_2,\xi_2)$ means $(x_1,\xi_1)\sim
(x_2,\xi_2)$ and $(x_1,\xi_1)$ comes after (before) $(x_2,\xi_2)$
w.r.t.~the time parameter of the bicharacteristic curve.\\
$C\setminus \Delta_N = C^+\dot{\cup}C^-$ is a special case of an orientation
of $C$:
\begin{dfn}
An {\rm\bf orientation of $C$} is a splitting of $C\setminus\Delta_N$
in a disjoint union of open subsets $C\setminus \Delta_N= C^1\dot{\cup}
C^2$ which are inverse relations (i.e. $\x\in C^1 \Leftrightarrow
(x_2,\xi_2;x_1,\xi_1)\in C^2$).
\end{dfn}
For the operator $P$ there are exactly 4 orientations, which we want
to calculate now:\\
$N$ has two connected components, namely
\begin{eqnarray}
N_+'&:=& \{(x,\xi)\in N;\; \xi^0>0\} \nonumber\\
N_-'&:=& \{(x,\xi)\in N;\;\xi^0<0\}. \label{1.33a}
\end{eqnarray}
We define
\[
\begin{array}{rlrl}
N_1^1 := & N_+'\dot{\cup}N_-' =N & N_1^2:= &\emptyset \\
N_2^1 :=& N_+'                   & N_2^2 :=&N_-'      \\
N_3^1 :=& N_-'                   & N_3^2 :=&N_+'      \\
N_4^1 :=&\emptyset               & N_4^2 :=& N,
\end{array}
\]
hence $N=N_i^1\dot{\cup}N_i^2$ for $i= 1,2,3,4.$\\
Let $B(x,\xi)$ denote the bicharacteristic strip through $(x,\xi)$.
Putting
\[ C^\pm(x,\xi):=C^\pm\cap (B(x,\xi)\times B(x,\xi))
\]
we obtain 4 orientations by
\beqa
C_i^1 &:=& \l(\bigcup_{N_i^1}C^+(x,\xi)\r)\cup\l(\bigcup_{N_i^2}
C^-(x,\xi)\r) \label{1.34}\\
C_i^2 &:=& \l(\bigcup_{N_i^1}C^-(x,\xi)\r)\cup\l(\bigcup_{N_i^2}
C^+(x,\xi)\r),\;i=1,2,3,4. \nonumber
\eeqa
In particular, $C_1^1=C^+=C_4^2,\; C_4^1=C^-=C_1^2,\; C_2^1=C_3^2,\;
C_3^1=C_2^2.$\\
It is easy to see that they are inverse relations and that
$C\setminus \Delta_N = C_i^1\dot{\cup}C_i^2$ for $i=1,2,3,4.$
The sets $C_1^1=C^+,\;C_2^1,\;C_3^1,\;C_4^1 =C^-$ are schematically
depicted in Figure \ref{figure1.1}.\\
\begin{figure}[p]
  \begin{center}
    \leavevmode
\begingroup\makeatletter
\def\x#1#2#3#4#5#6#7\relax{\def\x{#1#2#3#4#5#6}}%
\expandafter\x\fmtname xxxxxx\relax \def\y{splain}%
\ifx\x\y   
\gdef\SetFigFont#1#2#3{%
  \ifnum #1<17\tiny\else \ifnum #1<20\small\else
  \ifnum #1<24\normalsize\else \ifnum #1<29\large\else
  \ifnum #1<34\Large\else \ifnum #1<41\LARGE\else
     \huge\fi\fi\fi\fi\fi\fi
  \csname #3\endcsname}%
\else
\gdef\SetFigFont#1#2#3{\begingroup
  \count@#1\relax \ifnum 25<\count@\count@25\fi
  \def\x{\endgroup\@setsize\SetFigFont{#2pt}}%
  \expandafter\x
    \csname \romannumeral\the\count@ pt\expandafter\endcsname
    \csname @\romannumeral\the\count@ pt\endcsname
  \csname #3\endcsname}%
\fi
\endgroup
\begin{picture}(0,0)%
\includegraphics{figure1.1.pstex}%
\end{picture}%
\setlength{\unitlength}{0.012500in}%
\begingroup\makeatletter
\def\x#1#2#3#4#5#6#7\relax{\def\x{#1#2#3#4#5#6}}%
\expandafter\x\fmtname xxxxxx\relax \def\y{splain}%
\ifx\x\y   
\gdef\SetFigFont#1#2#3{%
  \ifnum #1<17\tiny\else \ifnum #1<20\small\else
  \ifnum #1<24\normalsize\else \ifnum #1<29\large\else
  \ifnum #1<34\Large\else \ifnum #1<41\LARGE\else
     \huge\fi\fi\fi\fi\fi\fi
  \csname #3\endcsname}%
\else
\gdef\SetFigFont#1#2#3{\begingroup
  \count@#1\relax \ifnum 25<\count@\count@25\fi
  \def\x{\endgroup\@setsize\SetFigFont{#2pt}}%
  \expandafter\x
    \csname \romannumeral\the\count@ pt\expandafter\endcsname
    \csname @\romannumeral\the\count@ pt\endcsname
  \csname #3\endcsname}%
\fi
\endgroup
\begin{picture}(420,597)(93,238)
\put(255,778){\makebox(0,0)[lb]{\smash{\SetFigFont{12}{14.4}{rm}$(x_1,\xi_1)$}}}
\put(258,643){\makebox(0,0)[lb]{\smash{\SetFigFont{12}{14.4}{rm}$(x_1,\xi_1)$}}}
\put(192,694){\makebox(0,0)[lb]{\smash{\SetFigFont{12}{14.4}{rm}$(x_2,\xi_2)$}}}
\put(354,580){\makebox(0,0)[lb]{\smash{\SetFigFont{12}{14.4}{rm}$C_2^1
=WF'(\Delta_R)\setminus\Delta^\ast$}}}
\put(357,280){\makebox(0,0)[lb]{\smash{\SetFigFont{12}{14.4}{rm}$C_4^1=WF'(\Delta_{\bar{F}})\setminus\Delta^\ast$}}}
\put(225,580){\makebox(0,0)[b]{\smash{\SetFigFont{12}{14.4}{rm}$C_1^1
=WF'(\Delta_F)\setminus\Delta^\ast$}}}
\put(225,280){\makebox(0,0)[b]{\smash{\SetFigFont{12}{14.4}{rm}$C_3^1
=WF'(\Delta_A)\setminus\Delta^\ast$}}}
\put(192,733){\makebox(0,0)[lb]{\smash{\SetFigFont{12}{14.4}{rm}$(x_2,\xi_2)$}}}
\end{picture}

  \end{center}
  \caption{The sets $C_1^1,\;C_2^1,\;C_3^1,\;C_4^1$ making up the
   orientations of $C$ and the wavefront sets of distinguished
   parametrices of the Klein-Gordon operator $P$.}
  \label{figure1.1}
\end{figure}
%
%
%
%
%
The importance of these orientations lies in the fact that they
determine uniquely (up to ${\cal C}^\infty$) distinguished
parametrices of $P$:
\begin{thm}[Theorem 6.5.3. of \cite{DH72}]\label{theorem1.16}
Let $P$ be the Klein-Gordon operator on a globally hyperbolic
manifold $(\M,g)$.\\
For every orientation $C\setminus \Delta_N = C_i^1\dot{\cup}C_i^2,\;
i=1,2,3,4,$ one can find parametrices $E_i^1$ and $E_i^2$ of $P$ with
\[
WF'(E_i^1)=\Delta^\ast \cup C_i^1,\;WF'(E_i^2)=\Delta^\ast\cup C_i^2
\]
where $\Delta^\ast$ is the diagonal in $(T^\ast\M\setminus\{0\})
\times (T^\ast\M\setminus\{0\})$.
Any right or left parametrix $E$ with $WF'(E)$ contained in
$\Delta^\ast\cup C_i^1$ resp.~$\Delta^\ast\cup C_i^2$ must be equal
to $E_i^1$ resp.~$E_i^2$ modulo a smooth kernel. ($WF'$ was defined
in \rf{1.26}.)
\end{thm}
Since $C_2^1$ and $C_3^1$ are nonempty only for $x_1$ in the future
of $x_2$, resp.~$x_1$ in the past of $x_2$, the corresponding
parametrices $E_2^1$ and $E_3^1$ must be (up to ${\cal C}^\infty$)
the retarded and advanced fundamental solutions $\Delta_R,\;\Delta_A$
of the inhomogeneous Klein-Gordon equation: $E^1_2=\Delta_R,\;
E^1_3=\Delta_A$.
Duistermaat and H\"ormander \cite[section 6.6]{DH72} gave $E_1^1$ and
$E_4^1$ the {\it names} Feynman and anti-Feynman propagator,
$\Delta_F$ and $\Delta_{\bar{F}}$, respectively. It was Radzikowski's
discovery \cite{Radz92} that these are indeed the
(anti-)Feynman-distributions (up to ${\cal C}^\infty$) of a
Hadamard state of a linear Klein-Gordon quantum field propagating on a
curved spacetime. We will take up this remark in section \ref{section2.3}.\\
In the next chapter we need the wavefront sets of differences of
distinguished parametrices. Therefore we state
\begin{thm}\label{theorem1.17}
The following holds\\
a) $WF'(\Delta_R-\Delta_A) = C, $\\
b) $WF'(\Delta_R-\Delta_F) = C\cap(N_-'\times N_-')=
    \{\x\in C;\;\xi^0_1<0,\xi^0_2<0\}$,\\
c) $WF'(\Delta_F-\Delta_A) = C\cap (N_+'\times N_+')=
   \{\x\in C;\; \xi^0_1>0,\xi^0_2>0\}$.
\end{thm}
{\sc Proof}:\\
a) From the singular support properties of $\Delta_R$ and $\Delta_A$
(Theorem~\ref{theorem1.16}, see Figure \ref{figure1.1}) we see that
\begin{eqnarray*}
\mbox{for}\;x_1^0>x_2^0: \quad WF'(\Delta_R-\Delta_A) &=& WF'(\Delta_R)
 = C^1_2,\\
\mbox{for}\;x_1^0<x_2^0:\quad WF'(\Delta_R-\Delta_A) &=& WF'(\Delta_A)
 = C_3^1.
\end{eqnarray*}
To determine $WF'(\Delta_R-\Delta_A)$ on the diagonal $x_1=x_2$
we use that
\[ P(\Delta_R-\Delta_A) = 0\:(mod\:{\cal C}^\infty) = (\Delta_R-\Delta_A)P,
\]
so by Theorem~\ref{theorem1.11a}
the singular directions are parallelly transported
along the bicharacteristic curves, hence
\[\Delta_N\subset WF'(\Delta_R-\Delta_A)
\]
and therefore
\[ WF'(\Delta_R-\Delta_A)=C^1_2\cup C^1_3\cup\Delta_N =C. \]
b) Again, by Theorem~\ref{theorem1.16} and Figure \ref{figure1.1}
we have that
\begin{eqnarray*}
\mbox{for}\;x_1^0<x_2^0:\;WF'(\Delta_R-\Delta_F) &=&
  WF'(\Delta_F) = C_1^1|_{x_1^0<x_2^0} = \\
 &=& \{\x\in C;\;x_1^0<x_2^0, \xi_1^0<0,\xi_2^0<0\}.
\end{eqnarray*}
To determine $WF'(\Delta_R-\Delta_F)$ for $x_1^0\geq x_2^0$ we again
use the fact that
\[ P(\Delta_R -\Delta_F) = 0 \:(mod \:{\cal C}^\infty)= (\Delta_R-\Delta_F)
P, \]
thus, by Theorem~\ref{theorem1.11a}, $WF'(\Delta_R-\Delta_F)|_{x_1^0\geq
x_2^0}$ contains the points in $N\times N$ that can be propagated
along the bicharacteristic strips from points in $WF'(\Delta_R-\Delta_F)
|_{x_1^0<x_2^0},$ hence
\[
WF'(\Delta_R-\Delta_F)= \{\x\in C;\;\xi_1^0<0,\xi_2^0<0\}=C\cap(N_-'
\times N_-').
\]
c) goes as b)\hfill{$\bf\Box$}\\
\begin{cor}\label{cor1.18}
It holds \[\Delta_R+\Delta_A=\Delta_F+\Delta_{\bar{F}}\:(mod\:{\cal
  C}^\infty). \]
\end{cor}
{\sc Proof}:\\
Consider $E:=\Delta_R+\Delta_A-\Delta_F.$\\
Since $PE = P\Delta_R+P\Delta_A-P\Delta_F = I\:(mod\:{\cal C}^\infty)=EP$,
$E$ is a parametrix of $P$. By the pseudolocal property (Theorem
\ref{theorem1.10}) of pseudodifferential operators
\[ WF'(E)\supset WF'(PE) =WF'(I) =\Delta^\ast.\]
On the other hand, by \rf{1.15a} and Theorem~\ref{theorem1.16}
\begin{eqnarray*}
WF'(E)&\subset& WF'(\Delta_R)\cup WF'(\Delta_A)\cup WF'(\Delta_F)=\\
&=&\Delta^\ast \cup C^1_2\cup C^1_3\cup C_1^1,
\end{eqnarray*}
but $C_i^1 =\emptyset$ on the diagonal $x_1=x_2$ for $i=1,2,3,4$,
hence we have on the diagonal $WF'(E)|_{x_1=x_2}=\Delta^\ast$.\\
To determine $WF'(E)$ outside the diagonal we use the results of
Theorem~\ref{theorem1.17}:
\begin{eqnarray*}
\mbox{for}\;x_1^0< x_2^0:\quad WF'(E) &=& WF'(\Delta_A-\Delta_F)=
      C\cap(N_+'\times N_+'), \\
\mbox{for}\;x_1^0>x_2^0:\quad WF'(E) &=& WF'(\Delta_R-\Delta_F)=
       C\cap(N_-'\times N_-').
\end{eqnarray*}
Thus, altogether, $WF'(E)=\Delta^\ast\cup C_4^1 =
WF'(\Delta_{\bar{F}})$, and from the uniqueness of distinguished
parametrices (Theorem~\ref{theorem1.16}) it follows $E=\Delta_{\bar{F}}
\:(mod\: {\cal C}^\infty)$. \hfill{$\bf\Box$}\\
\chapter{ Quasifree
quantum states of linear scalar fields on curved spacetimes}
\section{The Klein-Gordon field in globally hyperbolic spacetimes}
\label{section2.1}
In this work, we are concerned with the quantum theory of the linear
Klein-Gordon field in globally hyperbolic spacetimes. We first
present the properties of the classical scalar field in order to
introduce the phase space that underlies the quantization procedure.
Then we construct the Weyl algebra and define the set of quasifree
states on it. The material in this section is based on the papers
\cite{MV68,Dimock80,KW91}. Here, all function spaces are considered
to be spaces of {\it real} valued functions.\\
Let us start with the Klein-Gordon equation
\beq
(\Box_g+\mu^2)\Phi =0 \label{2.1}
\eeq
for a scalar field $\Phi:\M\to {\bf R}$ on a globally hyperbolic
spacetime $(\M,g)$ (see equ.~\rf{1.28}). Since \rf{2.1} is a
hyperbolic differential equation the Cauchy problem on a globally
hyperbolic space is well-posed. As a consequence, there are two unique
continuous linear operators
\[ \Delta_{R,A}:\D{\M}\to {\cal E}(\M)
\]
with the properties
\beqa
&&(\Box_g+\mu^2)\Delta_{R,A} f=\Delta_{R,A} (\Box_g+\mu^2)f =f
\label{2.2}\\
&&supp\:(\Delta_A f)\subset J^-(supp\:f)\nonumber\\
&&supp\:(\Delta_R f)\subset J^+(supp\:f)\nonumber
\eeqa
for $f\in\D{\M}$. They are called the advanced ($\Delta_A$) and
retarded ($\Delta_R$) fundamental solutions of the Klein-Gordon
equation \rf{2.1}. They are equal (up to smooth kernels) to the
distinguished parametrices $E_3^1$ and $E_2^1$ of Theorem~\ref{theorem1.16}.
$E:=\Delta_R-\Delta_A$ is called the fundamental solution or
propagator of \rf{2.1}. It has the properties
\beqa
&&(\Box_g+\mu^2)Ef=E(\Box_g+\mu^2)f=0\label{2.3}\\
&&supp\:(Ef)\subset J^+(supp\:f)\cup J^-(supp\:f)\nonumber
\eeqa
for $f\in\D{\M}$. We remind ourselves that the wavefront set of $E$
was computed in Theorem~\ref{theorem1.17} to be $WF'(E)=C$ with $C$
given in equ.~\rf{1.30}.\\
$\Delta_R,\Delta_A$ and $E$ can be continuously extended to the
adjoint operators
\[ \Delta_R', \Delta_A', E':{\cal E}'(\M)\to\Dp{\M}
\]
by $\Delta_R'=\Delta_A,\; \Delta_A'=\Delta_R,\;E'=-E$ (this means for
the kernel of $E$: $E(x_1,x_2)=-E(x_2,x_1)$).\\
Let $\Sigma$ denote a given Cauchy surface of $\M$ with
future-directed unit-normalfield $n^\alpha$. Then there are the
restriction operators
\beqa
\rho_o: &{\cal E}(\M)&\to {\cal E}(\Sigma) \nonumber\\
 &f& \mapsto f|_\Sigma \nonumber\\
\rho_1:&{\cal E}(\M)&\to {\cal E}(\Sigma) \label{2.4}\\
&f&\mapsto (n^\alpha \nabla_\alpha f)|_\Sigma,\nonumber
\eeqa
which have adjoints $\rho_o',\rho_1'$ mapping ${\cal E}'(\Sigma)$ to
${\cal E}'(\M)$. Dimock \cite{Dimock80} proves the following
existence and uniqueness result for the Cauchy problem:
\begin{thm}\label{theorem2.1}
a) $E\rho_o', E\rho_1'$ restrict to continuous operators from
$\D{\Sigma}$ ($\subset {\cal E}'(\Sigma)$) to ${\cal E}(\M)$ ($\subset
\Dp{\M}$) and the unique solution of the Cauchy problem \rf{2.1}
with initial data $u_o,u_1 \in \D{\Sigma}$ is given by
\beq
u=E\rho_o'u_1-E\rho_1'u_o. \label{2.5}
\eeq
b) Furthermore, \rf{2.5} also holds in the sense of distributions,
i.e.~given $u_o,u_1\in\Dp{\Sigma}$, there exists a unique distribution
$u\in\Dp{\M}$ which is a (weak) solution of \rf{2.1} and has initial
data $u_o=\rho_ou, u_1=\rho_1u$ (the restrictions in the sense of
Theorem~\ref{theorem1.14}). It is given by
\beq
u(f)=-u_1(\rho_oEf)+u_o(\rho_1Ef)\label{2.6}
\eeq
for $f\in\D{\M}$.
\end{thm}
Applying $\rho_o$ and $\rho_1$ to the identity \rf{2.5} we immediately
obtain:
\beq
\begin{array}{lcl}\rho_oE\rho_0'=0 & &\rho_oE\rho_1'=-1\\
                 \rho_1E\rho_o'=1 &  &\rho_1E\rho_1' =0 \end{array}
\label{2.7}
\eeq
which reads in a more conventional notation
\[\begin{array}{lcl}
E(t,\vec{y}_1;t,\vec{y}_2)=0 & & (1\otimes {\cal L}_n)E(t,\vec{y}_1;
t,\vec{y}_2)=-\delta(\vec{y}_1,\vec{y}_2)\\
({\cal L}_n\otimes 1)E(t,\vec{y}_1;t,\vec{y}_2)=\delta(\vec{y}_1,\vec{y}_2)
& & ({\cal L}_n\otimes {\cal L}_n)E(t,\vec{y}_1;t,\vec{y}_2)=0
\end{array}
\]
in local coordinates, where $\Sigma$ is given by $t=const.$,
$\vec{y}\in\Sigma$ and ${\cal L}_n:=n^\alpha\nabla_\alpha$
is the Lie derivative in direction $n^\alpha$.
Inserting $u=Ef$ into both sides of equ.~\rf{2.5} we get the identity
\beq
E= E\rho_o'\rho_1 E -E\rho_1'\rho_o E. \label{2.7a}
\eeq
Theorem~\ref{theorem2.1} allows us to formulate the classical phase
space of the field theory in terms of initial data on a Cauchy
surface, we need not consider the solutions themselves. This is
of great advantage for the quantum field theory as we will see soon.\\
Let $\Sigma$ be a Cauchy surface for $(\M,g)$ with volume element
$d^3\sigma$. Then we define the classical phase space of the
Klein-Gordon field as the real linear symplectic space $(\Gamma,\sigma)$,
where $\Gamma:=\D{\Sigma}\oplus\D{\Sigma}$ is the space of initial
data with compact support and $\sigma$ is the symplectic bilinear
form
\beqa
\sigma:  \Gamma\times\Gamma &\to& {\bf R} \nonumber\\
(F_1,F_2) &\mapsto&- \int_\Sigma [u_1p_2-u_2p_1]\,d^3\sigma \label{2.8}
\eeqa
for $F_i:={u_i\choose p_i}\in\Gamma, i=1,2$.\\
\rf{2.8} is independent of the choice of Cauchy surface: If $\Sigma_1$
and $\Sigma_2$ are two Cauchy surfaces (enclosing the volume
$V\subset \M$) and
$\Phi_1,\Phi_2$ the solutions of \rf{2.1} to the initial data
$F_1,F_2$ on $\Sigma_1$ (with compact supports) then we can write
\rf{2.8} as
\begin{eqnarray*}
- \sigma(\Phi_1,\Phi_2) &=& \int_{\Sigma_1}[\Phi_1\nabla_\alpha\Phi_2
-\Phi_2\nabla_\alpha\Phi_1]n^\alpha\,d^3\sigma  \\
&\equiv& \int_{\Sigma_1}j_\alpha n^\alpha\,d^3\sigma  \\
&=& \int_V(\nabla^\alpha j_\alpha)\,d^4\mu+\int_{\Sigma_2}j_\alpha
n^\alpha\,d^3\sigma \\
&=&\int_{\Sigma_2}j_\alpha n^\alpha \,d^3\sigma \\
&\equiv&\int_{\Sigma_2}[\Phi_1\nabla_\alpha \Phi_2 -\Phi_2\nabla_\alpha
\Phi_1]n^\alpha\,d^3\sigma
\end{eqnarray*}
since $j_\alpha$ is the conserved current ($\nabla^\alpha j_\alpha=0$)
of the Klein-Gordon field.\\  \\
Now, to the symplectic space ($\Gamma,\sigma$) there is associated
(uniquely up to unitary equivalence) a Weyl algebra ${\cal A}[\Gamma,
\sigma]$, which is a simple $C^*$-algebra generated by the elements
$W(F),\;F\in\Gamma$, that satisfy
\beqa
&&W(F)^*=W(F)^{-1}=W(-F)\quad\mbox{(unitarity)}\nonumber\\
&&W(F_1)W(F_2)=e^{-\frac{i}{2}\sigma(F_1,F_2)}W(F_1+F_2)\quad
\mbox{(Weyl relations)} \label{2.9}
\eeqa
for all $F_1,F_2\in\Gamma$. We can think of the elements $W(F)$ as
the exponentiated field operators $e^{i\hat{\Phi}(f)}$, smeared
with testfunctions $f\in\D{\M}$, where $F={\rho_oEf\choose\rho_1Ef}$.
\rf{2.9} then corresponds to the canonical commutation relations.\\
A local algebra ${\cal A (O)}$ (${\cal O}$ an open bounded subset of
$\M$) is the $C^*$-algebra generated by the elements $W(\rho_oEf,
\rho_1Ef)$ with $supp\:f\subset{\cal O}$. It is the algebra of quantum
observables measurable in the spacetime region ${\cal O}$.
Then ${\cal A}[\Gamma,\sigma]=\ol{\bigcup_{{\cal O}}{\cal A(O)}}^{C^*}$.\\
Dimock \cite{Dimock80} has shown that ${\cal O}\mapsto {\cal A(O)}$
is a net of local observable algebras in the sense of Haag and Kastler
\cite{HK64}, i.e.~it satisfies\\
i) ${\cal O}_1\subset{\cal O}_2\Rightarrow {\cal A(O}_1)\subset
{\cal A(O}_2)$ (isotony).\\
ii) ${\cal O}_1$ spacelike separated from ${\cal O}_2 \Rightarrow
[{\cal A(O}_1),{\cal A(O}_2)]=\{0\}$ (locality).\\
iii) There is a faithful irreducible representation of ${\cal A}$
(primitivity).\\
iv) ${\cal O}_1\subset D({\cal O}_2)\Rightarrow {\cal A(O}_1)\subset
{\cal A(O}_2)$.\\
v) For any isometry $\kappa:(\M,g)\to(\M,g)$ there is an isomorphism
$\alpha_\kappa:{\cal A}\to{\cal A}$ such that $\alpha_\kappa[{\cal
A(O)}]={\cal A}(\kappa({\cal O}))$ and $\alpha_{\kappa_1}\circ
\alpha_{\kappa_2}=\alpha_{\kappa_1\circ\kappa_2}$ (covariance).\\
The states on an observable algebra ${\cal A}$ are the linear
functionals $\omega:{\cal A}\to {\bf C}$ satisfying $\omega({\bf 1})
=1$ (normalization) and $\omega(A^*A)\geq 0\;\forall A\in\A$
(positivity). The set of states on our Weyl algebra
$\A[\Gamma,\sigma]$ is by far too large to be tractable in a concrete
way. Therefore, for linear systems, one usually restricts oneself to
the quasifree states, all of whose truncated n-point functions vanish for
$n\not=2$:
\begin{dfn}\label{dfn2.2}
Let $\mu:\Gamma\times\Gamma\to{\bf R}$ be a real scalar product
satisfying
\beq
\frac{1}{4}|\sigma(F_1,F_2)|^2\leq \mu(F_1,F_1)\mu(F_2,F_2)
\label{2.10}
\eeq
for all $F_1,F_2\in\Gamma$.
Then the {\rm\bf quasifree state} $\omega_\mu$ associated with $\mu$
is given by
\beq
\omega_\mu(W(F))=e^{-\frac{1}{2}\mu(F,F)}.\label{2.11}
\eeq
If $\omega_\mu$ is pure it is called a {\rm\bf Fock state}.
\end{dfn}
The connection between this algebraic notion of a quasifree state
and the usual notion of ``vacuum state'' in a Hilbert space is
established by the following theorem which we cite from \cite{KW91}:
\begin{thm}\label{theorem2.3}
Let $\omega_\mu$ be a quasifree state on $\A[\Gamma,\sigma]$.
\begin{enumerate}
\item[a)] Then there exists a {\rm\bf one-particle Hilbertspace structure},
i.e.~a Hilbert space ${\cal H}$ and a real-linear map $k:\Gamma\to
{\cal H}$ such that \\
i) $k\Gamma +ik\Gamma$ is dense in ${\cal H}$,\\
ii) $\mu(F_1,F_2)={\rm Re}\la kF_1,kF_2\ra_{\cal H}\;\forall
F_1,F_2\in \Gamma$,\\
iii) $\sigma(F_1,F_2)=2{\rm Im}\la kF_1,kF_2\ra_{\cal H}\;
\forall F_1,F_2\in\Gamma$.\\
Moreover, the pair $(k,{\cal H})$ is uniquely determined up to
unitary equivalence.\\ It holds: $\omega_\mu$ is pure
$\Leftrightarrow\;k(\Gamma)$ is dense in ${\cal H}$.
\item[b)] The GNS-triple $({\cal H}_{\omega_\mu},\pi_{\omega_\mu},\Omega_{
\omega_\mu})$ of the state $\omega_\mu$ can be represented as
$({\cal F}^s({\cal H}),\rho_\mu,\Omega^{{\cal F}})$, where\\
i) ${\cal F}^s({\cal H})$ is the symmetric Fock space over the
one-particle Hilbert space ${\cal H}$,\\
ii) $\rho_\mu [W(F)]=\exp\{-i\ol{[a^*(kF)+a(kF)]}\}$, where $a^*$
and $a$ are the standard creation and annihilation operators on
${\cal F}^s({\cal H})$ satisfying
\[[a(u),a^*(v)]=\la u,v\ra_{\cal H}\;\mbox{and}\;a(u)\Omega^{\cal F}=0
\]
for $u,v\in{\cal H}$ (the bar denotes the closure of the operator).\\
iii) $\Omega^{\cal F}:={\bf 1}\oplus{\bf 0}\oplus{\bf 0}\oplus\ldots$
is the (cyclic) Fock vacuum.\\
It holds: $\omega_\mu$ is pure $\Leftrightarrow$ $\rho_\mu$ is irreducible.
\end{enumerate}
\end{thm}
Thus, $\omega_\mu$ can also be represented as $\omega_\mu(W(F))=
\exp\{-\frac{1}{2}||kF||^2_{\cal H}\}$ (in case a)) or
$\omega_\mu(W(F))=\la \Omega^{\cal F},\rho_\mu (F)\Omega^{\cal F}\ra$
(in case b)). $\hat{\Phi}(F):=a^*(kF)+a(kF)$ is the usual field
operator on ${\cal F}^s({\cal H})$ and we can determine the
(``symplectically smeared'') two-point function as
\beqa
\lambda^{(2)}(F_1,F_2)&=& \la \Omega^{\cal F},
\hat{\Phi}(F_1)\hat{\Phi}(F_2)\Omega^{\cal F}\ra  \nonumber\\
&=& \la kF_1,kF_2\ra_{\cal H}  \label{2.13}\\
&=& \mu(F_1,F_2)+\frac{i}{2}\sigma (F_1,F_2) \nonumber
\eeqa
for $F_1,F_2\in \Gamma$,
resp.~the ``four-smeared'' (Wightman) two-point distribution as
\beq
\Lambda^{(2)}(f_1,f_2)=\lambda^{(2)}\l({\rho_oEf_1\choose \rho_1Ef_1}
,{\rho_o Ef_2\choose \rho_1Ef_2}\r)\label{2.14}
\eeq
for $f_1,f_2\in\D{\M}$.
The fact that the antisymmetric (= imaginary) part of $\lambda^{(2)}$
is the symplectic form $\sigma$ implies for $\Lambda^{(2)}$:
\beqa
{\rm Im}\Lambda^{(2)}(f_1,f_2)&=&-\frac{1}{2}\int_\Sigma
[f_1 E'\rho_o'\rho_1 E f_2-f_1E'\rho_1'\rho_o E f_2]\,d^3\sigma \nonumber\\
&=& \frac{1}{2}\la f_1,Ef_2\ra \label{2.14a}
\eeqa
by equ.~\rf{2.7a}.
All the other n-point functions can also be
calculated, one finds that they vanish if n is odd and that the
n-point functions for n even are sums of products of two-point
functions.\\
We want to stress that the restriction to quasifree states is a priori
not physically motivated but by the fact that they are exclusively
determined by their two-point function and therefore easily tractable.
Nevertheless, one gets a large class of states including e.g.~the
usual vacuum state on stationary spacetimes or the so-called
``frequency splitting vacua'' obtained by mode-decomposition of the
field operators (see e.g.~\cite{BD82}), but it also contains all
sorts of unphysical states. Therefore, as was discussed in the
introduction, we have to impose certain selection criteria even on
this restricted class of states. This will be done in
sections~\ref{section2.3} and \ref{section2.5} where we introduce
quasifree Hadamard states and adiabatic vacuum states and show that
they in fact select the same local folium of states. Since the
folium of a quasifree state also contains states that are not
quasifree one can in principle get statements about a larger class of
states than merely the quasifree ones. Recently, Kay \cite{Kay93}
 considered also states that allow for a non-vanishing one-point
 function.\\  \\
A curved spacetime does in general not possess any isometries, hence
property v) above of the net of local algebras is in general empty.
But there is the important class of spacetimes possessing a timelike
Killing vectorfield that deserves further attention.\\
Let $(\M,g)$ be a globally hyperbolic manifold, foliated into
spacelike Cauchy surfaces $\M={\bf R}\times\Sigma,\;\Sigma_t
=\{t\}\times\Sigma$, and possessing a one-parameter group of
isometries $\tau_t:\M\to\M,t\in{\bf R}$, such that
$\tau_t(\Sigma_{t_o})=\Sigma_{t_o+t}$. Define ${\cal T}(t):\Gamma\to
\Gamma$ by
\[{\cal T}(t)F_{t_o}:=F_{t_o+t}
\]
where $F_{t_o}, F_{t_o+t}$ are the Cauchy data of a solution of
\rf{2.1} taken on Cauchy surfaces $\Sigma_{t_o}$ resp.~$\Sigma_{t_o+t}$.
Since the symplectic form $\sigma$ is invariant under the action of
${\cal T}(t)$ and since ${\cal T}(t){\cal T}(s)={\cal T}(t+s)\;
\forall t,s\in {\bf R},\;{\cal T}(t)$ is a one-parameter group
of symplectic transformations (also called Bogoliubov
transformations). It gives rise to a group of automorphisms $\alpha (t),
t\in{\bf R}$, (Bogoliubov automorphisms) on the algebra ${\cal A}$ via
\[\alpha(t) W(F)=W({\cal T}(t) F).
\]
In this case, there exists a preferred class of states on ${\cal A}$,
namely those invariant under $\alpha(t)$. A quasifree state $\omega_\mu$
will be invariant under this symmetry if and only if
\[\mu({\cal T}(t)F_1,{\cal T}(t)F_2)=\mu(F_1,F_2)\;\forall t\in{\bf R}
\;\forall F_1,F_2\in\Gamma.
\]
The automorphism group $\alpha(t)$ can be unitarily implemented in the
one-particle Hilbertspace structure $(k,{\cal H})$ of an invariant state
$\omega_\mu$, i.e.~there exists a unitary group $U(t), t\in{\bf R},$ on
${\cal H}$ satisfying
\beqa
U(t)k &=& k{\cal T}(t) \label{2.15}\\
U(t)U(s) &=& U(t+s). \nonumber
\eeqa
(This follows easily from the uniqueness statement of
Theorem~\ref{theorem2.3}a).)
If $U(t)$ is strongly continuous it takes the form $U(t)=e^{-iht}$ for some
self-adjoint operator $h$ on ${\cal H}$.\\
We are now ready to define two particularly important classes of states
(we follow \cite{Kay85b}):
\begin{dfn}\label{dfn2.4}
Let the phase space $(\Gamma,\sigma,{\cal T}(t))$ be given.
\begin{enumerate}
\item[a)] A quasifree {\rm\bf ground state} is a quasifree state over ${\cal A}
[\Gamma,\sigma]$ with one-particle Hilbertspace structure $(k,{\cal H})$
and a strongly continuous unitary group $U(t)=e^{-iht}$ (satisfying
\rf{2.15}) such that $h$ is a positive operator (the ``one-particle
Hamiltonian'').
\item[b)] A quasifree {\rm\bf KMS-state} is a quasifree state over ${\cal A}
[\Gamma,\sigma]$ with one-particle Hilbertspace structure
($k^\beta,\tilde{{\cal H}}$) and a strongly continuous unitary group
$U(t)=e^{-i\tilde{h}t}$ (satisfying \rf{2.15}) such that the ``one-particle
KMS-condition'' is satisfied, namely $\forall u,v\in k^\beta\Gamma\;\forall
t\in{\bf R}$:
\beq
\la e^{-it\tilde{h}} u,v\ra_{\tilde{\cal H}} = \la e^{-\frac{\beta
\tilde{h}}{2}} v,e^{-it\tilde{h}} e^{-\frac{\beta\tilde{h}}{2}}
u\ra_{\tilde{\cal H}}.\label{2.16}
\eeq
\end{enumerate}
\end{dfn}
Although the definition of ground- and KMS-states can be given in a
very general context (see e.g.~\cite{Haag92}), we have restricted
ourselves to the case of quasifree states on the Weyl algebra
since this is the only situation we consider here. The physical
interpretation of these stationary states is the following: The ground
state is the state of lowest energy of the theory, it is closest to
what one would call the vacuum state in Minkowski space. It fulfills
the spectrum condition in that $h$ has positive spectrum. A KMS-state
is a thermodynamic equilibrium state of the theory at temperature
$1/\beta$. The KMS-condition \rf{2.16} is the generalization of the
Gibbs equilibrium condition to infinite systems. It was introduced
into quantum field theory by Haag, Hugenholtz and Winnink
\cite{HHW67}. We will present an explicit construction of ground- and
KMS-states on ultrastatic spacetimes in section~\ref{section2.4}.
\section{Hadamard states}\label{section2.3}
Now we introduce the notion of Hadamard states. Following Kay and Wald
\cite{KW91} we state the original definition of Hadamard states and
some of its consequences. Then we introduce Radzikowski's local
characterization of Hadamard states \cite{Radz92} and discuss its
relevance for quantum field theory on curved spacetimes.
First we need some preparatory definitions.
\begin{dfn}\label{dfn3.1}
Let $\Sigma$ be a spacelike Cauchy surface of $(\M,g)$.\\
A {\rm\bf causal normal neighborhood} $N$ of $\Sigma$ is an open
neighborhood of $\Sigma$ in $\M$ such that $\Sigma$ is a Cauchy
surface for $N$ and such that for all $x_1,x_2\in N$ with
$x_1\in J^+(x_2)$ there exists a convex normal neighborhood
which contains $J^-(x_1)
\cap J^+(x_2)$. (As a consequence, the squared geodesic distance
$\sigma(x_1,x_2)$ is then well defined and smooth for all causally
related pairs of points in $N$).
\end{dfn}
\begin{lemma}[Lemma 2.2 of \cite{KW91}]\label{lemma3.2}
For each spacelike Cauchy surface $\Sigma$ there exists a causal
normal neighborhood $N$.
\end{lemma}
We choose a preferred time orientation on ($\M,g$) and a smooth global
time function $T:\M\to {\bf R}$ which increases towards the future.
Let ${\cal O}\subset \M\times\M$ be an open neighborhood of the set of
causally related points $(x_1,x_2)$ such that $J^+(x_1)\cap J^-(x_2)$
and $J^-(x_1)\cap J^+(x_2)$ are contained within a convex normal
neighborhood and ${\cal O}'$ an open neighborhood in $N\times N$
of the set of causally related points such that $\overline{{\cal O}'}
\subset {\cal O}$.\\
Within ${\cal O}$ the squared geodesic distance $\sigma(x_1,x_2)$ is
well defined and we define for each $n\in {\bf N}$ a real function
$v^{(n)}\in {\cal C}^\infty({\cal O})$ as the power series
\beq
v^{(n)}(x_1,x_2) :=\sum_{m=0}^{n} v_m(x_1,x_2)\sigma^m \label{3.0}
\eeq
where the $v_m$ are uniquely determined by the Hadamard recursion
relations (see \cite{DB60}); note that the $v_m$ are solely
determined by the mass $\mu$ of the Klein-Gordon field and the
metric $g$ of the spacetime).\\
Let $\chi \in {\cal C}^\infty(N\times N)$ be a function with the
property that
\[ \chi(x_1,x_2) = \l\{ \begin{array}{cc}0,&\mbox{for}\;(x_1,x_2)
   \not\in {\cal O}\\
1,& \mbox{for}\;(x_1,x_2)\in {\cal O}'. \end{array}\r.
\]
For each $n\in {\bf N}$ and $\epsilon >0$ we define in ${\cal O}$
the (complex valued) function
\beq
G_\epsilon^{T,n}(x_1,x_2) := \frac{1}{(2\pi)^2}
\l(\frac{\Delta(x_1,x_2)^{1/2}}{\sigma+2i\epsilon (T(x_1)-T(x_2))+
\epsilon^2}+v^{(n)}(x_1,x_2)\ln (\sigma+2i\epsilon (T(x_1)-T(x_2))+
\epsilon^2)\r), \label{3.1}
\eeq
where $\Delta$ is the van Vleck-Morette determinant \cite{DB60}
and the branch-cut for the logarithm is taken to lie along the
negative real axis. Now we are ready to define:
\begin{dfn}\label{dfn3.3}
Let ($\M,g$) be a globally hyperbolic manifold, $\Sigma$ a Cauchy
surface of $\M$, $N$ a causal normal neighborhood of $\Sigma$ and
$\chi, T, G_\epsilon^{T,n}$ as above.\\
Then we call a quasifree state $\omega$ of the Weyl-algebra ${\cal A}$
of the Klein-Gordon field on ($\M,g$) a {\rm\bf (global) Hadamard state}
if its two-point distribution $\Lambda^{(2)}$ is such that there exists
a sequence of functions $H^n\in {\cal C}^n(N\times N)$ such that
 for all $f_1,f_2\in \co{N}$ and all $n\in {\bf N}$ we have
\beqa
\Lambda^{(2)}(f_1,f_2) &=& \lim_{\epsilon\to 0}\int_{N\times N}
\Lambda_\epsilon^{T,n}(x_1,x_2)f_1(x_1)f_2(x_2)\,d^4\mu(x_1)d^4\mu(x_2),
\label{3.2}\\
\mbox{where}\; \Lambda_\epsilon^{T,n}(x_1,x_2) &:=& \chi(x_1,x_2)
G_\epsilon^{T,n}(x_1,x_2)+H^{n}(x_1,x_2). \label{3.3}
\eeqa
\end{dfn}
Note that $\chi$ was chosen to be zero where $G_\epsilon^{T,n}$ was
not defined, so $\Lambda_\epsilon^{T,n}$ is well defined throughout
$N\times N$. Kay and Wald \cite{KW91} show that the definition
is actually independent of the choice of $N,\chi$ and $T$.
In \cite{FSW78} it was proved that the Hadamard property of a state
is preserved under Cauchy evolution, i.e.~if $\Lambda^{(2)}$ is of
the Hadamard form in a causal normal neighborhood $N$ of some Cauchy
surface $\Sigma$, then it is of Hadamard form in some causal normal
neighborhood $N'$ of any other Cauchy surface $\Sigma'$. This
implies in particular, that the Definition \ref{dfn3.3} above is
independent of the choice of $\Sigma$, too.\\
In \cite{FNW81} it is established that in any globally hyperbolic
spacetime there is always a class of quantum states, forming a
dense subspace of a Hilbert space, whose two-point functions have
the Hadamard singularity structure \rf{3.2}.\\
The essential ingredient in the Definition \ref{dfn3.3} is the
specification of the singularity structure \rf{3.1} of the
two-point distribution \rf{3.2}. It is the same for all Hadamard
states, whereas the smooth part $H^n$ in \rf{3.3} depends on the
respective state. This was the original motivation for the
consideration of Hadamard states because it allows for the
renormalization of the energy-momentum tensor. For a
Klein-Gordon field $\Phi$ the energy-momentum tensor is given by
\beq
T_{\mu\nu}[\Phi]=(\nabla_\mu\Phi)(\nabla_\nu\Phi)-\frac{1}{2}
g_{\mu\nu}(\nabla_\kappa \Phi \nabla^\kappa\Phi-\mu^2\Phi^2).
\label{3.4}
\eeq
In order that the semiclassical Einstein equations
\beq
R_{\mu\nu}-\frac{1}{2}g_{\mu\nu}R=8\pi
\l\la\hat{T}_{\mu\nu}(x)\r\ra_\omega \label{3.5}
\eeq
make sense, one must define the expectation value of the
energy-momentum operator $\hat{T}_{\mu\nu}$ of the quantum field
$\Phi$ in the state $\omega$ at a spacetime point $x$. One procedure
 which has been studied in great detail is the point-splitting
renormalization \cite{Christensen76,ALN77,AL78,Wald77,Wald78b}.
In this approach one regards initially $\la\hat{T}_{\mu\nu}(x,x')\ra
_\omega$ as a two-point distribution (note that $T_{\mu\nu}$ is
quadratic in $\Phi$ resp.~its derivatives), which, however, behaves
singular in the ``coincidence limit'' $x\to x'$. But if we admit only
states $\omega$ whose two-point distributions have the same
singularity structure, as is the case for Hadamard states, then
we can subtract from $\la\hat{T}_{\mu\nu}(x,x')\ra_\omega$
another distribution with this singularity structure and {\it define}
the renormalized value of $\la\hat{T}_{\mu\nu}(x)\ra_\omega$ as the
coincidence limit of this difference. Wald \cite{Wald77,Wald78b}
formulated a set of axioms that a physically reasonable
$\la\hat{T}_{\mu\nu}\ra$ should satisfy and showed that for Hadamard
states the point-splitting prescription gives a result consistent with
these axioms. However, there remains an ambiguity of adding local
curvature terms. (For a discussion of this non-uniqueness see
\cite{Fulling89}.)
Recently, a different renormalization scheme has been investigated
by K\"ohler \cite{Koehler95}. He utilizes the cancellation of
singularities in $\la\hat{T}_{\mu\nu}\ra$ that occur between two
scalar fields and one Dirac field in Hadamard states. This technique
yields results consistent with the point-splitting renormalization,
but works only on vacuum spacetimes. Thus, although it seems that
the last word on $\la\hat{T}_{\mu\nu}\ra$ has not yet been said, it
is clear that the Hadamard states play an important r\^{o}le in its
definition.\\
In \cite{Verch94}, Verch showed that any two Hadamard states for the
Klein-Gordon field on a globally hyperbolic manifold are locally
quasiequivalent. For a more restricted class of spacetimes, the
ultrastatic ones (see Definition \ref{dfn3.5} below), he proved
that the local von Neumann algebras are type $III_1$-factors and
that the Reeh-Schlieder property holds \cite{Verch93a}.
All these facts strongly suggest that the local quasiequivalence class
of states generated by the Hadamard states is a good candidate
for the set of physical states of scalar quantum fields on globally
hyperbolic manifolds.\\
However, the Hadamard states -- as they were defined above -- have
also some severe drawbacks. Firstly, the Definition \ref{dfn3.3} is
tailored for free fields. Note that the $v^{(n)}$ in \rf{3.1}
are constructed in such a way that the two-point function \rf{3.2}
is a distributional bi-solution of the Klein-Gordon equation. It is
not clear in this formulation how one could obtain a generalization
to nonlinear, interacting fields.\\
Secondly, the Definition \ref{dfn3.3} is a global definition in the
sense that the singularity structure has to be specified for spacetime
points $x_1, x_2$ ranging over a neighborhood of the whole Cauchy
surface; it states that there are no singularities at points
$(x_1,x_2) \in N\times N$ s.th.~$x_1$ is spacelike separated from
$x_2$. Our experience with local quantum physics and the very
concept of general relativity suggest that the physical information
of a theory is encoded locally, i.e.~in an arbitrarily small
neighborhood of each spacetime point. This was pointed
out by Fredenhagen and Haag \cite{FH87} and led Kay \cite{Kay87}
to conjecture that if already a local version of Definition
\ref{dfn3.3} is satisfied for a state then its two-point distribution
should not have singularities at spacelike separated points. Let us
give a precise definition of this ``local version'':
\begin{dfn}\label{dfn3.3a}
Let $(\M,g)$ be a globally hyperbolic manifold.\\
A two-point distribution $\Lambda^{(2)}\in \Dp{\M\times\M}$ is
said to be {\rm\bf locally Hadamard} if for each $x\in\M$ there is an
open neighborhood $U_x$ of $x$ s.th.~for each $n\in{\bf N}\;
\Lambda ^{(2)}|_{U_x\times U_x}$ coincides with $(\lim_{\epsilon\to
0}\Lambda_\epsilon^{T,n})|_{U_x\times U_x}$ (equ.~\rf{3.3}) of
a global Hadamard state when the Cauchy surface $\Sigma$, causal
normal
neighborhood $N$ and $U_x$ are chosen such that $U_x\subset N$.
\end{dfn}
That Kay's conjecture is indeed true was recently shown by Radzikowski
\cite{Radz92} in a very remarkable work. He realized that the local
information on the Hadamard property is encoded in the wavefront set
of the two-point distribution and proved the following important
theorem:
\begin{thm}[Theorem 2.6 of \cite{Radz92}, see also \cite{Koehler94}]
\label{theorem3.4}
A quasifree state of a Klein-Gordon quantum field on a globally
hyperbolic spacetime is a (global) Hadamard state if and only if its two-point
distribution $\Lambda^{(2)}$ possesses the following wavefront
set:
\beq
WF(\Lambda^{(2)})=\{(x_1,\xi_1;x_2,-\xi_2)\in T^*(\M\times\M)
\setminus\{0\};\;(x_1,\xi_1)\sim(x_2,\xi_2),\xi^0_1\geq 0\}.
\label{3.6}
\eeq
\end{thm}
The notation was introduced in section \ref{section1.3}:
$(x_1,\xi_1)\sim(x_2,\xi_2)$ means that $x_1$ and $x_2$ can be
connected by a null geodesic $\gamma$ such that $\xi_1^\mu$ is
tangential to $\gamma$ at $x_1$, and $\xi_2^\mu$ is the parallel
transport of $\xi_1^\mu$ along $\gamma$ at $x_2$. On the diagonal
$x_1=x_2$ $(x_1,\xi_1)\sim(x_2,\xi_2)$ means that $\xi_1=\xi_2,\;
\xi_1^2=0$.
The wavefront set \rf{3.6} is schematically depicted in Figure
\ref{figure3.1}.
\begin{figure}[htb]
  \begin{center}
  \leavevmode
\begingroup\makeatletter
\def\x#1#2#3#4#5#6#7\relax{\def\x{#1#2#3#4#5#6}}%
\expandafter\x\fmtname xxxxxx\relax \def\y{splain}%
\ifx\x\y   
\gdef\SetFigFont#1#2#3{%
  \ifnum #1<17\tiny\else \ifnum #1<20\small\else
  \ifnum #1<24\normalsize\else \ifnum #1<29\large\else
  \ifnum #1<34\Large\else \ifnum #1<41\LARGE\else
     \huge\fi\fi\fi\fi\fi\fi
  \csname #3\endcsname}%
\else
\gdef\SetFigFont#1#2#3{\begingroup
  \count@#1\relax \ifnum 25<\count@\count@25\fi
  \def\x{\endgroup\@setsize\SetFigFont{#2pt}}%
  \expandafter\x
    \csname \romannumeral\the\count@ pt\expandafter\endcsname
    \csname @\romannumeral\the\count@ pt\endcsname
  \csname #3\endcsname}%
\fi
\endgroup
\begin{picture}(0,0)%
\includegraphics{figure3.1.pstex}%
\end{picture}%
\setlength{\unitlength}{0.012500in}%
\begingroup\makeatletter
\def\x#1#2#3#4#5#6#7\relax{\def\x{#1#2#3#4#5#6}}%
\expandafter\x\fmtname xxxxxx\relax \def\y{splain}%
\ifx\x\y   
\gdef\SetFigFont#1#2#3{%
  \ifnum #1<17\tiny\else \ifnum #1<20\small\else
  \ifnum #1<24\normalsize\else \ifnum #1<29\large\else
  \ifnum #1<34\Large\else \ifnum #1<41\LARGE\else
     \huge\fi\fi\fi\fi\fi\fi
  \csname #3\endcsname}%
\else
\gdef\SetFigFont#1#2#3{\begingroup
  \count@#1\relax \ifnum 25<\count@\count@25\fi
  \def\x{\endgroup\@setsize\SetFigFont{#2pt}}%
  \expandafter\x
    \csname \romannumeral\the\count@ pt\expandafter\endcsname
    \csname @\romannumeral\the\count@ pt\endcsname
  \csname #3\endcsname}%
\fi
\endgroup
\begin{picture}(453,188)(107,532)
\put(280,685){\makebox(0,0)[lb]{\smash{\SetFigFont{12}{14.4}{rm}$(x_1,\xi_1)$}}}
\put(320,623){\makebox(0,0)[lb]{\smash{\SetFigFont{12}{14.4}{rm}$\bigcup$}}}
\put(486,623){\makebox(0,0)[lb]{\smash{\SetFigFont{12}{14.4}{rm}$(x,\xi;x,\xi)$}}}
\put(300,532){\makebox(0,0)[lb]{\smash{\SetFigFont{12}{14.4}{rm}$(x_1,\xi_1)$}}}
\put(218,622){\makebox(0,0)[lb]{\smash{\SetFigFont{12}{14.4}{rm}$(x_2,\xi_2)$}}}
\end{picture}

  \end{center}
\caption{The wavefront set $WF'(\Lambda^{(2)})$ of an Hadamard
  two-point distribution $\Lambda^{(2)}$.}
\label{figure3.1}
\end{figure}
%
%
%
%
%

Note that $\Lambda^{(2)}$ is a solution of the Klein-Gordon equation
in both arguments and that therefore $WF(\Lambda^{(2)})\subset
N\times N$ ($N$ was defined in equ.~\rf{1.28a}). Hence, the essential
physical content of Theorem~\ref{theorem3.4} is that singularities
only occur if $x_1$ and $x_2$ are lightlike connected and that the
singularities only have {\it positive} frequencies. This is the
remnant of the spectrum condition of quantum field theory on Minkowski
space. In fact, already in \cite{RSII} it was shown that the vacuum
state of a linear scalar quantum field on Minkowski space fulfills
\rf{3.6}. This was used in \cite{Radz92} and \cite{Koehler94} to
prove Theorem~\ref{theorem3.4}. \\
{}From Theorem~\ref{theorem3.4} it easily follows \cite{Radz92} that
the {\it\bf Feynman propagator}
\beq
E_F :=i\Lambda^{(2)}+\Delta_A \label{3.7}
\eeq
of a Hadamard state is equal (up to ${\cal C}^\infty$) to the
distinguished parametrix $\Delta_F\equiv E_1^1$ of Theorem~\ref{theorem1.16}.
With the new characterization of Hadamard states by their wavefront
set at his disposal Radzikowski was able to prove the following
``{\it\bf local-to-global-singularity theorem}'':
\begin{thm}[Theorem 3.3 of \cite{Radz92}]\label{theorem3.4a}
Let $(\M,g)$ be a globally hyperbolic spacetime.\\
Any quasifree state of a Klein-Gordon quantum field on $(\M,g)$ whose
two-point distribution is locally Hadamard (Definition \ref{dfn3.3a})
is a global Hadamard state (Definition \ref{dfn3.3}).
\end{thm}
This says that local physical information leads uniquely to global
physical information and verifies Kay's conjecture.\\
Since the wavefront set is a covariant and locally defined object on
a manifold (properties 1) and 2) of section~\ref{section1.2}) the
characterization of Hadamard states by the wavefront set of their
two-point distributions promises to be a very useful tool for quantum
field theory on curved spacetime, and from now on we will take
Theorem~\ref{theorem3.4} as a {\it definition} of Hadamard state.\\
A first important observation is that the concept of the wavefront
set is no longer tied to a certain linear field equation. Therefore
\rf{3.6} is a possible starting point for a characterization of the
physical states of an arbitrary interacting quantum field. The idea
is of course to specify the wavefront set of all the $n$-point
Wightman functions of the quantum fields. Radzikowski \cite{Radz92}
gave a first proposal of such a wavefront set spectrum condition, as
he called it. In \cite{BFK95} it is argued that this condition
is not consistent and a new one is suggested which is shown to be
satisfied for Wick ordered products of fields. In \cite{Koehler94}
the Theorems \ref{theorem1.13} and \ref{theorem3.4} were used to
define the product of two scalar fields in an Hadamard product state
as a new Wightman field on manifolds. In this case there may also appear
lightlike singularities, but at any rate no spacelike ones.\\
These first results indicate that the wavefront set is the relevant
mathematical object in order to investigate general properties of
quantum field theory on curved spacetimes.
\section{Ultrastatic ground- and KMS-states}\label{section2.4}
In this section we want to show that wavefront sets and pseudodifferential
operators also are a very useful tool for analysing concrete
physical states for linear fields on certain spacetime
models. Up to now there have
not many states been constructed which are known to be Hadamard
states, but a great many more which have not been characterized
as physical or not in such a way. Among the best investigated
examples are states for Klein-Gordon fields on ultrastatic
spacetimes, the Schwarzschild (resp.~Kruskal-)
spacetime and the Robertson-Walker
spacetimes. On Robertson-Walker spacetimes there exists the large
class of ``adiabatic vacua'' which we will investigate in detail
in the next section. On the Schwarzschild spacetime we explicitly
know the groundstate w.r.t.~the timelike Killing field (the
``Boulware vacuum'' \cite{Boulware75,Kay85}), the thermodynamic
equilibrium states (KMS-states \cite{Kay85}) and the Unruh state
\cite{Unruh76,DK87,DK86}, which describes the outflow of thermal
radiation from an eternal black hole. These are stationary quasifree states.
Kay and Wald \cite{KW91} showed that on spacetimes with bifurcate
Killing horizons (e.g.~the Kruskal extension of the
Schwarzschild metric) there can be at most one
stationary Hadamard state. This is the KMS-state at the Hawking
temperature $1/8\pi M$, where $M$ is the mass of the black hole
(the so-called Hartle-Hawking state \cite{HH76}). This result
indicates that it is the Hadamard condition which is responsible
for singling out the states that describe the thermal radiation of a
star collapsing to a black hole, which was discovered by Hawking
in his famous paper \cite{Hawking75}.
And indeed, in \cite{FH90} it is derived that in the gravitational
collapse of a spherically symmetric star all quantum states which
have the Hadamard property (in fact a somewhat weaker scaling
condition) at the intersection of the surface of the star and the
Schwarzschild radius are seen by asymptotic detectors (at large radial
distances and late times) as a thermal radiation (modified
by a gravitational barrier penetration effect) at the Hawking
temperature $1/8\pi M$. For a rotating black hole (Kerr metric)
a similar computation has been performed in \cite{Hessling92}.
On ultrastatic spacetimes it has been proven \cite{FNW81,Wald79}
that the ground state is a Hadamard state. We will now give a new
proof of this result. On the one hand, our proof is much more
transparent than that of \cite{FNW81,Wald79} since we use the
techniques of wavefront sets and pseudodifferential operators,
on the other hand it easily extends to the ultrastatic KMS-states and
also gives the essential ideas that are
used to prove the Hadamard property of the adiabatic vacua in the next
section.
\begin{dfn}\label{dfn3.5}
A spacetime $(\M,g)$ is {\rm\bf ultrastatic} if it possesses a
timelike
Killing field $t^\mu$ which is hypersurface-orthogonal and obeys
$g_{\mu\nu}t^\mu t^\nu =1$.
\end{dfn}
This is a slight specialization of a static spacetime where one
dispenses with the last condition. A static spacetime possesses a
global foliation $\M={\bf R}\times\Sigma$ into spacelike hypersurfaces
$\Sigma_t =\{t\}\times\Sigma, t\in {\bf R}$. If $\{x^i,i=1,2,3\}$ is a local
coordinate system for $\Sigma$ the static metric can be written as
\beq
ds^2=\alpha(\vec{x})^2 dt^2-h_{ij}(\vec{x})dx^i dx^j,\label{3.8a}
\eeq
where $\alpha\in {\cal C}^\infty (\Sigma)$ is the ``lapse function''
and $h_{ij}$ a Riemannian metric on $\Sigma$ (both of them independent
of $t$), and $t^\mu = (\partial/\partial t)^\mu$ is the timelike
Killing field orthogonal to $\Sigma_t$.\\
The ultrastatic case is now the special situation where $\alpha \equiv
1$, i.e.~the metric is
\beq
ds^2=dt^2-h_{ij}(\vec{x})dx^i dx^j \label{3.8}
\eeq
in a local coordinate system. In this case we have:
\begin{lemma}[see \cite{Kay78}]\label{lemma3.6}
Let $(\M,g)$ be an ultrastatic spacetime. Then the following
are equivalent:\\
1.) $(\M,g)$ is globally hyperbolic.\\
2.) $\Sigma_t$ is a Cauchy surface for each $t$.\\
3.) $(\Sigma,h)$ is geodesically complete.
\end{lemma}
Thus, by Theorem~\ref{theorem1.7}, for a globally hyperbolic
ultrastatic spacetime we can define $A:=\overline{-^{(3)}\Delta_h+
\mu^2}$ as a positive selfadjoint operator on $L^2_{\bf C}(\Sigma,
d^3\sigma)$ which is invertible for $\mu>0$, and likewise all its
powers ($^{(3)}\Delta_h$ denotes the Laplace-Beltrami operator on
$\Sigma$ w.r.t.~the metric $h_{ij}$, the bar denotes the closure of the
operator).
In this situation, we can define the ``canonical vacuum state'' on
the Weyl algebra of the Klein-Gordon field in the following way
\cite{Kay78}:\\
Let $(\Gamma,\sigma)$ denote the real symplectic space of initial
data $\Gamma:=\co{\Sigma_t}\oplus \co{\Sigma_t}$ on a Cauchy surface
$\Sigma_t$ with the symplectic form
\beq
\sigma(F_1,F_2):=-\int_{\Sigma_t}d^3\sigma\,[f_1p_2-p_1f_2],\label{3.9}
\eeq
where $d^3\sigma:=\sqrt{h}\,d^3x,\; h:=\det h_{ij},\;F_i:={f_i\choose
  p_i} \in \Gamma,\; i=1,2$.\\
Then for each $t\in {\bf R}$ we define a ``canonical vacuum state'' by
the one-particle Hilbert space structure (see Theorem~\ref{theorem2.3})
\beqa
k^t:\Gamma &\to& {\cal H}^t:=L^2_{\bf C}(\Sigma_t,d^3\sigma)\nonumber\\
(f,p)&\mapsto&
\frac{1}{\sqrt{2}}\l(A^{1/4}f-iA^{-1/4}p\r). \label{3.10}
\eeqa
{}From \cite{Kay78,Kay79} it follows:
\begin{thm}\label{theorem3.7}
For each $t\in{\bf R},\;(k^t,\Ht)$ defines a pure, quasifree state
$\omega_t$ on the Weyl algebra of the Klein-Gordon field on the
globally hyperbolic ultrastatic spacetime $(\M,g)$. $\omega_t$
is the unique quasifree ground state with respect to the time
translations $t\mapsto t+t_o$ (i.e.~in fact independent of $t$).
The one-particle Hamiltonian (Definition~\ref{dfn2.4}) is $h=A^{1/2}$.
\end{thm}
Let us calculate the (``four-smeared'') two-point function of
$\omega_t$:\\
For $h_1,h_2\in\D{\M}$, by equ.~\rf{2.14},
\beq
\Lambda^{(2)}_t(h_1,h_2) = \lambda_t^{(2)}\l({\rho_oEh_1\choose
\rho_1Eh_1 },{\rho_oEh_2\choose \rho_1Eh_2}\r),\label{3.11}
\eeq
where $E$ is the fundamental solution of the Klein-Gordon equation
in this spacetime, $\rho_o\Phi:=\Phi|_{\Sigma_t},\;\rho_1\Phi:=\frac
{\partial\Phi}{\partial t}|_{\Sigma_t}$, and the ``symplectically
smeared two-point function'' $\lambda_t^{(2)}$ is given on the initial
data $F_i={f_i\choose p_i}\in \Gamma$ by equ.~\rf{2.13},
\beqa
\lambda^{(2)}_t(F_1,F_2)&=& \l\la k^t F_1,k^tF_2\r\ra_\Ht \nonumber\\
&=&\frac{1}{2}\l\la A^{1/4}f_1-iA^{-1/4}p_1,A^{1/4}f_2-iA^{-1/4}p_2\r\ra
_\Ht  \nonumber\\
&=&\frac{1}{2}\l\la(A^{1/2}f_1-ip_1),
A^{-1/2}\l(A^{1/2}f_2-ip_2\r)\r\ra
_\Ht, \label{3.12}
\eeqa
since $A$ is selfadjoint. Combining \rf{3.11} and \rf{3.12} we obtain
\beq
\lt(h_1,h_2)=\frac{1}{2}\l\la\l(A^{1/2}\rho_o-i\rho_1\r)Eh_1,A^{-1/2}
\l(A^{1/2}\rho_o-i\rho_1\r)Eh_2\r\ra_\Ht \label{3.13}
\eeq
or in a more transparent integral representation (using $E'=-E$)
\beq
\lt(x_1,x_2)=-\frac{1}{2}\int_{\Sigma_t}d^3y\sqrt{h(\vec{y})}\,
\overline{E(x_1;t,\vec{y})\l(A^{1/2}-i\stackrel{\leftarrow}{\d{t}}\r)}
A^{-1/2}\l(A^{1/2}-i\stackrel{\rightarrow}{\d{t}}\r)E(t,\vec{y};x_2),
\label{3.14}
\eeq
where $A$ acts on $\vec{y}\in\Sigma_t$.
To show that $\omega_t$ is a Hadamard state we only have to prove that
the two-point distribution \rf{3.14} has the wavefront set \rf{3.6}
(Theorem~\ref{theorem3.4}). To this end, let us be slightly more
general than necessary at the moment. This will pay off in the
next section where we can apply the same method of proof to the
adiabatic vacua.\\
Let $K_1^t\in\Dp{\M\times \Sigma_t}$, $K_2^t\in\Dp{\Sigma_t\times\M}$
and $\Lambda_t^{(2)}\in \Dp{\M\times\M}$ be given such that
\beqa
\lt &=& K_1^t\circ K_2^t,\label{3.15}\\
K_1^t(x_1,\vec{y})&:=&\frac{1}{\sqrt{2}}\overline{E(x_1;t,\vec{y})\l(B
-i\stackrel{\leftarrow}{\d{t}}\r)} \nonumber\\
K_2^t(\vec{y},x_2)&:=& -\frac{1}{\sqrt{2}}A\l(B-i
\d{t}\r)E(t,\vec{y};x_2). \label{3.14a}
\eeqa
To calculate the wavefront set of the composition \rf{3.15} of two
distributions we prove a lemma which amounts to a proof of Theorem
\ref{theorem1.15} under somewhat different assumptions:
\begin{lemma}\label{lemma3.8}
If $A(t)$ and $B(t)$ are pseudodifferential operators on $\Sigma_t$
and $\lt, K_1^t, K_2^t$ are given as above, then
\beq
WF'(\lt)\subset WF'(K_1^t)\circ WF'(K_2^t). \label{3.15a}
\eeq
\end{lemma}
{\sc Proof}:\\
i) Our aim is to apply Theorem~\ref{theorem1.15}.
But we cannot do so directly since our
distributions $K_1^t$ and $K_2^t$ are not properly supported. However,
since $A\l(B-i\d{t}\r)$ is a pseudodifferential
operator we have by the pseudolocal property (Theorem
\ref{theorem1.10}) and Lemma \ref{lemma1.12a}a)
\beqa
WF'(K_1^t)&\subset& -WF'(E|_{\M\times\Sigma_t})\subset -\varphi^t_{2*}(C)
\nonumber\\
WF'(K_2^t)&\subset& WF'(E|_{\Sigma_t\times\M})\subset
\varphi^t_{1*}(C), \label{3.16}
\eeqa
where $C=WF'(E)$ (see Theorem~\ref{theorem1.17}a)) is given by equ.
\rf{1.30} and the second inclusion follows from Theorem
\ref{theorem1.14}, $\varphi_1^t,\varphi_2^t$ being the imbeddings
$\varphi_1^t:\Sigma_t\times\M\to \M\times\M$ and $\varphi_2^t:\M\times
\Sigma_t\to \M\times\M$.\\
Now we observe from \rf{3.16}
that $K_1^t$ and $K_2^t$ are properly supported w.r.t.~their
{\it singular} support. Therefore we can choose a properly
supported $\chi\in {\cal C}^\infty (\M\times \Sigma_t)$ with
$0\leq  \chi\leq 1,\;\chi=1$ in a neighborhood of
$singsupp\,E|_{\M\times \Sigma_t}$ and $\psi\in {\cal
    C}^\infty(\Sigma_t \times\M)$ with
  $\psi(\vec{y},x):=\chi(x,\vec{y}),\; x\in \M,\;\vec{y}\in\Sigma_t$.\\
Then $K_1^t\chi$ and $\psi K_2^t$ are properly supported, whereas
$K_1^t(1-\chi)$ and $(1-\psi)K_2^t$ are smooth. We decompose
\begin{eqnarray*}
\lt &=& (K_1^t\chi)\circ (\psi K_2^t) + K_1^t(1-\chi)\circ(1-\psi)K_2^t\\
& & +(K_1^t\chi)\circ (1-\psi)K_2^t + K_1^t(1-\chi)\circ(\psi K_2^t) \\
&=:& I_1 +I_2 +I_3 +I_4
\end{eqnarray*}
and claim that
\beq
WF(\lt)=WF(I_1) \label{3.17}
\eeq
(which, then, can be calculated from Theorem~\ref{theorem1.15}).
Since $I_2$ is a composition of smooth distributions it does not
contribute to the wavefront set. Let us consider
\[ I_4(x_1,x_2) = \int_{\Sigma_t}d^3y\sqrt{h(\vec{y})}\,
\l(K_1^t(1-\chi)\r)(x_1,\vec{y})\l(\psi K_2^t\r)(\vec{y},x_2).
\]
Localizing $I_4$ around $x_1$ and $x_2$ and taking the Fourier transform
we obtain
\beq
\hat{I}_4(\xi_1,\xi_2)\int d^3\eta\,[K_1^t(1-\chi)]\hat{\;}(\xi_1,-\eta)
(\widehat{\psi K_2^t})(\eta,\xi_2). \label{3.18}
\eeq
Since $\psi K_2^t$ is properly supported the integration over $\vec{y}\in
\Sigma_t$ is only over a compact set (which we can assume to be covered
by one coordinate patch), hence $\widehat{\psi K_2^t}(\eta,\xi_2)$ is
polynomially bounded in $|\eta|$, whereas $[K_1^t(1-\chi)]\hat{\;}
(\xi_1,-\eta)$ falls off rapidly in $|\eta|$ because $K_1^t(1-\chi)$
is smooth (which shows the existence of the integral).
The integrand of \rf{3.18} can be estimated by
\[ C_N(1+|\xi_1|+|\eta|)^{-N}(1+|\eta|+|\xi_2|)^k
\]
for arbitrary $N$ and some fixed $k$. If $\epsilon >0$ it follows
that $\hat{I}_4(\xi_1,\xi_2)$ is rapidly decreasing when either
$|\xi_1|>\epsilon |\xi_2|$ or $|\eta| >\epsilon|\xi_2|$.
Thus, by Theorem~\ref{theorem1.12}, the wavefront set of $I_4$ is
certainly contained in the set where this is not the case, namely
\begin{eqnarray*}
WF(I_4)&\subset& \{(x_1,0;x_2,\xi_2)\in T^*(\M\times\M);\;
(y,0;x_2,\xi_2)\in WF(E|_{\Sigma_t\times\M})\;\mbox{for some}\;
y\in \Sigma_t\}\\
&=&\M\times WF_\M(E|_{\Sigma_t\times\M})
\end{eqnarray*}
(see equ.~\rf{1.26}) which is however empty, as we shall find out in a
moment. Hence $I_4$ and (analogously) $I_3$ do not contribute to
$WF(\lt)$, i.e.~it holds \rf{3.17} as claimed.\\
ii) Since $I_1$ is the composition of two properly supported
distributions having the same singular points as $K_1^t$ and $K_2^t$
we obtain from \rf{3.17} and Theorem~\ref{theorem1.15}
\begin{eqnarray*}
WF'(\lt)&=& WF'(I_1) \\
 &\subset& WF'(K_1^t)\circ WF'(K_2^t)\cup (WF_\M(K_1^t)\times\M)
\cup(\M\times WF'_\M(K_2^t)).
\end{eqnarray*}
Now, by the defining equ.~\rf{1.26} and \rf{3.16},
\begin{eqnarray*}
WF_\M(K_1^t)&=&\{(x_1,\xi_1)\in T^*\M;\;(x_1,\xi_1;y,0)\in
WF(K_1^t)\;\mbox{for some}\; y\in \Sigma_t\}\\
&\subset& \{(x_1,\xi_1)\in T^*\M;\;(x_1,\xi_1;y,0)\in -\varphi^t_{2*}
(C)\;\mbox{for some}\;y\in\Sigma_t\}\\
&=&\emptyset
\end{eqnarray*}
by inspection of $C$ (equ.~\rf{1.30}) and similarly,
\[ WF'_\M(K_2^t)=\emptyset,
\]
hence $WF'(\lt)\subset WF'(K_1^t)\circ WF'(K_2^t)$ which was to be
proved.\hfill${\bf\Box}$\\
Now, with formula \rf{3.15a} at hand, we can calculate the wavefront
set of $\lt$. The next theorem contains one of the main
results of this work. It gives a sufficient criterion for a quasifree
state to be an Hadamard state. The idea is to use -- instead of
positive frequency solutions of the Klein-Gordon equation -- a separation
of the Klein-Gordon operator into first order factors that project out
the positive frequency parts of the fundamental solution:
\begin{thm}\label{theorem3.9}
Let $A(t)$ be an elliptic pseu\-do\-dif\-ferential operator on
$\Sigma_t$.
Let $B(t)$ be a pseu\-do\-dif\-ferential operator on $\Sigma_t$ such that
there exists a pseudodifferential operator $Q$ on $\M$ which has the
property $Q(B-i\partial_t)=\Box_g+\mu^2$ and possesses a principal
symbol $q$ with
\beq
q^{-1}(0)\setminus \{0\} \subset \{(x,\xi)\in T^*\M;\;\xi^0>0\}.
\label{3.19a}
\eeq
Let the two-point distribution $\lt\in\Dp{\M\times\M}$ of a quasifree
state be given by
\beq
\lt(h_1,h_2) =\frac{1}{2}\la(B-i\partial_t)Eh_1,
A(B-i\partial_t)Eh_2 \ra_{L^2_{\bf C}(\Sigma_t,d^3\sigma)}\label{3.19b}
\eeq
or in integral representation
\beq
\lt(x_1,x_2)=-\frac{1}{2}\int_{\Sigma_t}d^3y\sqrt{h(t,\vec{y})}\,
\ol{E(x_1;t,\vec{y})\l(B-i\stackrel{\leftarrow}{\partial_t}
\r)}A\l(B-i\stackrel{\rightarrow}{\partial_t}\r)E(t,\vec{y};x_2)
\label{3.19}
\eeq
where $A,B$ act on $\vec{y}\in
\Sigma_t$.\\
Then the wavefront set of $\lt$ is given by that of an Hadamard
distribution, namely
\[
WF(\lt)=\{(x_1,\xi_1;x_2,-\xi_2)\in T^*(\M\times\M)\setminus
\{0\};\;(x_1,\xi_1)\sim(x_2,\xi_2), \xi^0_1\geq0\}
\]
(see Theorem~\ref{theorem3.4}).
\end{thm}
{\sc Proof}:\\
i) Since $\lt$ is the two-point distribution of a quasifree state its
imaginary part must be proportional to the fundamental solution $E$ (by
equ.~\rf{2.14a}), and hence
\[
singsupp\:E\subset singsupp\:\lt,
\]
i.e.~$WF(\lt)$ is not empty.\\
ii) $\lt$ is Hermitean in the sense that $\lt(x_1,x_2)=\overline{\lt
(x_2,x_1)}$. Therefore (by Lemma \ref{lemma1.12a}b))
the wavefront set must be Hermitean in the sense
\[
\x\in WF(\lt)\Leftrightarrow (x_2,-\xi_2;x_1,-\xi_1)\in WF(\lt),
\]
i.e.~$WF'(\lt)$ must be symmetric.\\
iii) $\lt$ is a solution of the Klein-Gordon equation in both
arguments, i.e.~$\forall h_1,h_2\in \D{\M}$
\[
\lt((\Box_g+\mu^2)h_1,h_2)=\lt(h_1,(\Box_g+\mu^2)h_2)=0,
\]
therefore we can apply Theorem~\ref{theorem1.11a} to conclude that
\[
WF(\lt)\subset N\times N,
\]
where $N:=\{(x,\xi)\in T^*\M\setminus \{0\};\;g^{\mu\nu}\xi_\mu
\xi_\nu =0\}$ was already defined in equ.~\rf{1.28a}, and
$WF(\lt)$ must be invariant under the Hamiltonian vectorfield
\rf{1.29} (propagation of singularities), i.e.
\begin{eqnarray*}
\l.\begin{array}{c} \x\in WF(\lt)\\(x_2,\xi_2)\sim(x_2',\xi_2')
   \end{array}\r\}&\Rightarrow& (x_1,\xi_1;x_2',\xi_2')\in
 WF(\lt) \\
\l. \begin{array}{c} \x\in WF(\lt)\\(x_1,\xi_1)\sim (x_1',\xi_1')
   \end{array} \r\}&\Rightarrow& (x_1',\xi_1';x_2,\xi_2)\in WF(\lt).
\end{eqnarray*}
iv) To see that singularities can only occur on the lightcone let us
look at the initial data of $\lt$ on $\Sigma\times \Sigma$. Using
equations \rf{2.7} we calculate from \rf{3.19}
\beqa
\lt|_{\Sigma_t\times\Sigma_t}(f_1,f_2)&=&\frac{1}{2}\la f_1, Af_2\ra
\label{3.20}\\
\l(\d{x_1^0}\otimes{\bf 1}\r)\lt|_{\Sigma_t\times\Sigma_t}(f_1,f_2)&=&
\frac{i}{2}\la f_1,BA f_2\ra\nonumber\\
\l({\bf 1}\otimes \d{x_2^0}\r)\lt|_{\Sigma_t\times\Sigma_t}(f_1,f_2)&=&
\frac{i}{2}\la f_1,AB f_2\ra \nonumber\\
\l(\d{x_1^0}\otimes\d{x_2^0}\r)\lt|_{\Sigma_t\times\Sigma_t}(f_1,f_2)&=&
\frac{1}{2}\la f_1, BAB f_2\ra \nonumber
\eeqa
for $f_1,f_2\in\D{\Sigma_t}$, i.e.~the initial data are (proportional
to) the kernel distributions of pseu\-do\-dif\-ferential operators
and hence, by Lemma \ref{lemma1.5},
singular only on the diagonal of $\Sigma_t\times\Sigma_t$.
Therefore, by the propagation of singularities iii), singularities
can only occur if $x_1$ and $x_2$ are lightlike connected.\\
Summarizing i)--iv) we can conclude that $WF(\lt)$ must be a
non-empty subset of $\{\x\in N\times N;\;x_1
\;\mbox{and}\;x_2 \;\mbox{are lightlike connected}\}$ which is
invariant under the Hamiltonian vectorfield and Hermitean.\\
v) Now we are going to apply Lemma \ref{lemma3.8}.\\
Let $K_1^t$ and $K_2^t$ be defined as in \rf{3.14a}. First note,
that by Theorem~\ref{theorem1.11}, Theorem~\ref{theorem1.14} and
Lemma \ref{lemma1.12a}
\beqa
WF'(K_1^t)&=&WF'(\overline{(B-i\partial_t)E|_{\M\times \Sigma_t}})\subset
  - \varphi^t_{2*} WF'(({\bf 1}\otimes P)E) \nonumber\\
WF'(K_2^t) &=& WF'((B-i\partial_t)E|_{\Sigma_t\times\M})\subset
    \varphi^t_{1*} WF'((P\otimes {\bf 1})E) \label{3.21}
\eeqa
because $A$ is an elliptic pseudodifferential operator (we have
set $P:=B-i\partial_t$). Then decompose $E$ into
distinguished parametrices (see Theorem~\ref{theorem1.16})
\begin{eqnarray*}
E&=& \Delta_R-\Delta_A \\
&=& (\Delta_F-\Delta_A)+(\Delta_R-\Delta_F) \\
&=:& E^+ +E^-,
\end{eqnarray*}
where, by Theorem~\ref{theorem1.17} and the pseudolocal property
(Theorem~\ref{theorem1.10}),
\beqa
WF'((P\otimes {\bf 1})E^-) &\subset& WF'(E^-) = C\cap (N_-'\times
N_-') \nonumber\\
WF'(({\bf 1}\otimes P)E^+) &\subset& WF'(E^+) = C\cap (N_+'\times
N_+'). \label{3.22}
\eeqa
Now the essential observation is that
\beqa
(QP\otimes {\bf 1})E^-&=& ((\Box_g+\mu^2)\otimes {\bf 1})E^-\nonumber\\
&=& ((\Box_g+\mu^2)\otimes{\bf 1})(\Delta_R-\Delta_F) = 0\:(mod\:
{\cal C}^\infty) \label{3.24}\\
({\bf 1}\otimes QP)E^+ &=& ({\bf 1}\otimes(\Box_g+\mu^2))
(\Delta_F-\Delta_A) = 0\:(mod\:{\cal C}^\infty), \nonumber
\eeqa
since $\Delta_F, \Delta_R$ and $\Delta_A$ are parametrices of $(\Box_g+
\mu^2)$. Therefore, by Theorem~\ref{theorem1.11a} and \rf{3.22},
and using the assumption $q^{-1}(0)\setminus\{0\}\subset \{(x,\xi)\in
T^*\M;\;\xi^0>0\}$,
\beqa
WF'((P\otimes{\bf 1})E^-) &\subset&C\cap (N_-'\times N_-')\cap
(q^{-1}(0)\times T^*\M) = \emptyset\nonumber\\
WF'(({\bf 1}\otimes P)E^+) &\subset&C\cap (N_+'\times N_+')\cap
(T^*\M\times -q^{-1}(0)) = \emptyset,\label{3.25}
\eeqa
whereas
\beqa
WF'((P\otimes{\bf 1})E^+) &\subset&C\cap (N_+'\times N_+')\nonumber\\
WF'(({\bf 1}\otimes P)E^-) &\subset&C\cap (N_-'\times N_-')\label{3.26}
\eeqa
by the pseudolocal property.
Now we apply Lemma \ref{lemma3.8} to obtain from
equations \rf{3.21}, \rf{3.25} and \rf{3.26}
\begin{eqnarray*}
WF'(\lt)&\subset& WF'(K_1^t)\circ WF'(K_2^t) \\
&\subset& -\varphi^t_{2*}WF'(({\bf 1}\otimes P)(E^+ + E^-))
\circ \varphi^t_{1*} WF'((P\otimes {\bf 1})(E^+ +E^-))\\
&\subset&\varphi^t_{2*} (C\cap (N_+'\times N_+'))\circ
  \varphi^t_{1*}(C\cap(N_+'\times N_+'))\\
&\subset& C\cap (N_+'\times N_+').
\end{eqnarray*}
Together with i)--iv) we obtain $WF'(\Lambda_t^{(2)})= C\cap(N_+'
\times N_+')$, which is the wavefront set of an Hadamard distribution
as was to be proven.\hfill${\bf\Box}$\\  \\
As an immediate consequence of Theorem~\ref{theorem3.9} we can now
prove that the ultrastatic vacua defined by \rf{3.10} are Hadamard
states:
\begin{cor}\label{cor3.10}
Let $(\M,g)$ be an ultrastatic globally hyperbolic spacetime which
is foliated by compact spacelike Cauchy surfaces $\Sigma_t$.\\
Let $\omega$ be the quasifree ground state of the Klein-Gordon
quantum field on $(\M,g)$ as defined by \rf{3.10} and Theorem
\ref{theorem3.7}.\\
Then $\omega$ is an Hadamard state.
\end{cor}
{\sc Proof}:\\
Looking at the two-point distribution \rf{3.14} of $\omega$ it is
clear that it only remains to be shown that $A^{\pm 1/2}$ has the
properties which are demanded in Theorem~\ref{theorem3.9}.\\
But since $A=\overline{-^{(3)}\Delta_h +\mu^2}$ is an elliptic,
selfadjoint operator and $\Sigma_t$ compact we can apply Theorem~
\ref{theorem1.8} to conclude that $A^{\pm 1/2}$ is an elliptic
pseudodifferential operator with principal symbol
$(h^{ij}\xi_i\xi_j)^{\pm 1/2}$ of order 1. Furthermore $Q:=A^{1/2}+i
\partial_t$ is a pseudodifferential operator with principal symbol
$q=(h^{ij}\xi_i\xi_j)^{1/2} -\xi_0$, i.e.~it satisfies \rf{3.19a},
and $Q(A^{1/2}-i\partial_t)=A+\partial_t^2 = \Box_g+\mu^2$, since,
on an ultrastatic spacetime, $A$ is independent of $t$.
\hfill ${\bf \Box}$\\
{\bf Remarks}:
\begin{enumerate}
\item The restriction in the corollary to compact Cauchy surfaces has
the technical reason that only for those Theorem~\ref{theorem1.8}
is formulated in the mathematical literature. In fact, Corollary
\ref{cor3.10} also holds in the non-compact case as was proven in
\cite{FNW81}.
\item The statement of the corollary also holds for static spacetimes
\rf{3.8a}
on which the norm of the timelike Killing field $t^\mu$ is bounded from
below by a positive constant $\epsilon$:
\beq
\alpha(\vec{x})^2 =g_{\mu\nu}(x) t^\mu t^\nu \geq \epsilon >0 \;\forall\;
x \in \M. \label{3.27}
\eeq
In this case, the two-point function of the groundstate reads
\cite{Kay78}
\[ \Lambda^{(2)}(h_1,h_2) = \frac{1}{2}\la
(B-\frac{i}{\alpha}\partial_t) Eh_1, B^{-1}
(B-\frac{i}{\alpha}\partial_t) Eh_2\ra_\Ht \]
with
\begin{eqnarray*}
B &:=& \alpha^{-1/2}(\alpha^{1/2} \bar{A}\alpha^{1/2})^{1/2}\alpha^{-1/2}\\
A &:=& -(\partial^i \alpha)\partial_i +\alpha(- ^{(3)}\Delta_h +\mu^2)
\end{eqnarray*}
and it holds
\[\frac{1}{\alpha}(i\partial_t+B\alpha)(B-\frac{i}{\alpha}\partial_t)=
\frac{1}{\alpha^2}\partial_t^2 +\frac{1}{\alpha} A = \Box_g+\mu^2. \]
Since pseudodifferential operators commute in highest order the
principal symbol of $B$ is again $(h^{ij}\xi_i \xi_j)^{1/2}$ and the
proof from above carries over immediately.
\item Condition \rf{3.27} however cannot be relaxed further, since
e.g.~the ground state w.r.t.~the static Killing field in the
Schwarzschild metric (``Boulware vacuum'') is known to be not a
Hadamard state \cite{KW91}. There, \rf{3.27} is violated
on the horizon, and the mathematical analysis from above breaks down
since the corresponding operator is no longer elliptic.
\item Similar statements for stationary spacetimes (i.e.~spacetimes
with a timelike Killing field which is not necessarily
hypersurface-orthogonal) are not known since the two-point
function of the ground state (see \cite{Kay78}) cannot be
represented by such an explicit formula like \rf{3.14}.
\end{enumerate}
Let us consider next the KMS-states arising when one ``heats up''
the ground state \rf{3.10} to a temperature $T=1/\beta$.
Kay \cite{Kay85b} shows how one can construct quasifree KMS-states
from a given quasifree groundstate:\\
Let $(k,{\cal H}, e^{-iht})$ be a one-particle Hilbertspace structure
of a quasifree groundstate (Definition~\ref{dfn2.4})
over a phase space $(\Gamma, \sigma, {\cal T}(t)$) such that
$k\Gamma \subset \D{h^{-1/2}}$ (what Kay calls ``regularity condition'').
Then a one-particle structure for a quasifree KMS-state at temperature
$T=1/\beta$ is given by $(k^\beta, \tilde{{\cal H}}, e^{-i\tilde{h}t})$
as follows:
\beqa
k^\beta: \Gamma &\to& \tilde{{\cal H}}:={\cal H}\oplus {\cal H}\nonumber\\
 F &\mapsto& C(\sinh Z^\beta)kF\oplus (\cosh Z^\beta)kF,\label{3.27a}\\
e^{-i\tilde{h}t} &=& \l(\begin{array}{cc} e^{ith} & 0\\0&e^{-ith}
\end{array} \r), \nonumber
\eeqa
where $Z^\beta$ is implicitly defined by $\tanh Z^\beta=e^{-\beta h}$,
i.e.
\[ \sinh Z^\beta = \frac{e^{-\beta h/2}}{(1-e^{-\beta h})^{1/2}},\;
   \cosh Z^\beta = \frac{1}{(1-e^{-\beta h})^{1/2}},
\]
and $C:{\cal H}\to {\cal H}$ is a complex conjugation such that
$C e^{-ith} = e^{ith} C$. Since $\D{h^{-1/2}}\subset \D{\sinh Z^\beta},
\D{\cosh Z^\beta}$ (see \cite{Kay85b}) the regularity condition
guarantees that \rf{3.27a} is well defined.\\
In our case of the ultrastatic groundstate \rf{3.10} the regularity
condition is satisfied since $A^{1/4} \co{\Sigma_t}\subset A^{1/4}
\D{A^{1/2}}\subset \D{A^{-1/4}} = \D{h^{-1/2}}$ and $A^{-1/4}
\co{\Sigma_t} \subset A^{-1/4}\D{A^{-1/2}}\subset \D{A^{-1/4}}=
\D{h^{-1/2}}$. Representing $C$ by the ordinary complex conjugation
on ${\cal H}=L^2_{\bf C}(\Sigma_t,d^3\sigma)$ we can compute the
two-point function \rf{2.13}, \rf{2.14} of an ultrastatic KMS-state as
\beqa
\lambda^{(2)}_\beta (F_1,F_2) &=& \la k^\beta F_1,k^\beta F_2\ra_{
\tilde{{\cal H}}} \nonumber\\
&=& \frac{1}{2}\l\{\la C(\sinh Z^\beta )(A^{1/4} f_1-i A^{-1/4}p_1),
C(\sinh Z^\beta )(A^{1/4}f_2 -i A^{-1/4}p_2)\ra_{\cal H}\r.\nonumber\\
& &\l. +\la(\cosh Z^\beta)(A^{1/4}-i A^{-1/4}p_1), (\cosh Z^\beta)
(A^{1/4}f_2-i A^{-1/4}p_2)\ra_{\cal H}\r\}\nonumber\\
&=& \frac{1}{2}\l\{\la (A^{1/2}f_1+ip_1), (\sinh^2 Z^\beta)A^{-1/2}
(A^{1/2}f_2 +i p_2)\ra_{\cal H} \r.\nonumber\\
& & \l.+ \la (A^{1/2}f_1-ip_1),(\cosh^2 Z^\beta)A^{-1/2}(A^{1/2}f_2
-ip_2)\ra_{\cal H}\r\} \label{3.27b}
\eeqa
for $F_i = (f_i,p_i)\in \Gamma = \co{\Sigma_t}\oplus \co{\Sigma_t}$,
using the selfadjointness of $A$, and
\beqa
\Lambda^{(2)}_\beta(h_1,h_2) &=& \frac{1}{2}\l\{\la (A^{1/2}+i\partial_t
)Eh_1,(\sinh^2 Z^\beta) A^{-1/2}(A^{1/2}+i\partial_t)Eh_2\ra_{\cal H}\r.
\nonumber\\
& &+\l. \la (A^{1/2}-i\partial_t)Eh_1,(\cosh^2 Z^\beta)A^{-1/2}(A^{1/2}-i
\partial_t)Eh_2\ra_{\cal H}\r\} \label{3.27c}
\eeqa
for $h_i\in \D{\M}, i=1,2$, and we claim:
\begin{cor}\label{cor3.10a}
Let ($\M,g$) be an ultrastatic globally hyperbolic spacetime with
compact spacelike Cauchy surfaces $\Sigma_t$. Let $\omega_\beta$ be
the quasifree KMS-state ($\beta>0$) of the Klein-Gordon quantum field
on $(\M,g)$ as defined by \rf{3.27a}.\\
Then $\omega_\beta$ is an Hadamard state.
\end{cor}
{\sc Proof}:\\
Let us look at the two-point function \rf{3.27c}. The second term in
\rf{3.27c} is again of the form \rf{3.19b}, but with the operator $A$
in the middle replaced by
\[(\cosh^2 Z^\beta) A^{-1/2}=\frac{A^{-1/2}}{1-e^{-\beta A^{1/2}}},\]
which is, by Theorem \ref{theorem1.8}, an elliptic pseudodifferential
operator. Hence the second term contributes to $\Lambda_\beta^{(2)}$
the wavefront set of an Hadamard distribution, whereas the first term
is in fact smooth:\\
Using \rf{2.7} we can compute the distributional initial data of the
first term in \rf{3.27c} obtaining distributions which have as kernels
the pseudodifferential operators
$(\sinh^2 Z^\beta)A^{1/2},\;\sinh^2 Z^\beta$ and $(\sinh^2 Z^\beta)
A^{-1/2}$. But noting that
\[\sinh^2Z^\beta = \frac{e^{-\beta A^{1/2}}}{1-e^{-\beta A^{1/2}}}\]
and using again Theorem \ref{theorem1.8} we see that the principal
symbols of these operators fall off faster than any inverse
power in $\xi$. Therefore, by Lemma \ref{lemma1.5}b), these
distributions have smooth kernels, and, consequently, so has the first
term of \rf{3.27c}.\hfill{$\bf\Box$}\\
{\bf Remark:}\\
This result seems to be new. As for the groundstate, it also immediately
extends to the case of a static metric satisfying \rf{3.27}.
\section{Adiabatic vacua on Robertson-Walker spaces}\label{section2.5}
Adiabatic vacua were introduced by Parker \cite{Parker69,Parker71}
in order to investigate the particle production in the expanding
universe. A mathematically precise definition was given by L{\"u}ders
and Roberts \cite{LR90}. In the following we will define
adiabatic vacua following \cite{LR90} and review this paper as far as
it is necessary for our purposes. Then we will state and prove one of the
main results of this work, namely that all adiabatic vacuum states are
Hadamard states.\\
The homogeneous and isotropic spacetimes are the Lorentz manifolds
of the form $\M^\kappa={\bf R}\times \Sigma^\kappa,\;\kappa=-1,0,+1$,
endowed with the {\bf Robertson-Walker metrics}
\beq
ds^2= dt^2-a(t)^2\l[\frac{dr^2}{1-\kappa
  r^2}+r^2(d\theta^2+\sin^2\theta d\varphi^2)\r]\label{3.28}
\eeq
($\varphi\in [0,2\pi], \theta\in [0,\pi], r\in [0,\infty)$ for
$\kappa=0,-1,\;r\in [0,1)$ for $\kappa=+1$),
where $a$ is a strictly positive smooth function and $\Sigma^\kappa$
a homogeneous Riemannian manifold with constant negative ($\kappa
=-1$), positive ($\kappa=+1$) or zero ($\kappa=0$) curvature.
Choosing the simplest topologies for $\Sigma^\kappa$ we can regard
$\Sigma^\kappa$ as being embedded in ${\bf R}^4$:
\beqa
\Sigma ^+ &=&\{x\in{\bf R}^4;\;(x^0)^2+\sum_{i=1}^3(x^i)^2=1\},\nonumber\\
\Sigma^0 &=&\{x\in {\bf R}^4;\;x^0=0\},\label{3.29}\\
\Sigma^- &=&\{x\in {\bf R}^4;\;(x^0)^2-\sum_{i=1}^3(x^i)^2=1,x^0>0\}.
 \nonumber
\eeqa
Since $\Sigma^+$ is compact $\M^+$ is called a ``closed'' universe,
whereas $\M^0$ and $\M^-$ are models for ``open'' universes having
noncompact spatial sections. (For an interpretation of these as
cosmological models see e.g.~\cite{Wald84}.)
The Riemannian metric
\beq
s_{ij}^\kappa =\l( \begin{array}{ccc}\frac{1}{1-\kappa r^2}& & \\
              & r^2 & \\
              &     & r^2 \sin^2\theta \end{array}\r)\label{3.30}
\eeq
is induced on $\Sigma^\kappa$ by the Euclidean metric on ${\bf R}^4$
for $\kappa =+1,0$ and the Minkowski metric for $\kappa =-1$.
These spaces are homogeneous for the rotation group SO(4) ($\kappa=+1$),
the Euclidean group E(3) ($\kappa=0$), resp.~the Lorentz group
${\cal L}_+^\uparrow(4)$ ($\kappa=-1$).\\
The $\M^\kappa$ are globally hyperbolic and
the hypersurfaces $\Sigma_t^\kappa:=\{t\}\times \Sigma^\kappa$ are
Cauchy surfaces of $\M^\kappa$ with 3-metric $h_{ij}^\kappa = a^2(t) s_{ij}^
\kappa$. Their future-directed normal field is given by $n^\alpha =(1,0,0,0).$
It is geodesic ($n^\alpha\nabla_\alpha n^\beta =0$). The exterior
curvature $K$ of $\Sigma_t^\kappa$ is
\beq
K=\nabla_\alpha n^\alpha = 3\frac{\dot{a}(t)}{a(t)}.\label{3.31}
\eeq
In what follows we will omit the index $\kappa$ unless it is necessary
to specify one of its values.\\
We want to consider linear scalar fields on these spaces and therefore have to
study the Klein-Gordon equation in the background
\rf{3.28}
\beq
(\Box_g+\mu^2)\Phi =\frac{\partial^2\Phi}{\partial
  t^2}+3\frac{\dot{a}}{a}\frac{\partial\Phi}{\partial t}
+(-^{(3)}\Delta_h+\mu^2)\Phi =0, \label{3.32}
\eeq
where $^{(3)}\Delta_h$ is the Laplace-Beltrami operator on $\Sigma_t$,
\beqa
^{(3)}\Delta_h &=& \frac{1}{a^2}\l\{(1-\kappa r^2)\dd{r}+\frac{2-3\kappa
r^2}{r}\d{r}+\frac{1}{r^2}\Delta(\theta,\varphi)\r\}\label{3.33}\\
\Delta(\theta,\varphi) &:=& \frac{1}{\sin\theta}\l[\d{\theta}
\l(\sin\theta \d{\theta}\cdot\r)+\frac{1}{\sin\theta}\dd{\varphi}\cdot
\r].\nonumber
\eeqa
The partial differential equation \rf{3.32} can be separated by
\[\Phi(t,r,\theta,\varphi) = \int d\mu(\vec{k})\,T_{\vec{k}}(t)\phi_{\vec{k}}
(r,\theta,\varphi)
\]
into an ordinary differential equation for the time dependent part
$T_{\vec{k}}(t)$
\beq
\ddot{T}_{\vec{k}}+3\frac{\dot{a}}{a}\dot{T}_{\vec{k}}+\omega_k^2
T_{\vec{k}}
=0 \label{3.34}
\eeq
where
\beq
\omega_k^2(t) :=\frac{E(k)}{a^2(t)} +\mu^2 \label{3.35}
\eeq
and eigenfunctions $\phi_{\vec{k}}(r,\theta,\varphi)$ of the
Laplace-Beltrami operator on a hypersurface $\Sigma_t$:
\beq
^{(3)}\Delta_h \ph = -\frac{E(k)}{a^2}\ph.\label{3.36}
\eeq
(Note that $^{(3)}\Delta_h$, equ.~\rf{3.33}, is of the form
$\frac{1}{a^2}\tilde{\Delta}$, where $\tilde{\Delta}$ is the
Laplace-Beltrami
operator of $s_{ij}$, equ.~\rf{3.30}, so the $\ph$ live in fact on
$\Sigma$ and are independent of $t$.) The notation we have used is the
following:
\beqa
\dm &:=& \sum_{k=0}^\infty \sum_{l=0}^k \sum_{m=-l}^l,\;\vec{k}:=
(k,l,m),\;E(k):=k(k+2)\;\mbox{for}\;\kappa=+1 \nonumber\\
\dm &:=& \int_{{\bf R}^3}d^3k,\;\vec{k}:=(k_1,k_2,k_3)\in{\bf R}^3,\;
k:=|\vec{k}|,\;E(k):=k^2\;\mbox{for}\;\kappa=0 \label{3.37}\\
\dm &:=&\int_{{\bf R}^3}d^3k,\;\vec{k}\in {\bf R}^3,\;k:=|\vec{k}|,\;
E(k):=k^2+1\;\mbox{for}\;\kappa=-1.\nonumber
\eeqa
The (generalized) eigenfunctions $\ph(r,\theta,\varphi)$ are
$\frac{1}{(2\pi)^{3/2}}e^{i\vec{k}\vec{x}}$ for $\kappa=0$,
$A_{kl}\Pi^+_{kl}(r)Y_{lm}(\theta,\varphi)$ for $\kappa =+1$
(where the $Y_{lm}$ are the spherical harmonics on the two-sphere,
$\Pi_{kl}^+(r)$ are related to the Gegenbauer polynomials and
$A_{kl}$ are normalization constants) and
$\frac{1}{(2\pi)^{3/2}}(x\xi)^{-1+ik}$ for $\kappa=-1$ (where
$\xi:=(1,\vec{\xi}),\;\vec{\xi}:=\frac{\vec{k}}{|\vec{k}|},\;
x\xi=x^0-\vec{x}\vec{\xi}=\sqrt{1+r^2}-\vec{x}\vec{\xi}$).
For details see \cite{LR90}.
In each case the system of eigenfunctions is orthonormal and
complete, i.e.~we can define a generalized Fourier transform
by
\beqa
\tilde{\;}:L^2(\Sigma) &\to& L^2(\tilde{\Sigma}) \nonumber\\
h&\mapsto& \tilde{h}(\vec{k}):= (\ph,h)\equiv \int_\Sigma d^3\sigma
\,\overline{\ph(\vec{y})}h(\vec{y}), \label{3.38}
\eeqa
with $d^3\sigma :=\sqrt{|s|}d^3y = \frac{1}{\sqrt{1-\kappa r^2}}
r^2 dr\sin\theta d\theta d\varphi$ and $\tilde{\Sigma}$ the
momentum space associated to $\Sigma$ (i.e.~the range of values of
$\vec{k}$ equipped with the measure $d\mu(\vec{k})$).
The inverse is given by
\beq
h(\vec{y}) = \dm \ph(\vec{y})\tilde{h}(\vec{k}). \label{3.39}
\eeq
(Note that \rf{3.38} is defined on $\Sigma$ and not on the Cauchy
surface $\Sigma_t$.)\\
We consider the phase space $(\Gamma,\sigma)$ of initial data
$\Gamma:=\co{\Sigma}\oplus a^3\co{\Sigma}$ on $\Sigma$ with the
symplectic form
\[
\sigma(F_1,F_2) = -a^3 \int_\Sigma d^3\sigma\, [f_1p_2-p_1f_2]
\]
for $F_i:={f_i\choose a^3 p_i}\in \Gamma,i=1,2,$ and the Weyl algebra of
the
linear scalar field associated with $(\Gamma,\sigma)$\footnote{We
insert the factor $a^3(t)$ in the second component of the initial
data to comply with the conventions of \cite{LR90}. We differ from
\cite{LR90} in so far as they use a phase space with elements
$F:=(a^3 p,-f)$.}.
\begin{thm}[L{\"u}ders and Roberts \cite{LR90}]\label{theorem3.11}
The homogeneous and isotropic Fock states for the free Klein-Gordon
field in a Robertson-Walker spacetime are given by the following
two equivalent constructions:
\begin{enumerate}
\item[a)] a two-point function
\beqa
\lambda^{(2)}(F_1,F_2)&=&\dm\la \overline{\tilde{F}_1(\vec{k})}, S(k)
\tilde{F}_2(\vec{k})\ra \label{3.40}\\
S(k) &:=& \l(\begin{array}{cc} |p(k)|^2 & -q(k)\overline{p(k)}\\
                -\overline{q(k)}p(k) & |q(k)|^2
              \end{array}\r),\label{3.41}
\eeqa
where $p(k)$ and $q(k)$ are (essentially polynomially bounded
measurable) complex valued functions satisfying
\beq
\overline{q(k)}p(k)-q(k)\overline{p(k)}=-i.\label{3.42}
\eeq
\item[b)] a representation of the field operators
\beq
\hat{\Phi}(t,\vec{x})= \dm\,[a(\vec{k})\ph(\vec{x})
\ol{T_k(t)}+a^*\vk\ol{\ph(\vec{x})} T_k(t)]\label{3.43}
\eeq
on a bosonic Fock space with one-particle space $L^2(\tilde{\Sigma})$
and annihilation and creation operators $a$ and $a^*$,
\[
[a(\tilde{f}_1),a^*(\tilde{f}_2)]=\dm\,\ol{\tilde{f}_1\vk} \tilde{f}_2\vk
,\;\mbox{for}\;\tilde{f}_1,\tilde{f}_2\in L^2(\tilde{\Sigma}),
\]
where the complex valued functions $T_k(t)$ have to obey the
differential equation \rf{3.34} and the constraint
\beq
\ol{T_k}\dot{T}_k-T_k\ol{\dot{T}_k}=-\frac{i}{a^3}.\label{3.44}
\eeq
\end{enumerate}
\end{thm}
{\bf Remarks:}
\begin{enumerate}
\item That the such defined Fock states are homogeneous and isotropic
can be read off from the facts that they are constructed w.r.t.~the
homogeneous surfaces $\Sigma$ and that $S(k)$ resp.~$T_k(t)$ do not
depend on the full vectors $\vec{k}$, but only on the norm
$k=|\vec{k}|$ (as defined in \rf{3.37}).
\item The conditions \rf{3.42} resp.~\rf{3.44} guarantee that the
antisymmetric part of $\lambda^{(2)}$ is $i/2$ times the symplectic
form $\sigma$ (which is a necessary condition for two-point functions
of quasifree states according to equ.~\rf{2.13}).
\item The step from b) to a) is the usual way of constructing Fock
states via mode decomposition of the field operators (see e.g.~\cite
{BD82}). Since $T_k(t)$ obeys \rf{3.34} the field \rf{3.43} is a
(distributional) solution of the Klein-Gordon equation \rf{3.32}.
Therefore, starting from \rf{3.43} we may define a representation
of the field $\vartheta$ and its canonical conjugate momentum
$\pi:=\sqrt{|g|}\frac{\partial \Phi}{\partial t}$ on a
Cauchy surface $\Sigma_t$ and obtain for a testfunction $f\in \D{\M}$
(see Theorem~\ref{theorem2.1}b))
\[
\hat{\Phi}(f)=\vartheta(a^3\rho_1Ef)-\pi (\rho_oEf).
\]
Putting $F_i:={\rho_oEf_i\choose a^3\rho_1Ef_i}\in\Gamma, i=1,2$, and
defining the Fock state by $a\vk |0\ra = 0$ we can
calculate the two-point function as
\[
\Lambda^{(2)}(f_1,f_2)=\la 0|\Phi(f_1)\Phi(f_2)|0\ra =
\dm\,\l\la \ol{\tilde{F}_1\vk},\l(\begin{array}{cc}
 a^6\ol{\dot{T}_k}\dot{T}_k & -a^3 \ol{\dot{T}_k} T_k \\
 -a^3\ol{T_k} \dot{T}_k & \ol{T_k} T_k \end{array}\r)\tilde{F}_2\vk\r\ra.
\]
Comparing with \rf{3.41} we see that $p(k)$ and $q(k)$ are (proportional
to) the initial data of $T_k(t)$ on the Cauchy surface $\Sigma_t$.
\item The conclusion from a) to b) is proven in \cite{LR90}
imposing a certain continuity condition on the two-point function
$\lambda^{(2)}$. The Fock representation \rf{3.43} is constructed from
\rf{3.41} in such a way that $T_k(t)$ is a solution of \rf{3.34}
with initial data given by $p(k)$ and $q(k)$. \rf{3.44} follows
from \rf{3.42} initially at a time $t_o$, say. But then it holds for
all times, because putting $G(t):=\ol{T_k}\dot{T}_k-T_k\ol{\dot{T}_k}
+\frac{i}{a^3}$ we obtain from \rf{3.34} $\dot{G}_k(t)+3\frac{\dot{a}}{a}
G_k(t)=0$ which yields, together with $G_k(t_o)=0$, that $G_k(t)\equiv
0$ for all $t$.
\end{enumerate}
As we have seen, the only freedom in the choice of certain
homogeneous and isotropic Fock states is the choice of initial data
for the function $T_k(t)$. Parker's \cite{Parker69,Parker71} physical
motivation for the definition of the adiabatic vacua was to choose
$T_k(t)$ such as to minimize the particle creation in the expanding
universe. He achieved this by the use of a WKB-expansion around the
ultrastatic groundstate (which only exists if $a(t)=const.$).
The formal definition is according to \cite{LR90}:
\begin{dfn}\label{dfn3.12}
An {\rm\bf adiabatic vacuum state of order $n$} is a homogeneous, isotropic
Fock state whose two-point function \rf{3.40} is given by functions
$q(k):=T_k(t), p(k):=a^3\dot{T}_k(t)$ where $T_k(t)$ is a solution of
the differential equation \rf{3.34} with initial conditions at time
$t$
\beqa
T_k(t) &=& W_k^{(n)}(t) \nonumber\\
\dot{T}_k(t) &=& \dot{W}_k^{(n)}(t).\label{3.45}
\eeqa
Here,
\beq
W_k^{(n)}(t) := \frac{1}{a^{3/2}(t)\sqrt{2\Omega_k^{(n)}(t)}}
e^{-i\int_{t_o}^t\Omega_k^{(n)}(t')\,dt'}\label{3.46}
\eeq
is iteratively defined by
\beqa
(\Omega_k^{(0)})^2 &:=& \omega_k^2 = \frac{E(k)}{a^2}+\mu^2
\nonumber\\
(\Omega_k^{(n+1)})^2 &=& \omega_k^2-\frac{3}{4}\l(\frac{\dot{a}}{a}\r)^2
-\frac{3}{2}\frac{\ddot{a}}{a}+\frac{3}{4}\l(\frac{\dot{\Omega}_k^{(n)}}
{\Omega_k^{(n)}}\r)^2-\frac{1}{2}\frac{\ddot{\Omega}_k^{(n)}}
{\Omega_k^{(n)}}. \label{3.47}
\eeqa
\end{dfn}
{\bf Remarks:}
\begin{enumerate}
\item \rf{3.46}, \rf{3.47} is an iterative solution to \rf{3.34}.
If $a(t)=const.$ one obtains the ultrastatic groundstate.
The iteration procedure may break down yielding negative values
for $(\Omega_k^{(n+1)})^2$. But one can show \cite{LR90} that
for a finite time interval and sufficiently large $k$ $\Omega_k^{(n)}$
is always strictly positive. $\Omega_k^{(n)}$ can then be continued
(smoothly in $t$) to {\it all} values of $k$.
\item Condition \rf{3.44} (resp.~\rf{3.42}) is automatically satisfied
by the Ansatz \rf{3.46}.
\item An adiabatic vacuum state depends on
\begin{itemize}
\item the choice of initial time $t$ in \rf{3.45},
\item the order of iteration $n$,
\item the extrapolation of $\Omega_k^{(n)}$ to small values of $k$.
\end{itemize}
\end{enumerate}
L{\"u}ders and Roberts \cite{LR90} show that all adiabatic vacuum
states are locally quasiequivalent. To this end, they consider a
Bogoliubov transformation between two Fock states parametrized
by functions $(q(k),p(k))$ resp.~$(q'(k),p'(k))$
\beqa
q'(k)&=&\alpha(k)q(k) +\beta(k) \ol{q(k)}, \nonumber\\
p'(k) &=& \alpha(k)p(k) +\beta(k) \ol{p(k)}, \label{3.48}
\eeqa
where the Bogoliubov coefficients $\alpha$ and $\beta$ have to satisfy
\[
|\alpha(k)|^2-|\beta(k)|^2=1.
\]
Using \rf{3.42} one can solve for $\beta$ and gets
\beq
\beta(k) =- i(p'(k)q(k)-q'(k)p(k)).\label{3.49}
\eeq
The property of quasiequivalence of two such Fock states now depends
on the asymptotic behaviour of $\beta(k)$.
On the way to their main result they prove a necessary condition for
quasiequivalence which we state
(only for the case of the closed universe) for later reference:
\begin{lemma}[\cite{LR90}, p.47]\label{lemma3.13}
If two Fock states (parametrized by functions ($q(k),p(k)$) resp. ($
q'(k),p'(k)$)) of the Klein-Gordon quantum field on the closed
universe are locally quasiequivalent then
\beq
\sum_{k=0}^\infty (k+1)^2|\beta(k)|^2<\infty. \label{3.50}
\eeq
\end{lemma}
The main result of \cite{LR90} now reads:
\begin{thm}[Theorems 3.3 and 5.7 of \cite{LR90}]\label{theorem3.14}
a) In a closed Robertson-Walker spacetime any two adiabatic vacuum
states are unitarily equivalent.\\
b) In an open Robertson-Walker spacetime any two adiabatic vacuum
states of iteration order $n\geq 1$ are locally quasiequivalent.\\
c) If $\omega$ is an adiabatic vacuum state on an open Robertson-Walker
spacetime then $\pi_\omega |_{\cal O}$ is a factor when ${\cal O}=
D({\cal C})$ and ${\cal C}$ is an open bounded subset of some $\Sigma_t$
with smooth boundary.
\end{thm}
Thus, the class of adiabatic vacuum states satisfies the principle of
local definiteness. Furthermore, for these states the expectation
value of the energy-momentum tensor can be constructed using an
adiabatic regularization procedure due to Parker and Fulling
\cite{PF74}. So the adiabatic vacua seem to be as good a class of
physical states as the class of Hadamard states, and naturally the
question arises what the connection between Hadamard states and
adiabatic vacua might be.\\
Najmi and Ottewill \cite{NO85} show that a Hadamard state on
a Robertson-Walker space with flat spatial sections ($\kappa=0$)
has the same asymptotic behaviour in momentum space as an adiabatic
vacuum state of order 0. Bernard \cite{Bernard86} computes the high
energy behaviour of Hadamard states on a Bianchi type-I spacetime and
finds it in agreement with that of an adiabatic vacuum state of order
2. In \cite{MPC87} it is shown that the anticommutator function of
certain states which are constructed by a WKB-expansion very similar
to \rf{3.46}, \rf{3.47} has Hadamard singularities. These results
led L{\"u}ders and Roberts \cite{LR90} to conjecture that Hadamard states
and adiabatic vacua define the {\it same} class of physical states.
Pirk \cite{Pirk93} claims to have proven that in a spatially flat
Robertson-Walker spacetime an adiabatic vacuum state is an Hadamard
state if and only if it is of infinite order. The if-part of his
``proof'' is -- to say the least -- not trustworthy since he does not
control convergence of infinite series and there is a priori no reason
that an iteration like \rf{3.47} will converge, the only-if-part is
certainly false as we will see in a moment.\\
Now we formulate the first main physical result of this work in the
following theorem:
\begin{thm}\label{theorem3.15}
The adiabatic vacuum states of order $n\in{\bf N}_o$ of a linear
Klein-Gordon quantum field \rf{3.32} on the Robertson-Walker
spacetimes \rf{3.28} are Hadamard states.
\end{thm}
Before we prove the theorem we state an immediate consequence:
\begin{cor}\label{cor3.16}
All Hadamard states and adiabatic vacuum states of a linear
Klein-Gordon quantum field on the Robertson-Walker spacetimes
lie in the same local primary folium (quasiequivalence class) of states.
(This also extends the validity of Theorem~\ref{theorem3.14} to the
case of the adiabatic vacuum of order 0 on the open Robertson-Walker
spaces.)
\end{cor}
{\sc Proof} of Theorem~\ref{theorem3.15}:\\
The idea is again to compute the wavefront set of the two-point
function of an n-th order adiabatic vacuum state. Starting with
\rf{3.40}, \rf{3.41} and inserting Definition~\ref{dfn3.12} we obtain
for the two-point function:
\begin{eqnarray*}
\Lambda_n^{(2)}(f_1,f_2) &=& \lambda_n^{(2)} \l({\rho_o Ef_1\choose
a^3 \rho_1 Ef_1},{\rho_o Ef_2\choose a^3 \rho_1 Ef_2}\r) \\
&=&\dm\, \l\la \ol{{\widetilde{\rho_oEf_1}\choose
a^3\widetilde{\rho_1 Ef_1}}},
\l(\begin{array}{cc}a^6 |\dot{T}_k^{(n)}|^2 & -a^3T_k^{(n)}
\ol{\dot{T}_k^{(n)}}\\
-a^3 \ol{T_k^{(n)}}\dot{T}_k^{(n)} & |T_k^{(n)}|^2
\end{array}\r){\widetilde{\rho_oEf_2}\choose a^3\widetilde{\rho_1Ef_2}}\r\ra,
\end{eqnarray*}
for $f_1,f_2\in \D{\M}$, where $T_k^{(n)}(t)$ denotes the solution of
\rf{3.34} with initial conditions \rf{3.45}. We use the orthogonality
relation $(\ph,\phi_{\vec{l}})_{L^2(\Sigma)}= \delta(\vec{k},\vec{l})$
where $\delta$ is the delta function w.r.t.~the measure $d\mu(\vec{k})$:
\begin{eqnarray*}
\Lambda_n^{(2)}(f_1,f_2) &=& a^3 \dm \int d\mu(\vec{l})\,\delta(\vec{k},
\vec{l})\l\la\ol{{\rho_o(\widetilde{Ef_1})(\vec{k})\choose
\rho_1(\widetilde{Ef_1})(\vec{k})}}, \l(\begin{array}{cc}
a^3 \ol{\dot{T}_k^{(n)}}\dot{T}_l^{(n)} & -a^3\ol{\dot{T}_k^{(n)}} T_l^{(n)}\\
-a^3\ol{T_k^{(n)}}\dot{T}_l^{(n)} & a^3 \ol{T_k^{(n)}}T_l^{(n)}
\end{array}\r)\r. \\
& &\hspace{11cm}\l.{\rho_o(\widetilde{Ef_2})(\vec{l})\choose
  \rho_1(\widetilde{Ef_2})(\vec{l})}\r\ra \\
&=&a^3\int_\Sigma d^3\sigma_y
\dm\int d\mu(\vec{l})\,\ol{\ph(\vec{y}) (\widetilde{Ef_1})
(t,\vec{k})}\l[a^3\ol{\dot{T}_k^{(n)}}\dot{T}_l^{(n)} - \stackrel
{\leftarrow}{\partial_t}a^3\ol{T_k^{(n)}}\dot{T}_l^{(n)} \r.\\ \nopagebreak
& &\l.-a^3\ol{\dot{T}_k^{(n)}}T_l^{(n)}\stackrel{\rightarrow}{\partial_t}
+\stackrel{\leftarrow}{\partial_t}a^3\ol{T_k^{(n)}}T_l^{(n)}
\stackrel{\rightarrow}{\partial_t}\r] (\widetilde{Ef_2})(t,\vec{l})
\phi_{\vec{l}}(\vec{y}) =\\
&=&\int_{\Sigma_t}d^3y\sqrt{|h(t,\vec{y})|}
\overline{\l[\dm \ph(\vec{y})(\widetilde{Ef_1})(t,\vec{k})\l(
\frac{\dot{T}_k^{(n)}}{T_k^{(n)}}-\stackrel{\leftarrow}{\partial_t}\r)\r]}\\
& &\cdot \l[\int d\mu(\vec{l})a^{3}|T_l^{(n)}|^2\l(
\frac{\dot{T}_l^{(n)}}{T_l^{(n)}}-\stackrel{\rightarrow}{\partial_t}\r)
(\widetilde{Ef_2})(t,\vec{l}) \phi_{\vec{l}}(\vec{y})\r].
\end{eqnarray*}
Finally, noting that $E'=-E$, we can abbreviate the result in the form
\beqa
\Lambda_n^{(2)}(f_1,f_2)&=&(P_n Ef_1, A_n P_n Ef_2)_{L^2({\Sigma_t})}
\;\mbox{or} \nonumber\\
\Lambda_n^{(2)}(x_1,x_2) &=& -\int_{\Sigma_t}d^3y \sqrt{|h(t,\vec{y})|}\,
\ol{E(x_1;t,\vec{y})\stackrel{\leftarrow}{P_n}}A_n
\stackrel{\rightarrow}{P_n} E(t,\vec{y};x_2), \label{3.51}
\eeqa
where\footnote{Remember that \rf{3.51} is a somewhat sloppy
  notation. The product of the two distributions is properly defined by
  localization around $\Sigma_t$ and convolution of the Fourier
  transforms as in \rf{1.20} or \rf{3.18}. This is the reason why one
  has to know $T_k^{(n)}(t)$ in a whole (infinitesimal) neighborhood
  of $\Sigma_t$.}
$A_n,P_n$ shall denote the operators
\beqa
(A_nf)(t,\vec{y}) &:=& \dm a^{3}(t) |T_k^{(n)}(t)|^2
\tilde{f}(t,\vec{k})\ph(\vec{y}) \nonumber\\
(P_n f)(t,\vec{y}) &:=& (B_n(t)-\partial_t)f(t,\vec{y})\label{3.52}\\
(B_nf)(t,\vec{y}) &:=& \dm  \frac{\dot{T}_k^{(n)}}{T_k^{(n)}}
\tilde{f}(t,\vec{k}) \ph(\vec{y})\nonumber\\
\tilde{f}(t,\vec{k}) &=& \int_\Sigma d^3\sigma_z\,\ol{\ph(\vec{z})}
f(t,\vec{z})\;\mbox{for}\;f\in {\cal C}^\infty(\M).\nonumber
\eeqa
We recognize with satisfaction that the two-point function \rf{3.51}
is exactly of the form \rf{3.19} treated in connection with the
ultrastatic ground state, therefore we can apply Theorem~\ref{theorem3.9}
if we make sure that the operators $A_n$ and $P_n$ satisfy the
sufficient conditions of this theorem.
So we have to investigate the properties of $|T_k^{(n)}(t)|^2$ and
$\frac{\dot{T}_k^{(n)}}{T_k^{(n)}}$ as
operators. By the defining equ.~\rf{3.45},
\beqa
a_n(t,k)&:=& |T_k^{(n)}(t)|^2 = (a^3 2\Omega_k^{(n)}(t))^{-1}\nonumber\\
b_n(t,k) &:=& \frac{\dot{T}_k^{(n)}(t)}{T_k^{(n)}(t)}=
-\frac{3}{2}\frac{\dot{a}(t)}{a(t)}-\frac{1}{2}\frac{\dot{\Omega}_k^{(n)}
(t)}{\Omega_k^{(n)}(t)}-i\Omega_k^{(n)}(t) \label{3.53}
\eeqa
{\it on} a hypersurface $\Sigma_t$. Since $T_k^{(n)}(t)$ is a solution
of equation~\rf{3.34} we have (leaving for a moment the $k$'s and
$n$'s away)
\begin{eqnarray*}
\l(-\partial_t-3\frac{\dot{a}}{a}-\frac{\dot{T}}{T}\r)\l(\frac{\dot{T}}{T}
-\partial_t\r) &=& -\frac{\ddot{T}}{T}+\frac{\dot{T}^2}{T^2}-\frac{\dot{T}}{T}
\partial_t +\partial_t^2-3\frac{\dot{a}}{a}\frac{\dot{T}}{T}+3\frac{\dot{a}}
{a}\partial_t - \frac{\dot{T}^2}{T^2}+\frac{\dot{T}}{T}\partial_t\\
&=&\partial_t^2+3\frac{\dot{a}}{a}\partial_t-\frac{1}{T}(\ddot{T}+3\frac
{\dot{a}}{a}\dot{T}) \\
&=&\partial_t^2+3\frac{\dot{a}}{a}\partial_t +\omega^2 \\
&=& \Box_g+\mu^2,
\end{eqnarray*}
hence the operator $Q$ of Theorem~\ref{theorem3.9} reads in this case
\beqa
(Q_n f)(t,\vec{y}) &=& \dm\,q_n(t,k)\tilde{f}(t,\vec{k})\ph(\vec{y})\nonumber\\
q_n(t,k) &=& \partial_t+3\frac{\dot{a}}{a}+\frac{\dot{T}_k^{(n)}}{T_k^{(n)}}
\nonumber\\
&=& \partial_t+\frac{3}{2}\frac{\dot{a}}{a}-\frac{1}{2}\frac{\dot
{\Omega}_k^{(n)}}{\Omega_k^{(n)}} -i\Omega_k^{(n)}. \label{3.54}
\eeqa
We summarize the properties of $A_n,B_n$ and $Q_n$ in a lemma:
\begin{lemma}\label{lemma3.18}
For all $n\in{\bf N}_o$:\\
i) $A_n(t)\in L^{-1}_{1,0}$,
it is an elliptic pseudodifferential operator.\\
ii) $B_n(t)\in L^1_{1,0}$.\\
iii) $Q_n(t) \in L^1_{1,0}$ with principal symbol $q(t;\xi)$ such that
$q^{-1}(0)\setminus \{0\}\subset \{(x,\xi)\in
T^*\M;\;\xi_0>0\}$.
\end{lemma}
{\sc Proof}:\\
Looking at the iterative definition \rf{3.47} of $\Omega_k^{(n)}(t)$
and noting that
\begin{eqnarray*}
\dot{\omega}_k &=& \frac{\dot{a}}{a}\frac{\mu^2-\omega_k^2}{2\omega_k}\\
\ddot{\omega}_k &=&
\l(\frac{\dot{a}}{a}\r)^2\frac{\mu^4+2\mu^2\omega_k^2-3\omega_k^4}{4
\omega_k^3} +\frac{\ddot{a}}{a}\frac{\mu^2-\omega_k^2}{2\omega_k}\quad
\mbox{etc.}
\end{eqnarray*}
we see that $\Omega_k^{(n)}$ and $\dot{\Omega}_k^{(n)}/\Omega_k^{(n)}$
depend on $k$ only as rational functions of $\omega_k =(E(k)/a^2+\mu^2)^{1/2}$.
The analysis of the asymptotic behaviour of these functions
has been carried out by L{\"u}ders and Roberts
 \cite{LR90}. From their work one can conclude that
\begin{eqnarray}
\frac{\dot{\Omega}_k^{(n)}}{\Omega_k^{(n)}}&\in& S^0_{1,0} \nonumber\\
\Omega_k^{(n)}&\in& S^1_{1,0}\;\mbox{with leading term}\;
\omega_k\label{3.54a}
\end{eqnarray}
(for all $n\in {\bf N}_o$ and always uniformly on a bounded interval
in $t$).
Therefore, for $\kappa=0$ (in which case $\ph(\vec{y})=\frac{1}{(2\pi)^{3/2}}
e^{i\vec{k}\vec{y}}$), the lemma is proven, whereas for $\kappa =\pm
1$ it remains to be shown that $A_n,B_n, Q_n$ are indeed
pseudodifferential operators.\\
Let us consider as a prototype
\[ (Df)(t,\vec{y}) :=\dm\,\omega_k(t)\tilde{f}(t,\vec{k})\ph(\vec{y}).\]
Localizing in a coordinate neighborhood around $\vec{y}\in
\Sigma_t$ and denoting the Fourier transform (w.r.t.~$e^{i\vec{y}\vec{\xi}}$)
of $f$ by $\hat{f}$ we have
\begin{eqnarray*}
(Df)(t,\vec{y}) &=& \dm\,\omega_k(t)\ph(\vec{y})\int_\Sigma d^3\sigma_z\,
\ol{\ph(\vec{z})}f(t,\vec{z})\\
&=&\frac{1}{(2\pi)^{3/2}}\int_{{\bf R}^3}d^3\xi\,\hat{f}(t,\vec{\xi})
\dm\,\omega_k(t)\ph(\vec{y})\int_\Sigma d^3\sigma_z \,e^{i\vec{z}\vec{\xi}}
\ol{\ph(\vec{z})}\\
&=& \frac{1}{(2\pi)^{3/2}}\int_{{\bf R}^3}d^3\xi\,\hat{f}(t,\vec{\xi})
e^{i\vec{y}\vec{\xi}}\l\{e^{-i\vec{y}\vec{\xi}}\dm\,\omega_k(t)\ph(\vec{y})
\int_\Sigma d^3\sigma_z\,\ol{\ph(\vec{z})}e^{i\vec{z}\vec{\xi}}\r\}\\
&=:&\frac{1}{(2\pi)^{3/2}}\int_{{\bf R}^3}d^3\xi\,\hat{f}(t,\vec{\xi})
e^{i\vec{y}\vec{\xi}} d(t,\vec{y},\vec{\xi}).
\end{eqnarray*}
This is a pseudodifferential operator if $d(t,\vec{y},\vec{\xi})$
is a symbol. Using the fact that the $\ph(\vec{y})$ are a complete set
of eigenfunctions of the Laplace-Beltrami operator \rf{3.33} on
$\Sigma_t$ (which is selfadjoint w.r.t.~$d^3 \sigma_z$) we obtain
\begin{eqnarray*}
d(t,\vec{y},\vec{\xi}) &=& e^{-i\vec{y}\vec{\xi}}\dm\,\omega_k(t)
\ph(\vec{y})\int_\Sigma d^3\sigma_z\,\ol{\ph(\vec{z})}e^{i\vec{z}\vec{\xi}}\\
&=&e^{-i\vec{y}\vec{\xi}}\dm\,\ph(\vec{y})\int_\Sigma d^3\sigma_z\,
(-^{(3)}\Delta_z+\mu^2)^{1/2}\ol{\ph(\vec{z})}e^{i\vec{z}\vec{\xi}}\\
&=& e^{-i\vec{y}\vec{\xi}}\dm\,\ph(\vec{y})\int_\Sigma d^3\sigma_z\,
\ol{\ph(\vec{z})}(-^{(3)}\Delta_z+\mu^2)^{1/2} e^{i\vec{z}\vec{\xi}}\\
&=& e^{-i\vec{y}\vec{\xi}}(-^{(3)}\Delta_y+\mu^2)^{1/2} e^{i\vec{y}\vec{\xi}}\\
&=&\l\{\frac{|\vec{\xi}|^2}{a^2}
\l[1-\kappa \l(\frac{\vec{y}\vec{\xi}}{\xi}\r)^2
\r] +i\frac{|\vec{\xi}|}{a^2}3\kappa \frac{\vec{y}\vec{\xi}}{\xi}
+\mu^2\r\}^{1/2},
\end{eqnarray*}
which is a symbol of order 1 (note that $|\vec{y}|< 1$ for $\kappa =+1$),
 and consequently
$D\in L^1_{1,0}(\Sigma_t)$. A principal symbol of $D$ is given by
\[ \frac{|\vec{\xi}|}{a}\l[1-\kappa \l(\vec{y}\vec{\xi}/\xi\r)^2
\r]^{1/2}. \]
Now the lemma follows when we remember that $a_n$ and $b_n$ are
rational expressions in $\omega_k$ with leading terms $(a^3 2\omega_k)^{-1}$
resp.~$-i\omega_k$ and the correct asymptotic properties \rf{3.54a}.
\hfill ${\bf\Box}$\\  \\
Applying Theorem~\ref{theorem3.9} we can conclude that \rf{3.51}
has indeed wavefront set of an Hadamard state for all $n\in {\bf
  N}_o$. This proves the theorem.\hfill${\bf \Box}$\\
{\bf Remark:}\\
There are no obvious obstacles to extending our analysis to
spacetimes which are homogeneous, but not necessarily isotropic
(like the Bianchi-I-spacetime), or to the case of a Klein-Gordon
field coupled to the scalar curvature, i.e.~a field equation of the
form
\[ (\Box_g+\mu^2 +\xi R)\Phi=0.\]
\section{A counterexample}\label{section2.6}
The states we have presented in sections \ref{section2.4} and
\ref{section2.5} are constructed on very special spacetimes: in the
one case the spacetime possesses a static Killing vectorfield, in the
other there exists a preferred foliation of the spacetime into
homogeneous Cauchy surfaces such that the wave equation separates into
time- and space-dependent parts. It immediately arises the question
whether physical states can also be constructed on arbitrarily curved
globally hyperbolic spacetimes. Before we give our own solution of
this problem (in the next section) let us investigate
two proposals existing in
the literature for such a general construction, namely the method of
``Hamiltonian diagonalization'' (see the references in
\cite{Fulling79}) and the construction of ``energy states''
\cite{AM75,Chmielowski94}. Instead of presenting these constructions
in detail we want to pick out one particular example of a state that
lies in both classes and show that this state is in general physically
{\it not} acceptable.\\
Let $(\M,g)$ be a globally hyperbolic spacetime possessing a complete
Cauchy surface $\Sigma$
(with volume element $d^3\sigma$) and $A:=\ol{-\Delta
  +\mu^2}$ the (closure of the) Laplace-Beltrami operator on $\Sigma$
(with $\mu>0$).
Let $(\Gamma,\sigma)$ be the phase space of initial data of the
Klein-Gordon field on $\Sigma$ (equ.~\rf{3.9}). The idea is to mimic
the construction \rf{3.10} of the ultrastatic groundstate w.r.t.~the
Cauchy surface $\Sigma$, i.e.~we define a one-particle Hilbertspace
structure on $\Gamma$ by
\beqa
k^{\Sigma}:\Gamma &\to& {\cal H}:=L^2_{\bf C} (\Sigma,
d^3\sigma)\nonumber\\
(f,p) &\mapsto& \frac{1}{\sqrt{2}}(A^{1/4}f - iA^{-1/4}p).\label{3.55}
\eeqa
Of course, since $A$ depends on the chosen Cauchy surface each choice
of $\Sigma$ yields a different state, nevertheless \rf{3.55} is a well
defined one-particle Hilbertspace structure and therefore defines an
(in general not stationary) quasifree state of the Klein-Gordon
quantum field on $(\M,g$). So straightforward this construction may
appear, it does in general not yield reasonable physical states. To
show this we put the state \rf{3.55} on a closed Robertson-Walker
spacetime and prove that it does not lie in the folium of an Hadamard
state:
\begin{thm}\label{theorem3.19}
Let ($\M^+,g$) be the closed Robertson-Walker spacetime \rf{3.28}
($\kappa=+1$), $\Sigma_t$ a homogeneous Cauchy surface of $\M^+$ and
$\omega_t$ the quasifree state of the Klein-Gordon quantum field
defined
by $k^{\Sigma_t}$ (equ.~\rf{3.55}).\\
Then, $\omega_t$ is not quasiequivalent to an Hadamard state.
\end{thm}
{\sc Proof}:\\
The two-point function of \rf{3.55} reads
\begin{eqnarray}
\lambda^{(2)}( F_1,F_2) &=& \frac{1}{2}\la (A^{1/2} f_1-ip_1),(f_2
-iA^{-1/2}p_2)\ra_\Ht\nonumber\\
&=& \frac{1}{2}\la F_1, \l(\begin{array}{cc} A^{1/2} &-i\\ i& A^{-1/2}
                       \end{array}\r)F_2\ra_\Ht\nonumber\\
&=&\frac{1}{2}\l\la {f_1\choose a^3 p_1},\l(\begin{array}{cc}
A^{1/2} & -\frac{i}{a^3}\\\frac{i}{a^3}& \frac{1}{a^6}A^{-1/2}
\end{array}\r) {f_2\choose a^3 p_2}\r\ra_\Ht\nonumber\\
&=& \frac{1}{2}\dm\,\l\la \ol{{\tilde{f}_1\choose a^3\tilde{p}_1}},
\l(\begin{array}{cc} \omega_k & -\frac{i}{a^3}\\ \frac{i}{a^3}&
\frac{1}{a^6 \omega_k}\end{array}\r) {\tilde{f}_2\choose a^3 \tilde{p}_2}
\r\ra, \label{3.56}
\end{eqnarray}
where $f_i,p_i\in \co{\Sigma_t},\;\Ht:=L^2_{\bf
  C}(\Sigma_t,d^3\sigma),\; \omega_k(t)$ is given by \rf{3.35} and
the
(generalized) Fourier transform by \rf{3.38}. \rf{3.56} is of the form
\rf{3.40}, \rf{3.41} of a homogeneous, isotropic Fock state with
$p(k):=\sqrt{\omega_k(t)/2},\;q(k):=i(a^3
\sqrt{2\omega_k(t)})^{-1}$.\\
Let us compare this with the adiabatic vacuum state of order 0
(Definition \ref{dfn3.12}), where
\begin{eqnarray*}
 q^{(0)}(k) &=& W_k^{(0)}(t) = a^{-3/2} (2\omega_k)^{-1/2}
e^{-i\int_{t_o}^t \omega_k(t')\,dt'}\\
p^{(0)}(k) &=& a^3 \dot{W}_k^{(0)}(t) = - a^3\l[i\omega_k
+\frac{\dot{a}}{a}+
\frac{\dot{a}}{a}\frac{\mu^2}{2\omega_k^2}\r]W_k^{(0)}(t),
\end{eqnarray*}
and compute the Bogoliubov coefficient $\beta(k)$ between these two
states according to equ.~\rf{3.49}:
\[|\beta(k)| = \frac{1}{2\omega_k}\frac{1}{a^{3/2}}\l|\frac{\dot{a}}{a}\r|
\l[1+\frac{\mu^2}{2\omega_k^2}\r].
\]
Now, since $\omega_k^2=k(k+2)/a^2 +\mu^2$, we observe that
\[ \sum_{k=0}^\infty |\beta(k)|^2 (k+1)^2 =\infty, \]
for $\dot{a} \not= 0$, therefore, by Lemma \ref{lemma3.13}, these two
states cannot be quasiequivalent. Since the adiabatic vacuum state of
order 0 is an Hadamard state (Theorem \ref{theorem3.15}) our state
\rf{3.55} does not lie in the folium of Hadamard states. \hfill{$\bf\Box$}\\
{\bf Remark:}\\
Of course, on certain spacetimes (e.g.~the ultrastatic spacetimes, as
we have seen in section \ref{section2.4}) \rf{3.55} may be a well
behaved Hadamard state. The point we want to make here is just that
the seemingly very elegant and
general construction of \rf{3.55} (and similar ones
existing in the literature) does {\it in general} (on an arbitrarily curved
spacetime) not lead to a physical state. \\
Let us remark that this counterexample does not come as a surprise,
it only confirms the criticism already raised by Fulling
\cite{Fulling79} against the method of Hamiltonian diagonalization
(where states are constructed by selecting ad hoc the positive frequencies
on a single Cauchy surface as in our example),
which also applies to the states constructed in \cite{AM75}
and \cite{Chmielowski94}.
\section{Construction of Hadamard states}\label{section2.7}
We learnt from the example in the last section that the ``positive
frequencies'' in a non-stationary spacetime cannot be fixed on a
Cauchy surface, but must be dynamically determined off the Cauchy
surface. We saw in the proofs of Theorems \ref{theorem3.9} and
\ref{theorem3.15} that this is achieved by a separation of the
Klein-Gordon
operator into first-order factors that project out the correct
positive frequencies from the fundamental solution in an
(infinitesimal) neighborhood of the Cauchy surface. This was trivial
in the ultrastatic case where the Laplace-Beltrami operator is
time-independent. In the case of the adiabatic vacua an exact
factorization of the wave operator was accomplished by separating off
the time dependence, using a solution of the ordinary differential
equation \rf{3.34} and imposing initial conditions that enforce the
correct asymptotic behaviour.\\
On an arbitrarily curved globally hyperbolic spacetime such a method
is no longer possible. In this section we will present a general technique
for constructing Hadamard states by a local factorization of the wave operator
with the help of pseudodifferential operators.\\
First we give a clever parametrization of Fock states due to Deutsch
and Najmi \cite{DN83}:
\begin{thm}\label{theorem3.20}
Let ($\M,g$) be a globally hyperbolic spacetime with Cauchy surface
$\Sigma$.\\ Let ($\Gamma,\sigma$) be the phase space of initial data on
$\Sigma$ of the Klein-Gordon field (see section 3.1).\\
Let $R$ be a symmetric and $I$ a symmetric, positive and
invertible operator on $L^2_{\bf R}(\Sigma, d^3\sigma)$.\\
Then, with ${\cal H}= L^2_{\bf C}(\Sigma,d^3\sigma)$,
\beqa
k^\Sigma: \Gamma &\to& {\cal H} \nonumber\\
(f,p) &\mapsto& (2I)^{-1/2}\l[ (R-iI)f-p\r] \label{3.58}
\eeqa
is a one-particle Hilbertspace structure and defines a Fock state.
\end{thm}
{\sc Proof}:\\
For $F_i=(f_i,p_i)\in\Gamma,\;i=1,2,$
\begin{eqnarray*}
2 {\rm Im}\la k^\Sigma F_1,k^\Sigma F_2\ra_{\cal H} &=&
{\rm Im} \la (R-iI)f_1-p_1,I^{-1}\l[(R-iI)f_2-p_2\r]\ra_{\cal H}\\
&=& \la If_1,I^{-1}(Rf_2-p_2)\ra_{\cal H}-\la Rf_1-p_1,f_2\ra_{\cal H} \\
&=& -\la f_1,p_2\ra_{\cal H} +\la p_1,f_2\ra_{\cal H}\\
&=& \sigma(F_1,F_2),\\
2{\rm Re}\la k^\Sigma F_1, k^\Sigma F_2\ra_{\cal H} &=&
\la (Rf_1-p_1), I^{-1} (Rf_2-p_2)\ra_{\cal H}+\la I f_1,f_2\ra_{\cal
  H}\\
&=& \la (Rf_1-p_1),I^{-1}(Rf_2-p_2)\ra_{\cal H}+\la f_1,I f_2\ra_{\cal
  H}\\
&=:& 2\mu(F_1,F_2).
\end{eqnarray*}
$\mu$ is a scalar product since $I$ is positive and symmetric. \rf{2.10} is
automatically satisfied because
\begin{eqnarray*}
|{\rm Im}\la u,v\ra|^2 &\leq& |\la u,v\ra|^2 \leq \la u,u\ra \la v,v\ra\\
&=& ({\rm Re}\la u,u\ra +i{\rm Im}\la u,u\ra)({\rm Re}\la v,v\ra
+i{\rm Im} \la v,v\ra)\\
&=& {\rm Re}\la u,u\ra {\rm Re}\la v,v\ra
\end{eqnarray*}
is fulfilled for any scalar product $\la \cdot,\cdot\ra$.
Since ${\cal H}$ is the completion of $k^\Sigma \Gamma$,
$k^\Sigma$ describes a Fock state with one-particle Hilbertspace
${\cal H}$.\hfill${\bf \Box}$\\
    \\
As usual, we calculate the two-point function in the representation
\rf{3.58}:
\beqa
\Lambda_\Sigma^{(2)}(h_1,h_2) &=& \la k^\Sigma(\rho_oEh_1,\rho_1
Eh_1),k^\Sigma(\rho_oEh_2,\rho_1 Eh_2)\ra_{\cal H}\nonumber\\
&=& \frac{1}{2}\l\la I^{-1/2}\l[(R-iI)\rho_oEh_1-\rho_1 Eh_1\r],
I^{-1/2}\l[(R-iI)\rho_oEh_2-\rho_1Eh_2\r]\r\ra_{\cal H}\nonumber\\
&=& \frac{1}{2}\l\la (R-iI-n^\alpha \nabla_\alpha)Eh_1,
I^{-1}(R-iI-n^\alpha\nabla_\alpha)Eh_2\r\ra_{L^2_{\bf
    C}(\Sigma,d^3\sigma)}\label{3.59}
\eeqa
for $h_i\in \D{\M},\;i=1,2$, where $n^\alpha$ is the unit-normalfield
on $\Sigma$ and $E$ the fundamental solution of the Klein-Gordon
equation in $(\M,g)$. Comparing with the expressions
\rf{3.51}--\rf{3.53} we note that \rf{3.59} is of the same form as the
two-point function of the adiabatic vacua even if $(\M,g)$ is an
arbitrary spacetime. With this simple representation of Fock states
and the adiabatic vacua as a guiding example at
hand we have now a clear picture of how one can construct Hadamard
states on arbitrary curved spacetimes: we have to look for
pseudodifferential operators $R$ and $I$ (with the properties stated
in Theorem \ref{theorem3.20}, and $I$ elliptic) such that the
Klein-Gordon operator $\Box_g+\mu^2$ can be factorized into
$Q(R-iI-n^\alpha\nabla_\alpha)$ where $Q$ is a pseudodifferential
operator having the property stated in Theorem \ref{theorem3.9}.
Of course, in general we cannot expect to obtain such a factorization
exactly, but locally it can always be arranged modulo ``smoothing operators''
in $L^{-\infty}(\Sigma)$. This is sufficient for our argument in the
proof of Theorem \ref{theorem3.9} as the following little lemma shows:
\begin{lemma}\label{lemma3.21}
Let $R_1,R_2\in L^{-\infty}(\Sigma)$.\\
Then $(R_1 n^\alpha \nabla_\alpha +R_2)E$ is smooth.
\end{lemma}
{\sc Proof}:\\
Let us take a local coordinate system $X\subset \M$, where $\Sigma$ is
given by $x^0=0$ and let $R:=R_1\partial/\partial x^0 +R_2$.\\
We write
\beq
RE=R\chi(D) E +R(1-\chi(D))E\label{3.60}
\eeq
where $\chi(\xi)\in S^0_{1,0}({\bf R}^4)$ has support only in a small conic
neighborhood of the $\xi_0$-axis where no singular direction of $E$
lies and $\chi =1$ at infinity in a second, smaller conic neighborhood,
and $\chi(D)$ denotes the pseudodifferential operator with symbol
$\chi(\xi)$.\\
Then $R\chi(D)$ and $R(1-\chi(D))$ are 4-dim. pseudodifferential
operators and we can apply Theorem \ref{theorem1.15} to calculate the
wavefront set of \rf{3.60}. The first term is smooth since no singular
directions of $R\chi(D)$ and $E$ coincide, the second term is smooth
since $R(1-\chi(D))\in L^{-\infty}(X)$ (and $WF_{\M}(E)=\emptyset$ as
we have seen in the proof of Lemma \ref{lemma3.8}).
\hfill${\bf\Box}$\\
   \\
For the following let us fix a spacelike Cauchy surface $\Sigma$ of
($\M,g$) with induced metric $h_{ij}$ and unit-normalfield
$n^\alpha$. For reasons of notational simplicity  we work in {\bf Gau{\ss}ian
normal coordinates} in a neighborhood of $\Sigma$, i.e.~we choose a
local coordinate patch $S\subset\Sigma$ and consider for each point
$p=(x^1,x^2,x^3)\in S$ the geodesic that starts from $p$ with tangent
$\pm n^\alpha$. Then, each point in the neighborhood of $S$ can be
labeled by $(t,x^1,x^2,x^3)$ where $t$ is the proper time along the
geodesic on which it lies ($t< (>) 0$ for points that lie before
(after) $\Sigma$). The geodesics may cross after some time, but for
each $S$ there is certainly a finite interval $[-T,T]$ in which the
such constructed Gau{\ss}ian normal coordinates are well defined (for
details see \cite[p.42]{Wald84}). The hypersurface $S$ is then given
by $t=0$, the metric $g$ reads in these coordinates
\[ g_{\mu\nu} = \l(\begin{array}{cc}1& \\ & -h_{ij}(t,\vec{x})
\end{array}\r) \]
and the Klein-Gordon operator reduces to
\beqa
P\equiv \Box_g+\mu^2 &=& \frac{1}{\sqrt{|g|}}\partial_\mu(\sqrt{|g|}
g^{\mu\nu}\partial_\nu\cdot )+\mu^2\nonumber\\
&=& \partial_t^2 +K(t,\vec{x}) \partial_t - ^{(3)}\Delta_h+\mu^2\nonumber\\
&=& \frac{1}{\sqrt{h}}\partial_t(\sqrt{h}\partial_t \cdot)-
\frac{1}{\sqrt{h}}\partial_i(\sqrt{h}h^{ij} \partial_j\cdot)+\mu^2,
\label{3.61}
\eeqa
where $^{(3)}\Delta_h$ is the (time-dependent) Laplace-Beltrami
operator on $S$ and
\[ K(t,\vec{x}) =\partial_t \ln  \sqrt{h}=\nabla_\alpha n^\alpha =
g^{\alpha\beta}K_{\alpha\beta} \]
is the trace of the extrinsic curvature $K_{\alpha\beta}$ of the hypersurface
$t=const$.\\
In what follows we use the {\bf notation} $a(t,x,D)$ for the
pseudodifferential operator with symbol $a(t,x,\xi)$, we write
$a(t,x,\xi)\in S^m$ if $a(t,x,\xi)$ and $\partial_t^l a(t,x,\xi)
\in S^m_{1,0}(S\times{\bf R}^3), l=1,2,\ldots,$
uniformly in $t\in [-T,T]$ (similarly $a(t,x,D)\in L^m$), and here
is always $x\in S, \xi\in {\bf R}^3$ and $t\in[-T,T]$.\\
 We want to factorize \rf{3.61} into
\[ P=P_1\circ P_2 -r_1(t) \;\mbox{on}\;[-T,T] \]
with $r_1(t)\in L^{-\infty}$.
The principal symbol of $P$ is
\[ p(t,x,\tau,\xi) :=-\tau^2 +h^{kl}(t,x)\xi_k\xi_l\]
and its characteristic roots $\tau_\pm$ are given by
\beq
p(t,x,\tau_\pm,\xi)=0 \Leftrightarrow
\tau_\pm=\pm
\lambda(t,x,\xi):=\pm\l(h^{kl}(t,x)\xi_k\xi_l\r)^{1/2} \in S^1.
\label{3.62}
\eeq
It follows
\beq
|\lambda(t,x,\xi)|=\l(h^{kl}\xi_k\xi_l\r)^{1/2}\geq
C|\xi|\;\mbox{for}\;|\xi|\geq M \label{3.63}
\eeq
for some positive constants $C$ and $M$.
Let us modify the symbol $(h^{kl}\xi_k\xi_l)^{1/2}$ for $|\xi|<M$ such that
$|\lambda(t,x,\xi)|\geq\epsilon >0$ (by Lemma \ref{lemma1.2}d),
this changes $\lambda$ only modulo $S^{-\infty}$). We make the Ansatz
\beqa
P_1 &=& -a(t,x,D)-\frac{1}{\sqrt{h}}\partial_t \sqrt{h}\nonumber\\
P_2 &=& a(t,x,D)-\partial_t \label{3.64}
\eeqa
and determine $a$ by an asymptotic expansion of its symbol:
\beqa
a^{(n)}(t,x,\xi) &:=& \sum_{\nu=0}^{n} b^{(\nu)}(t,x,\xi)\nonumber\\
P_1^{(n)} &:=& -a^{(n)}(t,x,D)-\frac{1}{\sqrt{h}}\partial_t \sqrt{h}
\label{3.65}\\
P_2^{(n)} &:=& a^{(n)}(t,x,D)-\partial_t, \nonumber
\eeqa
where $b^{(\nu)}$ shall denote symbols in $S^{1-\nu}$.\\
For ${\bf n=0}$ we set
\beq
b^{(0)}(t,x,\xi) := -i \lambda(t,x,\xi)\quad \in S^1. \label{3.66}
\eeq
Then the principal symbol of $P_1\circ P_2 -P$ is
\[\lambda^2 -\tau\lambda+\lambda\tau-\tau^2-(-\tau^2
+h^{kl}\xi_k\xi_l) = 0\; (mod\:S^1),\]
i.e.~we have
\[P_1^{(0)}\circ P_2^{(0)}-P=r_1^{(0)}(t,x,D)\;\mbox{on}\;[-T,T]\]
with $r_1^{(0)}\in L^1$
($r_1^{(0)}(t,x,\xi)$ is an infinite asymptotic
sum deriving from the asymptotic expansion \rf{1.5b} of the product
$\lambda(t,x,D)\circ\lambda(t,x,D)$ which we do not want to give here).\\
Let us now iterate {\bf from $\bf n$ to $\bf n+1$} by assuming that
$b^{(\nu)}(t,x,\xi),\nu\leq n,$ are already determined such that
\[ P_1^{(n)}\circ P_2^{(n)}-P=r_1^{(n)}(t,x,D)\;\mbox{on}\;[-T,T] \]
for some $r_1^{(n)}\in L^{1-n}$.\\
We set
\beq
b^{(n+1)}(t,x,\xi) :=
-\frac{i}{2}\frac{r_1^{(n)}(t,x,\xi)}
{\lambda(t,x,\xi)}\quad\in S^{-n}.\label{3.67}
\eeq
This is well defined since we have modified $\lambda$ at small
values of $\xi$ such that $\lambda\not= 0$, and $b^{(n+1)}
\in S^{-n}$ because of \rf{3.63} and the assumption on $r_1^{(n)}$.
Then we have
\begin{eqnarray*}
P_1^{(n+1)}\circ P_2^{(n+1)}-P&=&(P_1^{(n)}\circ P_2^{(n)}-P)-
b^{(n+1)}(t,x,D)P_2^{(n)}\\
& & +P_1^{(n)} b^{(n+1)}(t,x,D)-b^{(n+1)}(t,x,D)b^{(n+1)}(t,x,D)\\
&=& r_1^{(n)}(t,x,D)-b^{(n+1)}(t,x,D)
(-i\lambda(t,x,D)-\partial_t)\\
& &+(i\lambda(t,x,D)-\frac{1}{\sqrt{h}}\partial_t \sqrt{h})
b^{(n+1)}(t,x,D)\;(mod\:L^{-n})
\end{eqnarray*}
and the principal symbol of this expression is
\begin{eqnarray*}
& &r_1^{(n)} -b^{(n+1)}(-i\lambda-i\tau)
+(i\lambda-i\tau)b^{(n+1)} =\\
&=& r_1^{(n)}+2i\lambda b^{(n+1)} =0\;(mod\:S^{-n}),
\end{eqnarray*}
i.e.~we have
\beq
P_1^{(n+1)}\circ P_2^{(n+1)}-P=r_1^{(n+1)}(t,x,D)
\;\mbox{on}\;[-T,T] \label{3.68}
\eeq
with $r_1^{(n+1)}\in L^{-n}$.\\
{}From Lemma \ref{lemma1.3} we can now conclude that
\beqa
s(t,x,\xi)&\sim& -\frac{1}{2i}\sum_{\nu=0}^\infty
\l[b^{(\nu)}(t,x,\xi)-\ol{b^{(\nu)}(t,x,-\xi)}\r]\quad
\in S^1\nonumber\\
r(t,x,\xi)&\sim& \frac{1}{2}\sum_{\nu=0}^\infty
\l[b^{(\nu)}(t,x,\xi)+\ol{b^{(\nu)}(t,x,-\xi)}\r]\quad
\in S^0 \label{3.69}
\eeqa
define symbols ($mod\:S^{-\infty}$) in the specified classes.
$s$ has principal symbol
\[-{\rm Im}\:b^{(0)}=\lambda(t,x,\xi)=\l(h^{kl}(t,x)\xi_k\xi_l\r)^{1/2}\]
which is elliptic and positive. Since the principal symbol is the same
for the adjoint (equ.~\rf{1.5d}), we can modify $s$ (smoothly in $t\in[-T,T]$)
modulo $S^{-\infty}$ such that
\beq
I(t):=\frac{1}{2}\l[s(t,x,D)+s(t,x,D)^t\r] \label{3.70}
\eeq
is a strictly positive, elliptic, symmetric pseudodifferential
operator in $S^1$ mapping real functions to real functions.
Similarly, we define
\beq
R(t):=\frac{1}{2}\l[r(t,x,D)+r(t,x,D)^t\r]\quad\in S^0\label{3.71}
\eeq
and claim that $I(t)$ and $R(t)$ realize via Theorem \ref{theorem3.20}
a quasifree Hadamard state. In fact, comparing the two-point function
\rf{3.59} with the one \rf{3.19b} dealt with in Theorem
\ref{theorem3.9}, we see that we only have to prove that the operator
$R-iI-\partial_t$ properly factorizes the wave operator:\\
The two-point function \rf{3.59} can be split into four terms
\beqa
\Lambda_\Sigma^{(2)}(h_1,h_2) &=& \frac{1}{2}\l\la(R-iI-\partial_t)Eh_1,
I^{-1}(R-iI-\partial_t)Eh_2\r\ra \nonumber\\
&=&\frac{1}{2}\l\la\l[\frac{1}{2}(r+r^t-is-is^t)-\partial_t\r]Eh_1,I^{-1}
\l[\frac{1}{2}(r+r^t-is-is^t)-\partial_t\r]Eh_2\r\ra \nonumber\\
&=& \frac{1}{8} \l\la
(r-is-\partial_t)Eh_1+(r^t-is^t-\partial_t)Eh_1,\r.\label{3.72}\\
& &\hspace{4cm}
\l.I^{-1}\l[(r-is-\partial_t)Eh_2+(r^t-is^t-\partial_t)Eh_2\r]\r\ra.
\nonumber
\eeqa
{}From equations \rf{3.64}--\rf{3.68} it follows after iteration
\[P_1(t)\circ P_2(t) =P + r_1(t,x,D)\;\mbox{on}\;[-T,T]\]
where
\begin{eqnarray*}
P_1(t) &=& -a(t,x,D)-\frac{1}{\sqrt{h}}\partial_t \sqrt{h}\\
P_2(t) &=& a(t,x,D)-\partial_t \\
a(t,x,\xi) &\sim&\sum_{\nu=0}^\infty b^{(\nu)}(t,x,\xi)
\sim r(t,x,\xi)-is(t,x,\xi)\\
r_1(t,x,D) &\in& L^{-\infty},
\end{eqnarray*}
and hence
\[(-r+is-\frac{1}{\sqrt{h}}\partial_t\sqrt{h})(r-is-\partial_t)=
P+r_1 \;\mbox{on}\;[-T,T].\]
By forming the adjoint one notes that this is equivalent to
\begin{eqnarray*}
& & -a\circ a -\frac{1}{\sqrt{h}}\partial_t\sqrt{h}a+a\partial_t = -\Delta
+\mu^2 +r_1 \\
&\Leftrightarrow& -a^t\circ a^t
-\frac{1}{\sqrt{h}}\partial_t\sqrt{h}a^t+a^t\partial_t = -\Delta
+\mu^2 +r_1^t\\
&\Leftrightarrow& (-r^t +is^t -\frac{1}{\sqrt{h}}\partial_t\sqrt{h})
(r^t-is^t-\partial_t) = P+r_1^t \;\mbox{on}\;[-T,T]
\end{eqnarray*}
with $r_1^t\in L^{-\infty}$. So let us define
\[Q_1:=-r+is-\frac{1}{\sqrt{h}}\partial_t\sqrt{h},
\quad Q_2:=-r^t+is^t-\frac{1}{\sqrt{h}}\partial_t\sqrt{h},\]
the principal symbol of both of them being
\[q(t,x,\xi_0,\xi)=i\lambda(t,x,\xi)-i\xi_0
=i\l(h^{kl}(t,x)\xi_k\xi_l\r)^{1/2}-i\xi_0,\]
i.e. $q^{-1}(0)\setminus\{0\}\subset \{(t,x;\xi_0,\xi)\in T^*([-T,T]
\times S);\;\xi_0>0\}$, as desired for the application of Theorem
\ref{theorem3.9}. Since
\begin{eqnarray*}
Q_1(r-is-\partial_t)E&=&(P+r_1)E=0\;(mod\:{\cal C}^\infty)\\
Q_2(r^t-is^t-\partial_t)E &=& (P+r_1^t)E =0\;(mod\:
{\cal C}^\infty)
\end{eqnarray*}
by Lemma \ref{lemma3.21}, the proof of Theorem \ref{theorem3.9} applies
also to this case where the wave operator factorizes only up to
smoothing operators.\\
Thus, summing up, we have proven the following theorem:
\begin{thm}\label{theorem3.22}
Let ($\M,g$) be a globally hyperbolic spacetime with spacelike Cauchy
surface $\Sigma$ and unit-normalfield $n^\alpha$.\\
In a neighborhood of $\Sigma$ let the pseudodifferential operators
$I$ and $R$ be given ($mod\:S^{-\infty}$) by the equations
\rf{3.69}--\rf{3.71}, \rf{3.66}--\rf{3.67} and \rf{3.62}.\\
Then
\[\Lambda_\Sigma^{(2)}(h_1,h_2):=\frac{1}{2}\l\la(R-iI-n^\alpha\nabla_\alpha)
Eh_1,I^{-1}(R-iI-n^\alpha\nabla_\alpha)Eh_2\r\ra_{L^2_{\bf
    C}(\Sigma,d^3\sigma)}\]
$(h_1,h_2\in \D{\M})$ defines a Hadamard Fock state of the
Klein-Gordon quantum field on $(\M,g)$.
\end{thm}
{\bf Remarks:}
\begin{enumerate}
\item Our construction depends on
\begin{itemize}
\item the choice of Cauchy surface $\Sigma$,
\item the choice of the vectorfield $n^\alpha$: in a general spacetime
  the choice of the normalfield on $\Sigma$ seems to be the most natural
  one, but on spacetimes with certain symmetries one might take other
  appropriate timelike vectorfields (on a stationary spacetime
  e.g.~the Killing field),
\item the modification of $s(t,x,\xi)$ in \rf{3.70}
  to make $I$ positive,
\item the choices of operators $R$ and $I$ which were only defined
  modulo $S^{-\infty}$.
\end{itemize}
\item If we break off the asymptotic expansion of the operators $s$ and
  $r$ in equ.~\rf{3.69} at a finite value of $\nu$, then our
  conjecture is that we obtain a state which is -- if not an Hadamard
  state -- at least locally quasiequivalent to one.
\item Our construction was purely local, we did not use any global
  properties of the spacetime ($\M,g$) apart from the existence of the
  fundamental solution $E$. Therefore, we can also construct local
  states in globally hyperbolic submanifolds of a non-globally
  hyperbolic spacetime $(\M,g$). We could e.g.~take a double-cone
region ${\cal N}$ with base $\Sigma$ such that ${\cal
  N}=D(\Sigma)$. Then, $({\cal N},g)$ is globally hyperbolic with
Cauchy surface $\Sigma$, the fundamental solution $E_{\cal N}$ of the
Klein-Gordon equation in ${\cal N}$ exists and our construction yields
states for observables localized in ${\cal N}$.
\end{enumerate}
Our construction is reminiscent of the adiabatic DeWitt-Schwinger expansion
of the Feynman propagator (see e.g.~\cite{BD82}) which is used for
regularizing the expectation value of the energy-momentum tensor, but
whereas there it is not clear that one ends up with a state (positivity
is not under control), one obtains here a manifestly positive
functional. It would be interesting to investigate whether our
treatment could be used to set up a similar regularization procedure
for the stress-energy tensor.\\
As we have already mentioned, not many explicit examples for Hadamard
states are known. The spacetimes most frequently considered
as backgrounds are the black hole spacetimes (Schwarzschild, Kerr,...).
Whereas for the Schwarzschild spacetime quite a lot is known (due to
the existence of a static Killing field) it seems that the only (rigorous)
result for the (non-stationary) Kerr spacetime is the statement
\cite{KW91} that no stationary Hadamard state can exist there. It should
have become
clear that our work supplies methods to deal with the problem of constructing
non-stationary states and it would be interesting to see how the
Hawking radiation comes out in this case.\\
Another possible direction for future research is the extension of the
whole story to spinor fields. But more interesting and promising seems
to be the investigation of the new wavefront set spectrum condition
proposed in \cite{BFK95} which is supposed to be a selection criterion
for interacting field theories. One could ask how a free state gets
perturbed when interaction is switched on and set up a perturbation
scheme within this frame. A possible application would be to
look how the thermal Hawking radiation changes in the presence of
interaction.\\
At any rate it should have become clear that the wavefront sets and
the mathematical microlocal techniques are the correct language to
describe physical phenomena of quantum fields in gravitational
background fields and we think that new and exciting results are to be
expected from these ideas.
\newpage
\centerline{\bf Acknowledgment}
I thank Prof. K. Fredenhagen for a careful reading of the manuscript
and useful discussions. I am particularly indebted to Marek
Radzikowski who made the results of his PhD thesis available prior to
publication and helped me during his visit in Hamburg a lot to
understand wavefront sets. These discussions finally lead to the
results of this work.\\
Thanks are also due to all my friends and colleagues in the
``algebraic group'', at the II. Institut, at DESY and outside, but in
particular to Martin K{\"o}hler and Rainer Verch for numerous
discussions and assistance in all mathematical or computer problems.\\
I feel deeply obliged to the Studienstiftung des deutschen Volkes
which supported me so generously over many years. Financial support
from the DFG is also gratefully acknowledged.\\
Finally I want to thank Cath\'erine for her encouragement and patience.
\begin{appendix}
\chapter{Notation and conventions}
In this appendix we want to fix our notation
and conventions used throughout this work.\\
$\Rn$: $n$-dim. real (Euclidean) space\\
$x=(x^1,\ldots,x^n),\;\xi=(\xi_1,\ldots,\xi_n)$: coordinates of $\Rn$
resp.~its dual space\\
$x\xi=x^1\xi_1+\ldots+x^n\xi_n$: scalar product in $\Rn$\\
$|\xi|=[\xi_1^2+\ldots+\xi_n^2]^{1/2}$\\
$\alpha = (\alpha_1,\ldots,\alpha_n)\in {\bf N}_o^n$: multiindex\\
$|\alpha|= \alpha_1+\ldots+\alpha_n$: length of multiindex $\alpha$\\
$\alpha ! = \alpha_1 !\cdot\ldots\cdot\alpha_n!$\\
$x^\alpha = (x^1)^{\alpha_1}\cdot\ldots\cdot(x^n)^{\alpha_n},\;
\xi^\alpha = (\xi_1)^{\alpha_1}\cdot\ldots\cdot(\xi_n)^{\alpha_n}$
for $x,\xi\in\Rn$\\
$D^\alpha_x = \l(\frac{1}{i}\d{x^1}\r)^{\alpha_1}\cdot\ldots\cdot
\l(\frac{1}{i}\d{x^n}\r)^{\alpha_n}$ with $i=\sqrt{-1}$\\
$D^\alpha_\xi = \l(\frac{1}{i}\d{\xi_1}\r)^{\alpha_1}\cdot\ldots\cdot
\l(\frac{1}{i}\d{\xi_n}\r)^{\alpha_n}$\\
$supp\:f$: the support of the function $f$, i.e.~the closure of the
set of points at which $f$ does not vanish\\
$X$ an open subset of $\Rn$\\
${\cal E}(X) \equiv {\cal C}^\infty(X)$: space of smooth (infinitely
differentiable) functions $f:X\to {\bf C}$ with locally convex
topology generated by the family of seminorms
$p_{n,K}(f):=\max_{x\in K}\sum_{|\alpha|\leq n}|D^\alpha_x f(x)|$
for compact subsets $K\subset X$.\\
${\cal E}'(X)$: dual space of ${\cal E}(X)$ w.r.t.~this topology,
i.e.~the continuous linear maps ${\cal E}(X)\to{\bf C}$. It is the
space of distributions with compact support.\\
$\D{X}\equiv \co{X}=\{f\in {\cal C}^\infty(X);\;supp\,f\;\mbox{compact}\}$:
space of testfunctions with the inductive limit topology induced
from ${\cal E}(K_i)$ for compact sets $K_i \to X$.\\
$\Dp{X}$: dual space of $\D{X}$, the space of distributions in $X$\\
${\cal S}(\Rn)$: Schwartz space of functions in ${\cal C}^\infty(\Rn)$
that are rapidly decaying, i.e.~for any pair of integers $n,N\geq0$
are the seminorms
$p_{n,N}(f)=\sup_{x\in\Rn}\l[(1+|x|)^N\sum_{|\alpha|\leq n}|D^\alpha_x
f(x)|\r]$ finite.\\
${\cal S}'(\Rn)$: the dual space of ${\cal S}(\Rn)$ w.r.t.~the
locally convex topology generated by these seminorms. It is the
space of tempered distributions.\\
Since $\D{\Rn} \subset {\cal S}(\Rn)\subset {\cal E}(\Rn)$ we have
${\cal E}'(\Rn)\subset {\cal S}'(\Rn)\subset \Dp{\Rn}$.\\
$singsupp\:u$: the singular support of $u\in\Dp{X}$, i.e.~the smallest
closed set in the complement of which $u$ is a ${\cal C}^\infty$-function.\\
The Fourier transformation $u(x)\mapsto
\hat{u}(\xi)=\frac{1}{(2\pi)^{n/2}} \int_\Rn e^{-ix\xi}u(x)\;d^nx$ is an
isomorphism of ${\cal S}(\Rn)$ onto ${\cal S}(\Rn)$, its inverse is
given by $\hat{u}(\xi)\mapsto u(x)=\frac{1}{(2\pi)^{n/2}}\int_\Rn
e^{ix\xi}\hat{u}(\xi)\;d^n\xi.$\\
${\cal E(M), \;D(M),\; E'(M),\;D'(M)}$ can be analogously defined on a
manifold $\M$ by localization,
whereas ${\cal S}$ and ${\cal S}'$ in general not.\\
$\M$ is a (4-dim.) paracompact ${\cal C}^\infty$-manifold endowed
with a Lorentzian metric $g$ of signature ($+---$). Units are chosen
such that $\hbar = c=G=k_B=1$.\\
$(\M,g)$ is time-orientable if there exists a smooth timelike
vector field on $\M$, i.e.~we can unambiguously distinguish the
future light cone from the past light cone throughout the manifold.\\
$(\M,g)$ is globally hyperbolic if it possesses a Cauchy surface $\Sigma$,
 i.e.~a 3-dim. hypersurface that is intersected by each inextendible
causal (null or timelike) curve in $\M$ exactly once.
Equivalently, ($\M,g$) possesses a foliation $\M={\bf R}\times \Sigma$
into hypersurfaces $\Sigma$ such that $\Sigma_t = \{t\}\times\Sigma$
is a spacelike Cauchy surface for ($\M,g$). In this work we always
understand a globally hyperbolic manifold as being time-orientable.\\
If $S$ is a subset of $\M$, we define $J^+(S)$ (resp.~$J^-(S)$)
to be the set of all points $x\in\M$ such that $x$ can be connected
to a point in $S$ by a future-directed (resp.~past-directed)
causal curve from $S$ to $x$ in $\M$, and $D^+(S)$ (resp.~($D^-(S)$)
the set of all points $x\in J^+(S)$ (resp.~$J^-(S)$) such that every
past- (resp.~future-) directed inextendible causal curve through $x$
passes through $S$. $D(S):=D^+(S)\cup D^-(S)$ is called the domain of
dependence of $S$.\\
A convex normal neighborhood in $\M$ is an open set $U\subset \M$
s.th.~for any two points $x_1,x_2\in\M$ there exists a unique
geodesic contained in $U$ which connects $x_1$ and $x_2$.\\
$\sigma(x_1,x_2)=\pm \l(\int_a^b\l|g_{\mu\nu}(x(\tau))\frac{d x^\mu}
{d \tau}\frac{d x^\nu}{d\tau}\r|^{1/2}\,d\tau\r)^2$ is the square
of the geodesic distance from $x_1$ to $x_2$ in a convex normal
neighborhood $U$, where $[a,b]\to\M,\tau\mapsto x(\tau)$ is
the unique geodesic curve in $U$ from $x_1$ to $x_2$ (+ ($-$) is
chosen if $x(\tau)$ is spacelike (timelike)).
\end{appendix}
%
%

%
\end{document}